\documentclass[a4paper, 11pt]{article}
\usepackage{amsthm,amsmath,amssymb,stackengine,comment}
\usepackage{amsfonts,mathrsfs,bm}
\usepackage{color}
\usepackage{enumerate}
\usepackage{booktabs,float,placeins}
\usepackage[flushleft]{threeparttable}
\usepackage{siunitx}
\usepackage{graphicx}
\usepackage{multirow}
\usepackage{diagbox}

\usepackage{pgfplots}
\usepackage[margin=1.09in]{geometry}

\usepackage[utf8]{inputenc}
\usepackage[T1]{fontenc}
\usepackage{microtype}

\usepackage{amsmath, amssymb, mathtools}
\usepackage{algorithm}
\usepackage{xcolor}
\definecolor{corrblue}{RGB}{74,114,165}
  \definecolor{zerogray}{RGB}{128,128,128}
  \definecolor{posred}{RGB}{168,79,61}
  \definecolor{negblue}{RGB}{74,114,165}
  \definecolor{panelbg}{HTML}{FAFAF8}
  \definecolor{uboxfill}{HTML}{F7E8E3}
  \definecolor{uboxline}{RGB}{168,79,61}
\definecolor{darkblue}{rgb}{0.0, 0.0, 0.55}
\definecolor{darkred}{rgb}{0.55, 0.0, 0.0}
\usepackage[colorlinks=true, 
            citecolor=darkblue, 
            linkcolor=black, 
            urlcolor=darkblue]{hyperref}

\usepackage{tikz}
\usetikzlibrary{arrows.meta,positioning,decorations.pathreplacing,calc}
\definecolor{rankRed}{RGB}{168, 79, 61}
\definecolor{rankBlue}{RGB}{74, 114, 165}
\definecolor{rankGrey}{RGB}{128, 128, 128}

\usepackage{enumitem} 
\usepackage{caption}
\usepackage{subcaption}

\usepackage{enumerate}
\usepackage{mathtools}
\usepackage{algorithm}
\usepackage{algorithmicx}
\usepackage[noend]{algpseudocode}
\usepackage{natbib}
\usepackage{url}
\usepackage{xr}

\newcommand{\blind}{1}

\theoremstyle{plain}

\newtheorem{assumption}{Assumption}

\newtheorem{theorem}{Theorem}[section]

\newtheorem{proposition}{Proposition}[section]
\newtheorem{lemma}{Lemma}[section]
\newtheorem{definition}{Definition}[section]
\newtheorem{remark}{Remark}
\theoremstyle{remark}

\DeclareMathOperator*{\argmin}{arg\,min}

\date{}

\newcommand{\R}{\mathbb{R}}

\newcommand{\Prob}{\mathbb{P}}
\newcommand{\expect}{\mathbb{E}}

\newcommand{\hbu}{\widehat{\bm{u}}}

\newcommand{\hbv}{\widehat{\bm{v}}}

\newcommand{\hbtheta}{\widehat{\bm{\theta}}}
\newcommand{\bthetastar}{\bm{\theta}^{*}}
\newcommand{\bu}{\bm{u}}
\newcommand{\bv}{\bm{v}}
\newcommand{\btheta}{\bm{\theta}}

\newcommand{\tbu}{\widetilde{\bm{u}}}
\newcommand{\tbv}{\widetilde{\bm{v}}}
\newcommand{\tu}{\widetilde{u}}

\newcommand{\tbtheta}{\widetilde{\btheta}}

\newcommand{\bz}{\bm{z}}

\newcommand{\W}{\bm{W}}
\newcommand{\K}{\bm{K}}
\newcommand{\Q}{\bm{Q}}

\newcommand{\bLambda}{\bm{\Lambda}}

\newcommand{\A}{\mathcal{A}}
\newcommand{\I}{\mathcal{I}}

\newcommand{\G}{\mathcal{G}}

\newcommand{\X}{\mathcal{X}}

\newcommand{\norm}[1]{\left\lVert#1\right\rVert}
\newcommand{\rank}{\mathrm{rank }}
\newcommand{\abs}[1]{\left\lvert#1\right\rvert}

\newcommand{\bDelta}{\bm\Delta}

\DeclareMathOperator{\diam}{diam}
\DeclareMathOperator{\diag}{diag}

\DeclareMathOperator{\supp}{supp}

\renewcommand{\arraystretch}{1.3}
\allowdisplaybreaks 

\begin{document}

	\def\spacingset#1{\renewcommand{\baselinestretch}%
		{#1}\small\normalsize} \spacingset{1}

	%%%%%%%%%%%%%%%%%%%%%%%%%%%%%%%%%%%%%%%%%%%%%%%%%%%%%%%%%%%%%%%%%%%%%%%%%%%%%%
	
	\if1\blind
	{
		\title{\bf High-dimensional covariate-assisted ranking}
		\author{
			{Yujia Liu\textsuperscript{1}, Yuanyuan Lin\textsuperscript{1}, Ruijian Han\textsuperscript{2} and Yiming Xu\textsuperscript{3} }\\ [0.1cm]
			{\small \textsuperscript{1}Department of Statistics and Data Science, The Chinese University of Hong Kong
			}\\
			{\small	 \textsuperscript{2} Department of Data Science and Artificial Intelligence, The Hong Kong Polytechnic University}\\
		    {\small	  \textsuperscript{3} Department of Mathematics, University of Kentucky}
			}
		\maketitle
	} \fi
	
	\if0\blind
	{
		\bigskip
		\bigskip
		\bigskip
		\begin{center}
			{\LARGE\bf High-dimensional covariate-assisted ranking}
		\end{center}
		\medskip
	} \fi
	
	\bigskip
	\begin{abstract}

Ranking data in modern applications often involves covariate vectors whose dimension grows with the number of objects. Identifying relevant covariates and integrating them with item-specific utilities can yield better predictive performance. In this work, we study a Plackett--Luce model that incorporates both item-specific utilities and high-dimensional covariate effects. Under a sparsity assumption on the covariate coefficients, we establish a necessary and sufficient condition for identifiability in terms of the Kruskal rank of the restricted covariate design. We then propose an $\ell_1$-regularized maximum likelihood estimator for the covariate coefficients. Under suitable design and connectivity conditions, we establish the existence and uniqueness of the estimator with probability tending to one, as well as its uniform consistency. With an additional minimum-signal condition, we further establish exact support recovery. Our analysis combines a hypergraph-based chaining technique with a profile-likelihood perturbation analysis, yielding estimation rates that explicitly capture the effects of the comparison hypergraph's connectivity and the underlying sparsity level. Numerical experiments using synthetic data and a horse racing dataset corroborate our theoretical findings.

	\end{abstract}
	
	\noindent%
	
	%\newpage
	%\spacingset{1.5}
	
%\tableofcontents
	
\section{Introduction}

Ranking data record the relative ordering of items and often reflect both item-specific and contextual effects.
In tennis matches, for instance, the outcome of a comparison between two players may depend not only on their abilities but also on factors such as court surface and weather conditions. Similar considerations arise in reinforcement learning from human feedback \citep{christiano2017deep, ouyang2022training, bai2022constitutional, rafailov2023direct, kaufmann2023survey, chaudhari2025rlhf}, recommendation systems \citep{liu2009learning}, and other sports tournaments \citep{cattelan2013dynamic, masarotto2012ranking}. Such data are commonly analyzed using classical ranking models that focus primarily on item-specific utilities or incorporate only limited covariate information. With the increasing availability of rich covariate data, effectively leveraging such information while accounting for item-specific effects offers opportunities to improve ranking and prediction. 

\begin{figure}[!htbp]
\centering
\includegraphics[width=\linewidth]{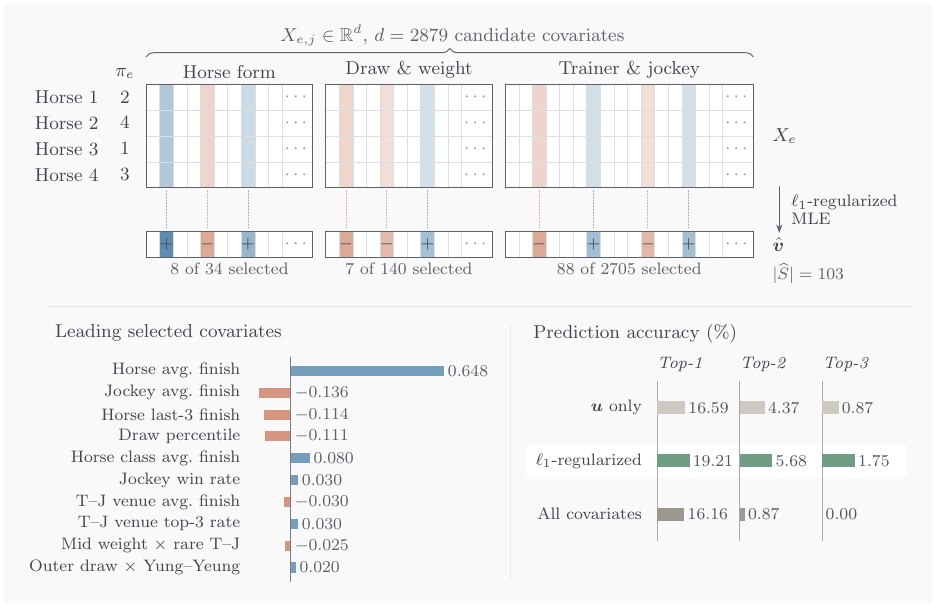}
\caption{Overview of the proposed framework. Rankings are modeled through a high-dimensional set of edge-dependent covariates, and the $\ell_1$-regularized MLE recovers the sparse subset that drives the outcomes (shaded; colour indicates sign).}
\label{fig:horseracing}
\end{figure}

In this article, we study a high-dimensional covariate-assisted Plackett--Luce (PL) model in which the number of covariates may grow with the number of compared objects and even exceed the number of observed comparisons. The classical PL model \citep{plackett1975analysis, luce1959individual} generates rankings based solely on item-specific utilities, while \citet{dong2026statistical} extend this framework by incorporating fixed-dimensional linear effects of comparison-dependent covariates; see Section~\ref{sec:21} for the mathematical formulation. A motivating application is horse racing, where ranking outcomes depend on both horse-specific abilities and numerous covariates related to race conditions and historical performance (Figure~\ref{fig:horseracing}). Selecting relevant covariates is crucial for both prediction and interpretation, as including all available covariates may lead to overfitting, while ignoring them may discard useful information. The growing covariate dimension poses challenges for identifiability, estimation, and variable selection, as covariate effects must be distinguished from item-specific utilities whose number also increases. We address these challenges by imposing sparsity on the covariate effects while leaving the item-specific utilities unregularized.

%\FloatBarrier

\subsection{Contributions}

Under a sparsity assumption, a natural approach is to estimate both the item-specific utilities and covariate coefficients jointly via an $\ell_1$-regularized maximum likelihood estimator (MLE), with regularization imposed only on the latter. We characterize the well-posedness of the model and establish the asymptotic theory for the regularized MLE. Our contributions are summarized below.

\begin{enumerate}

    \item We provide a necessary and sufficient condition for identifiability of the PL model with high-dimensional covariates under a sparsity assumption on the covariates. This condition takes the form of a Kruskal-rank condition on the covariate design after removing the utility component. It remains valid even when the covariate dimension exceeds the number of comparisons.

    \item We show that, with high probability, the $\ell_1$-regularized MLE exists uniquely, excludes all inactive covariates, and is uniformly consistent for both the item-specific utilities and covariate coefficients, with explicit convergence rates for each. Under an additional minimum-signal condition, the estimator also achieves exact support recovery.
    
\end{enumerate}

Our analysis combines the primal--dual witness framework \citep{wainwright2009sharp,ravikumar2010high} with a hypergraph-based chaining technique \citep{han2025unified} and substantially extends both. A central difficulty lies in the interaction between the growing number of unregularized item-specific utilities and the high-dimensional covariate effects, which complicates both estimation and support recovery. To address this, we first analyze the estimator restricted to the true active set. We extend the chaining technique to control the utility estimation error and establish preliminary bounds, and then develop a refined perturbation analysis based on the profile likelihood. This perturbation analysis is, to our knowledge, new. It captures the distinct roles of hypergraph connectivity and sparsity while avoiding potentially large inverse spectral-gap factors. This allows us to obtain sharp error bounds for the item-specific utilities and the shared covariate effects commensurate with their local and global nature.  We then establish support recovery for the regularized estimator by adapting the primal--dual witness argument to separate the utility and covariate components through the expected Hessian, leading to an incoherence condition that accommodates the growing unregularized component.

The proposed estimator is implemented through an alternating minimization algorithm. We also conduct numerical experiments to verify the convergence rates obtained in our analysis and show how support recovery varies with signal strength, the number of items, and sparsity.

\subsection{Related work}

\paragraph{Ranking models with covariates.}

The PL model \citep{plackett1975analysis} generalizes the classical Bradley--Terry (BT) model for pairwise comparisons \citep{zermelo1929berechnung, bradley1952rank} to multiway comparisons under Luce's choice axiom \citep{luce1959individual}. Extensive research has studied the asymptotic properties of the MLE \citep{simons1999asymptotics, chen2019spectral, han2020asymptotic, han2023general, fan2025ranking, fan2026spectral}, as well as computational methods \citep{hunter2004mm, maystre2015fast, agarwal2018accelerated, newman2023efficient, yeung2025efficient, han2026convergence}. Early extensions of the BT and PL models parameterize item-specific utilities fully through shared covariate effects with a fixed covariate dimension \citep{cattelan2012models, guo2018experimental, schafer2018dyad, zhao2022learning}. More recent studies incorporate both item-specific utilities and covariate effects, addressing identifiability through informative priors \citep{li2022bayesian}, orthogonality constraints \citep{fan2024uncertainty}, or sparsity assumptions on the item-specific utilities \citep{fan2024covariate}. In particular, \citet{fan2024covariate} consider sparse utilities and dense, low-dimensional covariate effects, whereas we impose sparsity on the covariate coefficients while leaving the utilities dense and unregularized. Closely related to our work is \citet{dong2026statistical}, who develop a unified framework incorporating context-dependent covariates in multiway comparisons. However, existing formulations typically assume a fixed covariate dimension and do not accommodate the high-dimensional regime in which the covariate dimension may grow with the number of items or even exceed the number of observed comparisons.

\paragraph{Sparse comparison hypergraphs.}
A related strand of the BT and PL literature studies how comparison topology affects statistical recovery. Classical asymptotic theory typically considers dense comparison graphs with $O(n^2)$ observed pairs \citep{simons1999asymptotics, yan2012sparse}. More recent studies have focused on sparse graphs and hypergraphs, under both Erd\H{o}s--R\'enyi-type random designs \citep{erdos1960evolution, negahban2017rank, chen2019spectral, han2020asymptotic, chen2021partial, gao2023uncertainty, fan2025ranking} and more general comparison topologies \citep{shah2016estimation, hendrickx2020minimax, li2022bound, han2023general, chen2023ranking, han2024statistical, yang2024top, han2025unified, fan2026spectral,han2026bradley, yang2026graph}. These studies illustrate how various graph-theoretic properties govern the estimation error for the item-specific utilities; see also \citet{fang2026recent} for a comprehensive review.

\paragraph{High-dimensional regression.}
$\ell_1$-regularization is a standard approach to variable selection in high-dimensional models, with classical theory for the lasso and related estimators \citep{tibshirani1996regression, zhao2006model, bickel2009simultaneous, wainwright2009sharp, buhlmann2011statistics} and a unified treatment of regularized $M$-estimators provided by \citet{negahban2012unified}. The primal-dual witness construction, in particular, underlies exact support recovery results for linear regression \citep{wainwright2009sharp}, Ising models \citep{ravikumar2010high}, and sparse Gaussian graphical models \citep{meinshausen2006high, ravikumar2010high}. 

\subsection{Organization and notation}

The rest of the paper is organized as follows. Section~\ref{background} introduces the background and problem formulation. Section~\ref{method} presents the proposed method. Section~\ref{theory} presents the main theoretical results. Section~\ref{sec:randomdesign} studies random hypergraph models and covariate designs under mild conditions. Section~\ref{simulation} provides simulation studies and a real-data analysis of Hong Kong horse racing. Section~\ref{conclusion} concludes the article.

\paragraph{Notation.}
For $\bv\in\mathbb R^d$, let $\supp(\bv)=\{j\in[d]:v_j\neq0\}$. For $S\subseteq[d]$, let $\bv_S$ be the subvector indexed by $S$ and define $V(S)=\{\bv\in\mathbb R^d:\supp(\bv)\subseteq S\}$. For a matrix $\bm A$, let $[\bm A]_S$ and $\bm A_{SS}$ denote its column and principal submatrices, respectively, and define $\|\bm A\|_\infty=\max_i\sum_j|A_{ij}|$ and $\|\bm A\|_{\infty,2}=\max_k\|\bm A_{k:}\|_2$. For the true support $S^*$, write $\mathcal I=[n]\cup\{n+j:j\in S^*\}$ and $\widetilde{\bm1}=(\bm1_n^\top,\bm0_{d_0}^\top)^\top$, where $\mathcal I^\complement$ is relative to $[n+d]$. For a finite set $e$, let $\mathfrak S(e)$ be the set of its permutations. Finally, $a\lesssim b$ means $a\le Cb$ for a dimension-independent constant $C>0$; define $\gtrsim$ analogously, and write $a\asymp b$ when both relations hold.

\section{Problem formulation}
\label{background}
We begin with a brief review of the PL model with dynamic covariates and then formulate the high-dimensional estimation problem.

\subsection{PL model with high-dimensional covariates}
\label{sec:21}

Let $\mathcal{V}=\{1,\dots,n\}$ denote the set of items. The comparison structure is represented by a hypergraph $\mathcal{H}=(\mathcal{V},\mathcal{E})$, where each hyperedge $e\in\mathcal{E}$ consists of the items involved in a single comparison. The total number of comparisons is denoted by $N=|\mathcal{E}|$. The PL model assigns each item $j$ a latent utility $u_j^*$, and models the probability of ranking $j_1 \succ \cdots \succ j_m$ on event $e = \{j_1, \dots, j_m\}$ as
\begin{equation*}
    \Prob_{\bu^*}\{j_1 \succ \cdots \succ j_m \mid e\} = \prod_{k=1}^{m} \frac{\exp(u_{j_k}^*)}{\sum_{t=k}^{m} \exp(u_{j_t}^*)},
\end{equation*}
with identifiability constraint $\mathbf{1}^\top \bu^* = 0$. \cite{dong2026statistical} augments each item's log-score with a comparison-dependent covariate effect. For item $j$ in event $e$, let $X_{e,j} \in \mathbb{R}^d$ be the covariate vector and $\bv^* \in \mathbb{R}^d$ the coefficient vector. The log-score of item $j$ is
\begin{equation*}
    s_j(\btheta^*; e) = u_j^* + X_{e,j}^\top \bv^*,
\end{equation*}
where $\btheta^* = [(\bu^*)^\top, (\bv^*)^\top]^\top \in \mathbb{R}^{n+d}$, and the ranking probability becomes
\begin{equation*}
    \Prob_{\btheta^*}\{j_1 \succ \cdots \succ j_m \mid e\} = \prod_{k=1}^{m} \frac{\exp(s_{j_k}(\btheta^*; e))}{\sum_{t=k}^{m} \exp(s_{j_t}(\btheta^*; e))}.
\end{equation*}

Identifiability of the PL model with fixed-dimensional covariates depends jointly on the comparison structure and the observed covariates and is typically ensured by a full-column-rank condition. However, this condition may fail as the covariate dimension grows and necessarily fails when it exceeds the number of observed comparisons. In this setting, additional assumptions are needed to ensure identifiability.

To establish identifiability in the high-dimensional setting, we impose sparsity on the covariate effects, assuming that only a subset of covariates influences comparison outcomes.

\begin{assumption}[Sparsity]
    \label{ass:sparsity}
    The true coefficient vector $\bv^* \in \R^d$ is $d_0$-sparse. Specifically, the active support set $S^* \coloneqq \supp(\bv^*)$ satisfies $0 < \abs{S^*} = d_0 \leq n \leq d$.
\end{assumption}

The sparsity assumption is widely adopted to ensure that a high-dimensional estimation problem is well-posed
\citep{tibshirani1996regression,wainwright2009sharp,buhlmann2011statistics, candes2005decoding, donoho2006compressed}. Under Assumption~\ref{ass:sparsity}, we impose the rank condition only on sparse covariates, leading to the following notion of sparse identifiability, which is analogous to the well-posedness condition in sparse recovery problems from compressed sensing \citep{donoho2003optimally, cohen2009compressed}.

\begin{definition}[Sparse model identifiability]
    \label{def:iden}
    Given a comparison hypergraph $\mathcal{H}(\mathcal{V}, \mathcal{E})$, the PL model with high-dimensional covariates is said to be 
    \textit{$d_0$-sparse identifiable} if the following holds: for any two parameter 
    vectors $\btheta_1, \btheta_2$ belonging to the restricted parameter space 
    $$\Theta_{d_0} = \left\{ (\bu^\top, \bv^\top)^\top \in \R^{n+d} : |\supp(\bv)| \leq d_0,\; \mathbf{1}^\top \bu = 0 \right\},$$ 
    the condition
    \begin{equation*}
        \Prob_{\btheta_1}\{\,\cdot \mid e\} \;=\; 
        \Prob_{\btheta_2}\{\,\cdot \mid e\} 
        \quad \text{for all } e \in \mathcal{E}
    \end{equation*}
    implies $\btheta_1 = \btheta_2$.
\end{definition}

Before stating the identifiability condition, we introduce some notation for the comparison hypergraph and the covariates. Fix an ordering of the items in $\mathcal{V}$ that is independent of the observed rankings. For each hyperedge $e_i$, write its elements as $e_i=(j_{i_1},\dots,j_{i_{m_i}})$ according to this ordering and designate $j_{i_1}$ as the reference item. We define the broken comparison hypergraph $\mathcal{H}_{\mathrm{br}}=(\mathcal{V},\mathcal{E}_{\mathrm{br}})$ by replacing each hyperedge $e_i$ with the $m_i-1$ directed edges from $j_{i_1}$ to the remaining items. Specifically,
\begin{equation*}
    \mathcal{E}_{\mathrm{br}} = \bigl\{(j_{i_1},j_{i_t}) : i\in[N],\ t=2,\dots,m_i\bigr\},
\end{equation*}
where edges are indexed by $(i,t)$ so that repeated item pairs are retained. Thus, $\mathcal{H}_{\mathrm{br}}$ has $N_{\mathrm{br}}=\sum_{i=1}^N(m_i-1)$
directed edges. Let $\Q_n\in\mathbb{R}^{n\times N_{\mathrm{br}}}$ denote the incidence matrix of $\mathcal{H}_{\mathrm{br}}$. For each directed edge $(j_{i_1},j_{i_t})$, the corresponding column of $\Q_n$ has $-1$ at vertex $j_{i_1}$, $1$ at vertex $j_{i_t}$, and zeros elsewhere.

For the covariates, define $\Delta X_i\in\mathbb{R}^{(m_i-1)\times d}$ as the matrix of covariate differences relative to the reference item $j_{i_1}$, whose $(t-1)$-th row is $X_{e_i,j_{i_t}}-X_{e_i,j_{i_1}}$ for $t=2,\dots,m_i$. Stacking these blocks over $i=1,\dots, N$ gives the aggregated covariate difference matrix $\K_n=[\Delta X_1^\top,\dots,\Delta X_{N}^\top]^\top\in\mathbb{R}^{N_{\mathrm{br}}\times d}$. We then define the full design matrix
\begin{equation*}
\W_n=[\Q_n^\top,\K_n]\in\mathbb R^{N_{\mathrm{br}}\times (n+d)}.
\end{equation*}
The identifiability condition is stated as a rank condition on a suitable submatrix of $\W_n$.

\begin{theorem}[Identifiability]
\label{thm:iden}
Under Assumption~\ref{ass:sparsity}, the PL model with high-dimensional covariates is $d_0$-sparse identifiable in the sense of Definition~\ref{def:iden} if and only if, for every support set $S \subseteq [d]$ with $|S| \leq 2d_0$,
\begin{equation}
    \label{eq:rank_condition}
    \rank\bigl([\bm{W}_n]_{\mathcal{I}(S)}\bigr) = n + |S| - 1,
\end{equation}
where $\mathcal{I}(S) \coloneqq [n] \cup \{n + j : j \in S\}$ denotes the index set of the utility parameters and the covariates indexed by $S$.
\end{theorem}

The rank condition in~\eqref{eq:rank_condition} is a Kruskal rank requirement on the covariate columns relative to the utility design space. Let $\bm{\Pi}_{\bm Q}^{\perp}$ denote the orthogonal projection onto the orthogonal complement of $\operatorname{range}(\bm Q_n^\top)$. Then \eqref{eq:rank_condition} is equivalent to
\begin{equation*}
\rank(\bm Q_n^\top)=n-1
\quad\text{and}\quad
\operatorname{krank} \bigl(\bm{\Pi}_{\bm Q}^{\perp}\bm K_n\bigr)\ge 2d_0,
\end{equation*}
where $\operatorname{krank}(\bm A)$, the Kruskal rank of $\bm A$, is the largest integer $r$ such that every $r$ columns of $\bm A$ are linearly independent. Such a condition typically holds under random designs with nondegenerate covariate distributions. The proof of Theorem~\ref{thm:iden} is deferred to Appendix~\ref{app:iden}.

\subsection{Comparison hypergraph designs}

Throughout the deterministic design analysis, for each $n$ we condition on $\mathscr D_n\coloneqq\sigma(\mathcal H_n,\mathcal X_n)$, the sigma-field generated by the fixed covariates and the deterministic comparison hypergraph $\mathcal H_n=(\mathcal V_n,\mathcal E_n)$, which may vary with $n$; unless stated otherwise, all expectations and probabilities are taken with respect to the ranking outcomes conditional on $\mathscr D_n$. In Section~\ref{sec:randomdesign}, we show that the conditions imposed below hold with high probability under a class of random comparison hypergraph and covariate designs.

\begin{assumption}[Bounded comparison size]
    \label{ass:size}
    The maximum size of any comparison set is uniformly bounded, that is, $M \coloneqq \sup_n \max_{e\in\mathcal{E}_n}|e| < \infty$.
\end{assumption}

This assumption is common in applications with local comparisons, such as sports competitions, where each match or race typically involves only a small subset of all teams or players \citep{masarotto2012ranking}. We next introduce the graph quantities used to state the connectivity assumption.

\begin{definition}
For the comparison hypergraph $\mathcal{H}_n=(\mathcal{V}_n,\mathcal{E}_n)$, define the following quantities.
\begin{enumerate}
    \item For any vertex subset $U\subset[n]$, define its hyperedge boundary by
    \begin{equation*}
        \partial U\coloneqq
        \{e\in\mathcal E_n:e\cap U\neq\emptyset,\ 
        e\cap U^\complement\neq\emptyset\}.
    \end{equation*}
    The unnormalized \textit{Cheeger constant} is
    \begin{equation*}
        h_{\mathcal{H}_n}\coloneqq
        \min_{\substack{U\subset[n],\\ U\neq\emptyset,\;U\neq[n]}}
        \frac{|\partial U|}{\min\{|U|,\,|U^\complement|\}}.
    \end{equation*}

    \item For $\eta \in (0,1)$, a sequence of vertex subsets $\bm A = \{A_j\}_{j=0}^{J}$ is said to be \textit{$\eta$-weakly admissible} if
    \begin{equation*}
        \emptyset \neq A_0 \subsetneq A_1 \subsetneq \cdots \subsetneq A_J \subseteq [n],
    \end{equation*}
    and, for every $0 \le j < J$, we have $\lvert \partial A_j \rvert > 0$ and
    \begin{equation*}
        \frac{\bigl\lvert \bigl\{ e \in \partial A_j \,:\, e \cap (A_{j+1} \setminus A_j) \neq \emptyset \bigr\} \bigr\rvert}
            {\lvert \partial A_j \rvert}
        \;\ge\; \eta .
    \end{equation*}
    Thus, at each step, at least an $\eta$-fraction of the boundary
    edges of $A_j$ intersects the newly added vertices.
    Let $\mathcal{A}_{\mathcal{H}_n}(\eta)$ denote the collection
    of all such sequences. Its diameter is
    \begin{equation*}
        \diam\bigl(\mathcal{A}_{\mathcal{H}_n}(\eta)\bigr)
        \coloneqq
        \max\Bigl\{J:
        \{A_j\}_{j=0}^J\in\mathcal{A}_{\mathcal{H}_n}(\eta)\Bigr\}.
    \end{equation*}
\end{enumerate}
\end{definition}

The quantity $\diam\bigl(\mathcal{A}_{\mathcal{H}_n}(\eta)\bigr)$ is nonincreasing in $\eta$ and measures the maximum length of an admissible sequence. It can be viewed as an analogue of the usual hypergraph diameter. For a fixed $\eta$, a smaller admissible diameter reflects stronger global connectivity.

\begin{assumption}[Hypergraph connectivity]
\label{ass:hypergraph_topology}
The comparison hypergraph $\mathcal H_n$ is connected.
For any fixed $\eta\in(0,1/2]$,
\begin{equation}
\label{eq:diam-condition}
    \frac{
        \diam^2\bigl(\mathcal{A}_{\mathcal{H}_n}(\eta)\bigr)
        \log n
    }{h_{\mathcal{H}_n}}
    \longrightarrow 0
    \qquad\text{as }n\to\infty.
\end{equation}
\end{assumption}

Assumption~\ref{ass:hypergraph_topology} requires the Cheeger
constant to grow sufficiently fast relative to the squared
admissible diameter. It ensures that there are
enough comparisons across every cut to control the accumulation
of estimation error along admissible sequences.

\section{Estimation}
\label{method}
We consider a setting in which the true coefficient vector $\bv^*$ is sparse, and aim to construct an estimator that simultaneously identifies the active covariates and consistently estimates all model parameters. To this end, we introduce an $\ell_1$-regularized MLE for PL model with high-dimensional covariates. Its computation is carried out via an alternating minimization algorithm, with a Minorization--Maximization (MM) step \citep{hunter2004mm} for updating $\bu$ and a FISTA \citep{beck2009fast} step for updating $\bv$. 

\subsection{\texorpdfstring{$\ell_1$}{l1}-regularized MLE}

Given a comparison hypergraph $\mathcal{H}_n(\mathcal{V}_n, \mathcal{E}_n)$, edge-dependent covariates $\X=\{X_{e,j}\}_{e\in\mathcal{E}_n,\,j\in e}$, and independent comparison outcomes $\{\pi_i\}_{i\in[N]}$ with $\pi_i\in\mathfrak{S}(e_i)$ and $|e_i|=m_i$, the normalized log-likelihood at $\btheta = (\bu^\top, \bv^\top)^\top$ can be written as
\begin{equation*}
    l_n(\btheta) = \frac{1}{N} \sum_{i \in [N]} \sum_{j=1}^{m_i} \left[ s_{ij} - \log \left( \sum_{t=j}^{m_i} \exp(s_{it}) \right) \right],
\end{equation*}
where $s_{ij} \coloneqq s_{\pi_i(j)}(\btheta; e_i) = u_{\pi_i(j)} + X_{e_i, \pi_i(j)}^\top \bv$, and $\pi_i(j)$ denotes the item in $e_i$ ranked in position $j$ under the observed outcome $\pi_i$. Here, the score consists of an item-specific utility component $u_{\pi_i(j)}$ and a shared linear covariate effect $X_{e_i, \pi_i(j)}^\top \bv$.

In the high-dimensional regime, the coefficient vector $\bv$ is assumed sparse. Accordingly, we penalize $\bv$ through its $\ell_1$ norm to induce sparsity and enable variable selection. Under the identifiability constraint $\bm{1}^\top \bu = 0$, we define the estimator $\hbtheta = (\hbu^\top, \hbv^\top)^\top$ as a minimizer of the regularized negative log-likelihood: 
\begin{equation}
  \label{eq:mle}
  \hbtheta \in \argmin_{\btheta \in \R^{n+d}: \, \bm{1}^{\top} \bu = 0}
  \left\{ -l_n(\btheta) + \lambda_n \norm{\bv}_1 \right\}.
\end{equation}
The parameter $\lambda_n$ controls the trade-off between fit and sparsity, with larger values favoring sparser estimates and smaller values reducing shrinkage. Before proceeding to the theoretical results of this regularized MLE, we first discuss its computation, provided it exists.

\subsection{Alternating minimization}
\label{sec:algorithm}
Because the PL likelihood is invariant under a common shift of $\bu$, the constraint $\bm{1}^\top\bu=0$ can be enforced by centering after each $\bu$-update without changing the objective value. The complete procedure is summarized in Algorithm~\ref{alg:am_l1}.

\begin{algorithm}[ht]
\small
\caption{Alternating minimization for the $\ell_1$-penalized MLE}
\label{alg:am_l1}
\begin{algorithmic}[1]
\Require Tuning parameter $\lambda_n > 0$; convergence tolerance $\varepsilon > 0$.
% \Ensure Estimator $(\hbu, \hbv)$.

\State \textbf{Initialize:} $\bu^{(0)} \in \mathbb{R}^n$ with $\bm{1}^\top \bu^{(0)} = 0$, and $\bv^{(0)} \in \mathbb{R}^d$; set $k \gets 0$.

\Repeat
    \State $k \gets k + 1$

    \State \textbf{$\bv$-update (FISTA):} solve the $\ell_1$regularized subproblem
    \Statex \qquad $\bv^{(k)} \gets \argmin_{\bv} \bigl\{ -l_n(\bu^{(k-1)}, \bv) + \lambda_n \norm{\bv}_1 \bigr\}$

    \State \textbf{$\bu$-update (MM):} solve the weighted PL subproblem
    \Statex \qquad $\tilde{\bu}^{(k)} \gets \argmin_{\bu} \bigl\{ -l_n(\bu, \bv^{(k)}) \bigr\}$

    \State \textbf{Centering:} enforce the identifiability constraint
    \Statex \qquad $\bu^{(k)} \gets \tilde{\bu}^{(k)} - \bigl(n^{-1}\bm{1}^\top \tilde{\bu}^{(k)}\bigr)\bm{1}$

\Until{the relative change in the objective is below $\varepsilon$}

\State \Return $(\hbu, \hbv) \gets (\bu^{(k)}, \bv^{(k)})$
\end{algorithmic}
\end{algorithm}

\section{Main results}
\label{theory}

In this section, we present the main theoretical guarantees for the $\ell_1$-regularized MLE. Under suitable conditions on the sample size and hypergraph connectivity, we establish uniform consistency and show that the estimator excludes all inactive covariates. Under an additional minimum signal condition, it further achieves exact support recovery.

\subsection{Regularity conditions}\label{sec:reg}

Before presenting the main results, we introduce the assumptions used in the theoretical analysis. The first assumption is a standard regularity condition that controls the scale of the parameter space and ensures that the score function remains well behaved.

\begin{assumption}[Boundedness]
\label{ass:unibound}
The true parameter $\bthetastar$ and covariates $X_{e_i,j}$ are uniformly bounded,
\begin{equation*}
    C \coloneqq \sup_{n\in \mathbb{N}} \max \left\{ \norm{\bu^*}_\infty,\, \norm{\bv^*}_1, \, \max_{i \in [N], j \in e_i} \norm{X_{e_i,j}}_\infty\right\} < \infty.
\end{equation*}
\end{assumption}

\begin{assumption}[Well-posedness]
\label{ass:wellposed}
The following conditions hold jointly.
\begin{enumerate}[label=(\roman*)]
    \item \label{ass:exist}
    With probability at least $1-n^{-3}$, for every nonzero $\btheta=(\bu^\top,\bv_{S^*}^\top)^\top\in\mathbb{R}^{n+d_0}$ satisfying $\bm{1}^\top\bu=0$, there exist $e\in\mathcal{E}$ and $k_1,k_2\in e$ such that $k_1\succ k_2$ on $e$ and
    $$ u_{k_1}+[X_{e,k_1}]_{S^*}^\top \bv_{S^*} < u_{k_2}+[X_{e,k_2}]_{S^*}^\top \bv_{S^*}. $$

    \item \label{subspace} \emph{(Subspace separation)} 
    There exists $\epsilon > 0$ such that
    \begin{equation*}
        \sup_{n \in \mathbb{N}}\; \sup_{\substack{ \bm{x} \in \operatorname{range}(\bm{Q}_n^\top),\; \bm{y} \in \operatorname{range}([\K_n]_{S^*}) \\ 
        \lVert \bm{x} \rVert_2 = \lVert \bm{y} \rVert_2 = 1 }} \bm{x}^\top \bm{y} \;<\; 1 - \epsilon.
    \end{equation*}

    \item \label{nondegeneracy} \emph{(Active-design spectral stability)}
    There exist constants $c_X,C_X>0$, independent of $n$, such that, for
    all sufficiently large $n$,
    \begin{equation*}
        c_X
        \le
        \lambda_{\min}\!\left(
        \frac1N[\K_n]_{S^*}^{\top}[\K_n]_{S^*}
        \right)
        \le
        \lambda_{\max}\!\left(
        \frac1N[\K_n]_{S^*}^{\top}[\K_n]_{S^*}
        \right)
        \le C_X.
    \end{equation*}
\end{enumerate}
\end{assumption}
Assumption \ref{ass:wellposed} ensures strict convexity and coercivity of the restricted negative log-likelihood, and hence existence and uniqueness of the restricted estimator. Specifically, Condition~\ref{ass:exist} is an existence requirement on the active covariate set $S^*$. Condition~\ref{subspace} is an asymptotic extension of the model identifiability condition. Condition~\ref{nondegeneracy} requires the active design to be uniformly well-conditioned, ruling out both asymptotically unidentifiable directions and diverging spectral scale. Roughly speaking, both hold when $[\K_n]_{S^*}$ has sufficient randomness.

For the next assumption, the subscripts $[n]$, $S^*$, and $S^{*\complement}$ index the utility, active-covariate, and inactive-covariate coordinates, respectively. Let $\bm M=\nabla^2l_n(\btheta^*)$ and $\bm M^*=\mathbb E[\bm M\mid\mathcal H_n,\mathcal X_n]$. Let $\bm M_{[n][n]}$ denote the principal submatrix of $\bm M$ corresponding to the utility coordinates. Under connectivity, $\bm M_{[n][n]}$ has kernel $\operatorname{span}(\bm1_n)$, and its Moore--Penrose inverse is
\begin{equation*}
\bm M_{[n][n]}^\dagger
=(\bm M_{[n][n]}+\bm1_n\bm1_n^\top/n)^{-1}
-\bm1_n\bm1_n^\top/n.
\end{equation*}
This inverse acts as the inverse of $\bm M_{[n][n]}$ on $\bm1_n^\perp$ and vanishes on $\operatorname{span}(\bm1_n)$. We use the same convention for $(\bm M^*_{[n][n]})^\dagger$. Define
\begin{equation*}
\bm H
=\bm M_{S^*S^*}
-\bm M_{S^*[n]}\bm M_{[n][n]}^\dagger\bm M_{[n]S^*},\qquad
\bm G 
=\bm M_{S^{*\complement} S^*}
-\bm M_{S^{*\complement}[n]}
 \bm M_{[n][n]}^\dagger\bm M_{[n]S^*}.
\end{equation*}
Define $\bm H^*$ and $\bm G^*$ analogously with $\bm M$ replaced by
$\bm M^*$. 

\begin{assumption}[Profiled incoherence]
\label{ass:incoherence}
The matrix $\bm H^*$ is nonsingular and, for some $\alpha\in(0,1)$,
    \begin{equation*}
        \norm{\bm G^*\bm H^{*-1}}_\infty\le 1-\alpha.
    \end{equation*}
\end{assumption}

Here $\bm H$ is the Hessian for the active covariates after profiling out the utilities, while $\bm G$ measures the corresponding dependence of the inactive covariates on the active ones. Subspace separation and active-design spectral stability in Assumption~\ref{ass:wellposed}\ref{subspace}--\ref{nondegeneracy}, together with bounded comparison size and boundedness, ensure that $-\bm H^*$ has eigenvalues bounded above and away from zero. Assumption~\ref{ass:incoherence} controls the remaining dependence between inactive and active covariates after population profiling.

\subsection{Estimation and selection consistency}

Under the conditions in Section~\ref{sec:reg}, we now state our main result. 

\begin{theorem}[Model selection consistency and estimation error]
\label{thm:main_result}
Suppose that Assumptions~\ref{ass:sparsity}--\ref{ass:incoherence} hold, $\max \{ \log d, \, d_0 \} \le n$, and the sample size satisfies $N \gg \operatorname{diam}^2(\mathcal{A}_{\mathcal{H}_n}(\eta))
\{ n d_0 + d_0^2 \log(n+d) \}$. Consider the estimator $\hbtheta$ defined in~\eqref{eq:mle} with regularization parameter
\begin{equation*}
\lambda_n = c_1 \max\biggl\{
    \sqrt{\frac{\log(n+d)}{N}},\;
    \frac{n\,\diam\bigl(\mathcal A_{\mathcal H_n}(\eta)\bigr)}{N}
\biggr\},
\end{equation*}
for a sufficiently large fixed constant $c_1>0$. Then, for all large $n$, the following properties hold with probability at least $1-O(n^{-2})$:
\begin{enumerate}[label={(\arabic*)}]
    \item \textbf{Unique existence:}
    the optimal solution $\hbtheta=(\hbu^\top,\hbv^\top)^\top$ exists and is unique;

    \item \textbf{Support inclusion:}
    $\supp(\hbv)\subseteq\supp(\bv^*)$;

    \item \textbf{Estimation error:}
    \begin{align*}
    \norm{\hbv-\bv^*}_2
    &\lesssim
    \max\biggl\{
        \sqrt{\frac{d_0\log(n+d)}{N}},\;
        \frac{\sqrt{d_0}\, n\,\diam\bigl(\mathcal A_{\mathcal H_n}(\eta)\bigr)}{N}
    \biggr\},\\[4pt]
    \norm{\hbu-\bu^*}_\infty
    &\lesssim
    \diam\bigl(\mathcal A_{\mathcal H_n}(\eta)\bigr)
    \max\biggl\{
        \sqrt{\frac{\log n}{h_{\mathcal H_n}}},\;
        d_0\sqrt{\frac{\log(n+d)}{N}},\;
        \frac{d_0\, n\,\diam\bigl(\mathcal A_{\mathcal H_n}(\eta)\bigr)}{N}
    \biggr\}.
    \end{align*}
\end{enumerate}
In particular, both estimation errors converge to zero.
\end{theorem}

\begin{remark}
Theorem~\ref{thm:main_result} can apply to the classical setting with a fixed number of active
covariates, $d_0 = d = O(1)$. In this case, the error bounds become
\begin{align*}
&\norm{\hbu-\bu^*}_\infty \lesssim \diam(\mathcal A_{\mathcal H_n}(\eta))
\max\biggl\{\sqrt{\frac{\log n}{h_{\mathcal H_n}}},\, \frac{n\,\diam(\mathcal A_{\mathcal H_n}(\eta))}{N}\biggr\},
\,\\
&\norm{\hbv-\bv^*}_2 \lesssim \max\biggl\{\sqrt{\frac{\log n}{N}},\, \frac{n\,\diam(\mathcal A_{\mathcal H_n}(\eta))}{N}\biggr\}.
\end{align*}
These bounds improve upon existing results. For example, our bound on $\hbv$ is smaller than that of \citet{dong2026statistical}, $\diam(\mathcal A_{\mathcal H_n}(\eta)\sqrt{\log n/h_{\mathcal H_n}}$, by a factor of at least $\min\{\sqrt{n},\sqrt{N/n}\}$, since $h_{\mathcal H_n} \lesssim N/n$.
\end{remark}

The assumptions $d_0 \le n$ and $d \le e^n$ are not essential and are imposed mainly to simplify the statements. The error bounds for both $\hbu$ and $\hbv$ consist of an intrinsic term and an interaction term, where the intrinsic terms coincide with the rates for the two classical submodels. Specifically, in the absence of covariates, the error bound for $\hbu$ reduces to $\diam(\mathcal A_{\mathcal H_n}(\eta))\sqrt{\log n/h_{\mathcal H_n}}$, which matches the rate for the standard PL model \citep{han2025unified}. Similarly, in the absence of item effects, the error bound for $\hbv$ reduces to $\sqrt{d_0\log(d)/N}$, the usual rate for $\ell_1$-regularized multinomial logistic regression \citep{negahban2012unified}.

The interaction terms reflect the price of jointly estimating $\bu$ and $\bv$. Specifically, estimating $\bv$ requires estimating all $n$ components of $\bu$, and the accumulated error in $\hbu$ contributes the term $\sqrt{d_0}\,n\,\diam(\mathcal A_{\mathcal H_n}(\eta))/N$. Conversely, the error in $\hbv$ shifts the utilities within each comparison by roughly $\sqrt{d_0}\,\norm{\hbv-\bv^*}_2$, and these shifts propagate along paths in the comparison graph. Consequently, the covariates have little effect on the error of $\hbu$ when $d_0$ is small, whereas this error grows linearly with $d_0$ once the interaction terms dominate; see Section~\ref{simulation} for numerical evidence.

\begin{proposition}[Exact support recovery]
\label{prop:exact_support_recovery}
Suppose the conditions of Theorem~\ref{thm:main_result} hold. If
\begin{equation*}
\min_{j\in S^*}|v_j^*|
\ge C_{\min}\max\biggl\{
        \sqrt{\frac{d_0\log(n+d)}{N}},\;
        \frac{\sqrt{d_0}\, n\,\diam\bigl(\mathcal A_{\mathcal H_n}(\eta)\bigr)}{N}
    \biggr\}
\end{equation*}
for a sufficiently large fixed constant $C_{\min}>0$, then, for all sufficiently large $n$, $\supp(\hbv)=S^*$ with probability at least $1-O(n^{-2})$.
\end{proposition}

Proposition~\ref{prop:exact_support_recovery} complements the support inclusion guarantee: when the smallest nonzero coefficient exceeds the estimation error bound, all active covariates are retained, yielding exact support recovery.

\section{Uniform consistency under random designs}
\label{sec:randomdesign}

Theorem~\ref{thm:main_result} establishes uniform consistency under fixed-design profile conditions. However, it is not immediately clear whether these conditions are satisfied in concrete settings. We next show that the same conclusions hold under two widely studied random hypergraph models.

The first model is the \textit{nonuniform random hypergraph model} (NURHM). Let $M$ be as in Assumption~\ref{ass:size}. Its edge set is given by
$\mathcal{E}=\bigsqcup_{m=2}^{M}\mathcal{E}^{(m)}$, where
$\mathcal{E}^{(m)}\subseteq\binom{\mathcal{V}}{m}$ and the random hypergraphs
$\{\mathcal{E}^{(m)}\}_{m=2}^{M}$ are mutually independent. For each fixed $m$, the edge indicators
$\{\mathbf{1}_{\{e\in\mathcal{E}^{(m)}\}}:e\in\binom{\mathcal{V}}{m}\}$
are independent Bernoulli random variables with inclusion probabilities
$p_{e,n}^{(m)}\in(0,1)$. Define the minimum and maximum $m$-edge inclusion probabilities by
\begin{equation*}
p_n^{(m)}\coloneqq\min_{e\in\binom{\mathcal{V}}{m}}p_{e,n}^{(m)},
\qquad
q_n^{(m)}\coloneqq\max_{e\in\binom{\mathcal{V}}{m}}p_{e,n}^{(m)}.
\end{equation*}
We further define
\begin{equation*}
\xi_{n,-}\coloneqq\sum_{m=2}^{M}n^{m-1}p_n^{(m)},
\qquad
\xi_{n,+}\coloneqq\sum_{m=2}^{M}n^{m-1}q_n^{(m)},
\end{equation*}
which characterize the lower and upper orders of the expected vertex degrees, respectively. The NURHM reduces to the $M$-uniform Erd\H{o}s--R\'enyi hypergraph model when
$p_{e,n}^{(m)}=p_n^{(M)}\mathbf{1}_{\{m=M\}}$ for all
$e\in\binom{\mathcal{V}}{m}$ and $m\in\{2,\ldots,M\}$.

The second model is the \textit{hypergraph stochastic block model} (HSBM), which captures latent community structure. We focus on the $M$-uniform HSBM with $L$ communities, where $L \geq 1$ is a fixed integer independent of $n$. The vertex set $\mathcal{V}$ is partitioned into $L$ disjoint subsets $\mathcal{V} = \bigsqcup_{l \in [L]} \mathcal{V}_l$, and the edge inclusion probability depends only on the community memberships of the constituent vertices:
\begin{equation*}
\Prob\{e \in \mathcal{E}\} =
\begin{cases}
\omega_{n,l} \, \mathbf{1}_{\{e \in \binom{\mathcal{V}}{M}\}}, & \text{if } e \subseteq \mathcal{V}_l, \; l \in [L], \\
\omega_{n,0} \, \mathbf{1}_{\{e \in \binom{\mathcal{V}}{M}\}}, & \text{otherwise.}
\end{cases}
\end{equation*}
The maximal and minimal orders of the expected number of edges are defined as
\begin{equation*}
\zeta_{n,+} \coloneqq n^{M-1} \max_{0 \le l \le L} \omega_{n,l}, \quad \zeta_{n,-} \coloneqq n^{M-1} \min_{0 \le l \le L} \omega_{n,l}.
\end{equation*}

\begin{assumption}[Random design]
\label{ass:random_design}
The comparison hypergraph sequence $\mathcal{H}_n$ and the covariate collection $\X$ are generated as follows.
\begin{enumerate}
    \item \textbf{Comparison hypergraphs.}
    The comparison hypergraph $\mathcal H_n$ follows either the NURHM or the HSBM. Assume that $\max \{ \log d, \, d_0 \} \le n$. Under the NURHM, assume
    \begin{equation*}
    \frac{\xi_{n,+}^2\log^2 n}{\xi_{n,-}^3}
    \biggl\{\log n+d_0+\frac{d_0^2\log(n+d)}{n}\biggr\}
  = o(1).
    \end{equation*}
    Under the HSBM, assume
    \begin{equation*}
    \frac{\log^2 n}{\zeta_{n,-}}
    \biggl\{\log n+d_0+\frac{d_0^2\log(n+d)}{n}\biggr\}
   = o(1).
    \end{equation*}

    \item \textbf{Active features.}
    The active covariates $\{[X_{e,j}]_{S^*}:e\in\mathcal{E}_n,\,j\in e\}$ are i.i.d., independent of $\mathcal{H}_n$, uniformly bounded in $\norm{\cdot}_\infty$, and have covariance matrices satisfying
    \begin{equation*}
    c_X\le\lambda_{\min}\bigl(\operatorname{Cov}([X_{e,j}]_{S^*})\bigr)
    \le\lambda_{\max}\bigl(\operatorname{Cov}([X_{e,j}]_{S^*})\bigr)
    \le C_X
    \end{equation*}
    for fixed constants $c_X,C_X>0$, uniformly in $n$. Moreover, for some constants $\delta,c_0>0$,
    \begin{equation*}
        \liminf_{n\to\infty} \min_{i\in[N]}
        \inf_{\substack{\bv_0\in\R^{d_0}\\ \norm{\bv_0}_2=1}} \min_{j\in[m_i-1]}
        \Prob\bigl\{ [\Delta X_i]_{jS^*}^\top\bv_0>\delta \bigr\} \ge c_0.
    \end{equation*}

    \item \textbf{Noise features.}
    Conditional on $\mathcal H_n$, the noise-feature vectors $\bigl\{[X_{e,j}]_{S^{*\complement}}: e \in \mathcal{E}_n,\ j\in e\bigr\}$ are independent across the comparison incidences $(e,j)$, coordinatewise mean-zero, and uniformly bounded in $\norm{\cdot}_\infty$. They are jointly independent of the active covariates and $\btheta^*$.
\end{enumerate}
\end{assumption}

\begin{theorem}[Uniform consistency under random designs]
\label{thm:randomcon}
Suppose that Assumptions~\ref{ass:sparsity}, \ref{ass:size}, \ref{ass:unibound}, and~\ref{ass:random_design} hold, and choose $\lambda_n$ as in Theorem~\ref{thm:main_result}. Then, eventually almost surely, the estimator exists uniquely, satisfies $\supp(\hbv)\subseteq S^*$, and obeys the following bounds.

\noindent\textbf{NURHM:}
\begin{align*}
\norm{\hbv-\bv^*}_2
&\lesssim
\max\biggl\{
    \sqrt{\frac{d_0\log(n+d)}{n\xi_{n,-}}},\;
    \frac{\sqrt{d_0}\,\xi_{n,+}\log n}{\xi_{n,-}^2}
\biggr\},\\[4pt]
\norm{\hbu-\bu^*}_\infty
&\lesssim
\frac{\xi_{n,+}\log n}{\xi_{n,-}}
\max\biggl\{
    \sqrt{\frac{\log n}{\xi_{n,-}}},\;
    d_0\sqrt{\frac{\log(n+d)}{n\xi_{n,-}}},\;
    \frac{d_0\,\xi_{n,+}\log n}{\xi_{n,-}^2}
\biggr\}.
\end{align*}

\noindent\textbf{HSBM:}
\begin{align*}
\norm{\hbv-\bv^*}_2
&\lesssim
\max\biggl\{
    \sqrt{\frac{d_0\log(n+d)}{n\zeta_{n,-}}},\;
    \frac{\sqrt{d_0}\,\log n}{\zeta_{n,-}}
\biggr\},\\[4pt]
\norm{\hbu-\bu^*}_\infty
&\lesssim
\log n
\max\biggl\{
    \sqrt{\frac{\log n}{\zeta_{n,-}}},\;
    d_0\sqrt{\frac{\log(n+d)}{n\zeta_{n,-}}},\;
    \frac{d_0\log n}{\zeta_{n,-}}
\biggr\}.
\end{align*}
Both errors tend to zero. Moreover, exact support recovery $\supp(\hbv)=S^*$ holds eventually almost surely if, for a sufficiently large fixed constant $C_{\min}>0$,
\begin{align*}
\textbf{NURHM:}\quad
\min_{j\in S^*}|v_j^*|
&\ge C_{\min}\max\biggl\{
    \sqrt{\frac{d_0\log(n+d)}{n\xi_{n,-}}},\;
    \frac{\sqrt{d_0}\,\xi_{n,+}\log n}{\xi_{n,-}^2}
\biggr\},\\[4pt]
\textbf{HSBM:}\quad
\min_{j\in S^*}|v_j^*|
&\ge C_{\min}\max\biggl\{
    \sqrt{\frac{d_0\log(n+d)}{n\zeta_{n,-}}},\;
    \frac{\sqrt{d_0}\,\log n}{\zeta_{n,-}}
\biggr\}.
\end{align*}

\end{theorem}

\section{Numerical experiments}
\label{simulation}

In this section, we evaluate the $\ell_1$-regularized MLE on both synthetic and real data. The simulation studies examine its estimation and selection performance across a range of $n$ and $d_0$, and the real-data application uses it to identify a sparse set of active covariates in a Hong Kong horse-racing dataset. Throughout, the estimator is computed by the alternating minimization algorithm described in Section~\ref{sec:algorithm}.

\subsection{Simulation studies}
\label{subsec:sim}

We first describe the setup shared by all experiments. The $n$ items are split into two communities of sizes $n_1=\lfloor n/3\rfloor$ and $n_2=n-n_1$, and $N=\lceil 0.5\,n(\log n)^3\rceil$ pairwise comparisons are sampled independently with replacement. Each comparison is drawn within community~1, within community~2, or across the two communities with probabilities proportional to $n$, $n$ and $(\log n)^3$, respectively, and its two items are then drawn uniformly from the corresponding communities. The resulting comparison graph is connected in all replications.

The item effects are drawn independently from $\operatorname{Unif}[-1,1]$ and centered so that $\mathbf{1}^\top\bu^*=0$. We set $d=1000$ and draw each coordinate of the covariate differences independently from $\mathcal{N}(0,2)$. For a given sparsity level $d_0$, the support of $\bv^*$ consists of the first $d_0$ coordinates of a fixed random permutation of $\{1,\ldots,d\}$, so that the supports are nested in $d_0$. The estimator is fitted over all $d$ coordinates with regularization parameter
\begin{equation*}
\lambda_n=\widehat C\sqrt{\frac{\log(n+d)}{N}},
\end{equation*}
where the choice of $\widehat C$ is specified in each experiment.

\subsubsection{Consistency}
\label{subsubsec:consistency}

We first examine the consistency of the $\ell_1$-regularized MLE as $n$ grows. We consider $n \in \{100, 200, 300, 400, 500\}$ and two sparsity levels, $d_0 = 2$ and $d_0 = 90$, representing a very sparse setting and a setting where $d_0$ is comparable to $n$, respectively. The active coefficients decay in blocks: $v_j^* = 0.2\, b_j^{-1.01}$ with $b_j = 1 + \lceil (j-50)_+/25 \rceil$. For each $n$ and replication, $\widehat C$ is selected from $\{0.1, 0.2, 0.3, 0.4, 0.5\}$ by minimizing the negative log-likelihood on a $20\%$ holdout sample, pooled across all $d_0$. We measure accuracy by $\norm{\hbu-\bu^*}_\infty$ and $\norm{\hbv-\bv^*}_2$.

\begin{figure}[!htbp]
    \centering
    \includegraphics[width=\linewidth]{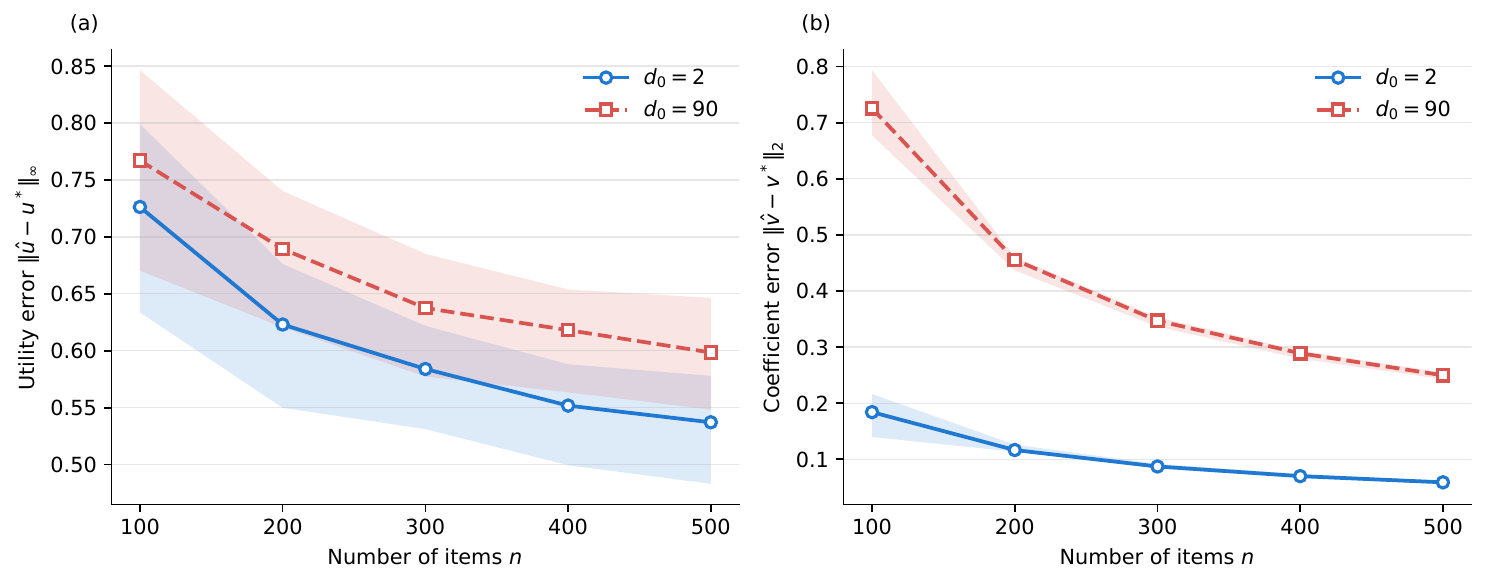}
    \caption{Estimation errors versus $n$ under the HSBM design.
    (a) $\norm{\hbu-\bu^*}_\infty$; (b) $\norm{\hbv-\bv^*}_2$.
    Lines show means over $300$ replications; shaded bands show interquartile ranges.}
    \label{fig:error-vs-n}
\end{figure}

As shown in Figure~\ref{fig:error-vs-n}, both errors decrease steadily as $n$ grows for each $d_0$, supporting the consistency of the estimator. In addition, both errors are generally larger for larger $d_0$, as more nonzero coefficients need to be estimated. 

\subsubsection{Effect of sparsity}
\label{subsubsec:sparsity}

\begin{figure}[!htbp]
    \centering
    \includegraphics[width=\linewidth]{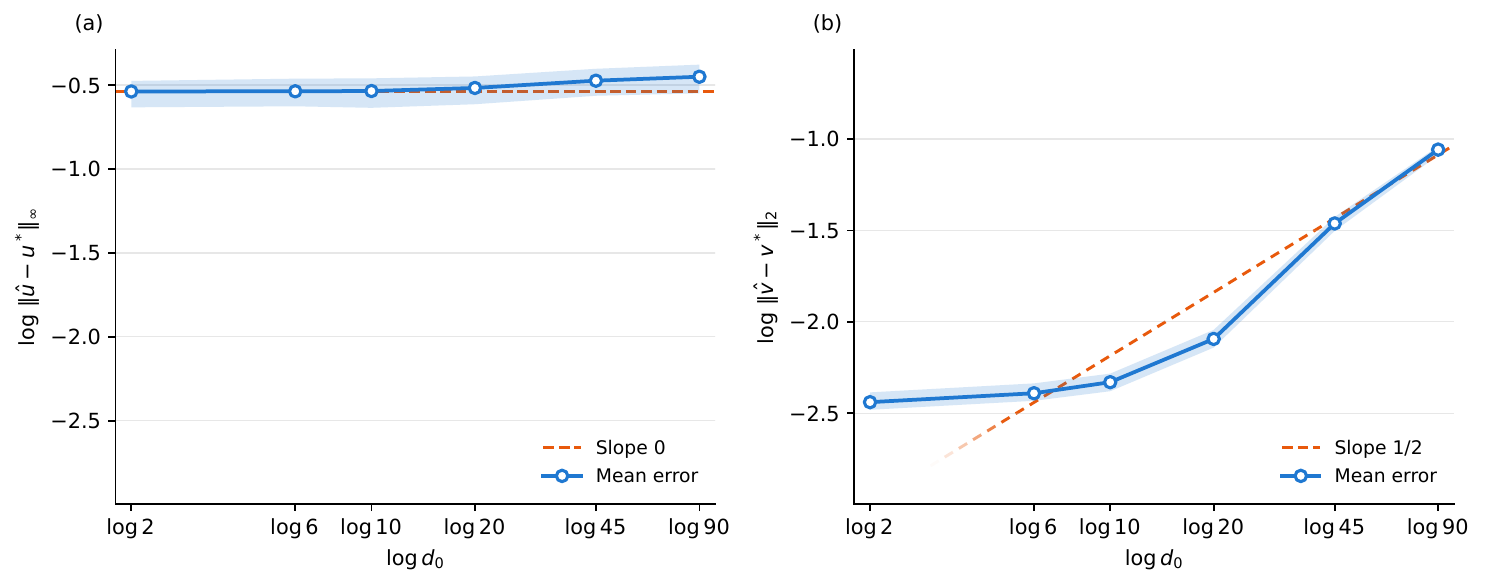}
    \caption{Log--log plot of the estimation errors against $d_0$ under the HSBM design with $n = 300$.
    (a) $\log\norm{\hbu-\bu^*}_\infty$, with a reference line of slope $0$; (b) $\log\norm{\hbv-\bv^*}_2$, with a reference line of slope $1/2$.
    Lines show means over $300$ replications; shaded bands show interquartile ranges.}
    \label{fig:error-vs-d0}
\end{figure}

We then explore how the estimation errors scale with the number of active covariates $d_0$. To this end, we fix $n = 300$ and keep the coefficient design $\bv^*$ above. Figure~\ref{fig:error-vs-d0} plots both errors against $d_0$ on a log--log scale. The slope of each curve shows how fast the error grows with $d_0$.

The two errors behave quite differently. In panel~(a), the error of $\hbu$ remains nearly constant for $d_0 \le 10$ and rises slightly only for $d_0 \ge 20$. This is consistent with Theorem~\ref{thm:main_result}, where the $d_0$-dependent term in the bound for $\hbu$ becomes dominant only when $d_0$ is large. In panel~(b), the error of $\hbv$ increases steadily with $d_0$, with a slope of about $1/2$ for $d_0 \ge 6$, in line with the $\sqrt{d_0}$ order given by Theorem~\ref{thm:main_result}.

\subsubsection{Variable selection}
\label{subsubsec:selection}

\begin{figure}[!htbp]
    \centering
    \includegraphics[width=\linewidth]{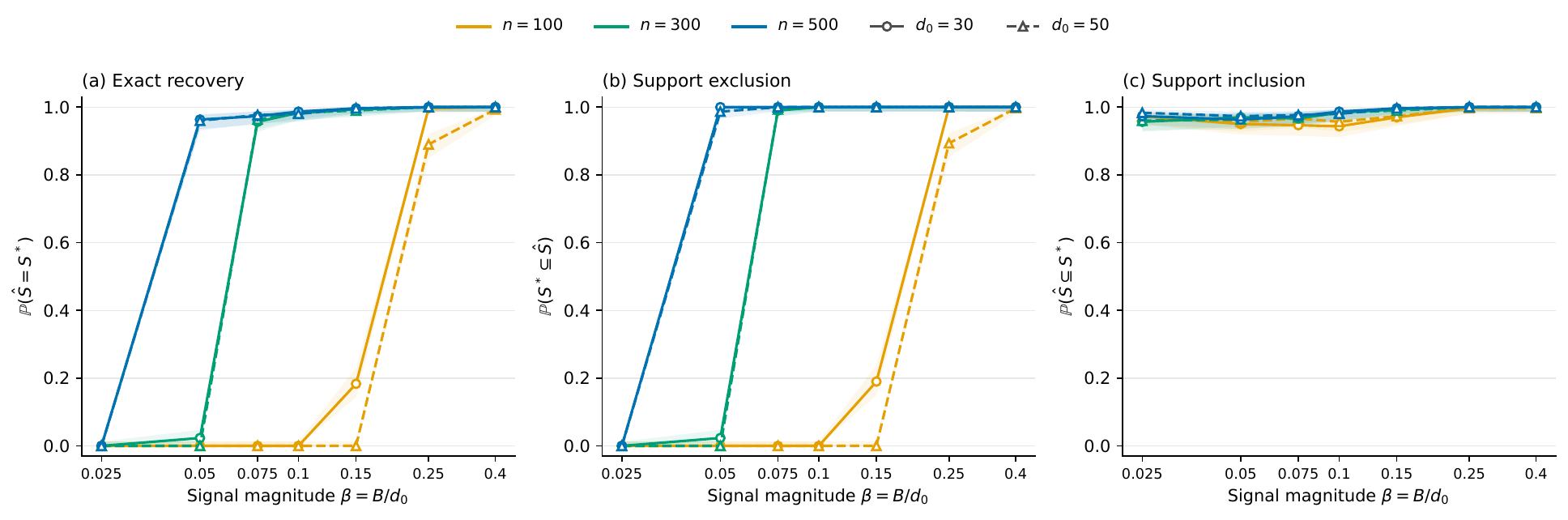}
    \caption{Variable selection versus the signal magnitude $\beta = B/d_0$ under the HSBM design, for $n \in \{100, 300, 500\}$ and $d_0 \in \{30, 50\}$.
    (a) $\mathbb{P}(\widehat S = S^*)$;
    (b) $\mathbb{P}(S^* \subseteq \widehat S)$;
    (c) $\mathbb{P}(\widehat S \subseteq S^*)$.
    Lines show empirical frequencies over 300 replications, and shaded bands show pointwise 95\% confidence intervals for these frequencies.}
    \label{fig:selection}
\end{figure}

We next study support recovery for $n\in\{100,300, 500\}$ and $d_0\in\{30,50\}$. We set $v_j^*=\beta$ for $j\in S^*$ and $v_j^*=0$ otherwise, so that $\norm{\bv^*}_1=d_0\beta=:B$, with $\beta\in\{0.025,0.05,0.075,0.1,0.15,0.25,0.4\}$. The multiplier is fixed at $\widehat C=1$. Within each replication, the supports for $d_0=30$ and $50$ are nested, and all other generated quantities are held fixed as $\beta$ varies, so that differences across $\beta$ reflect signal strength alone. Let $\widehat S=\{j:\widehat v_j\neq 0\}$. Figure~\ref{fig:selection} reports the empirical probabilities of exact recovery ($\widehat S=S^*$), no false negatives ($S^*\subseteq\widehat S$), and no false positives ($\widehat S\subseteq S^*$).

The probability of $\widehat S\subseteq S^*$ is close to one in all settings, so exact recovery is governed almost entirely by whether all active covariates are retained; accordingly, panels (a) and (b) nearly coincide. Both probabilities rise sharply from zero to one as $\beta$ grows, and this transition occurs at weaker signals for larger $n$ and at stronger signals for larger $d_0$. These patterns agree with Theorem~\ref{thm:main_result} and Proposition~\ref{prop:exact_support_recovery}: the estimator excludes inactive covariates with high probability, while exact recovery further requires a minimum signal strength that decreases with $n$ and increases with $d_0$.

\subsection{Real-data application: Hong Kong horse racing}
\label{sec:realdata}

\begin{figure}[t]
\centering
\includegraphics{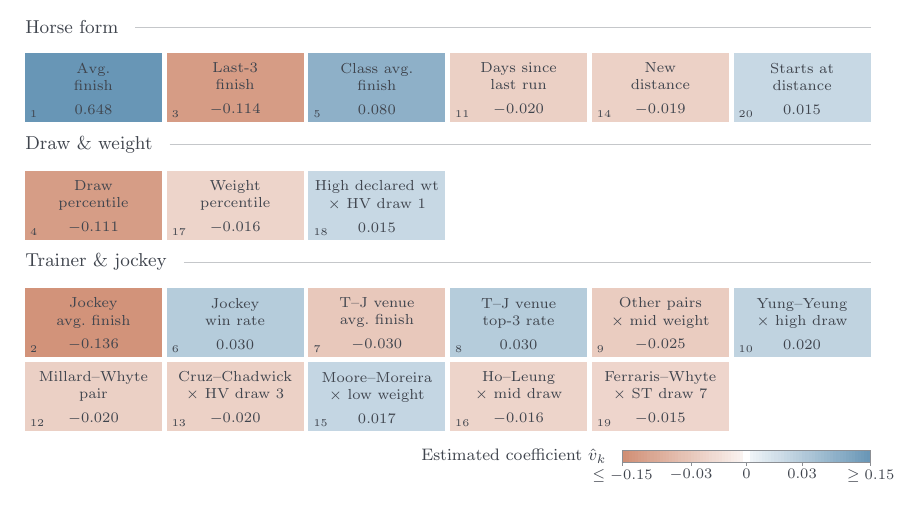}
\caption{The twenty covariates with the largest absolute estimated coefficients, grouped by type. Blue and orange indicate positive and negative coefficients, respectively; darker colours indicate larger magnitudes. Corner numbers indicate ranks by absolute coefficient. T--J denotes a trainer--jockey pair; ST and HV denote Sha Tin and Happy Valley.}
\label{fig:top20}
\end{figure}

We apply the proposed method to Hong Kong horse-racing data from June 2015 to July 2017, covering 1114 races and 1796 horses. Each race gives a ranking of its horses by finishing position. The $d=2879$ candidate covariates vary across races and include each horse's recent form, draw, carried weight, jockey and trainer--jockey performance, and their pairwise interactions. Rare categorical levels are pooled, and continuous covariates are standardized. Figure~\ref{fig:top20} shows the twenty covariates with the largest absolute coefficients.

\subsubsection{Predictive performance}
\label{sec:prediction}

We use the first 885 races for training and the remaining 229 for testing. All preprocessing, including covariate standardization, is fitted on the training data and applied unchanged to the test data. We compare three estimators, namely a fit with item effects only, the $\ell_1$-regularized fit, and an unregularized fit with all covariates ($\lambda = 0$). $\lambda$ is chosen by time-series cross-validation on the training data, which gives $\lambda = 0.2$.

We measure predictive accuracy by top-1 accuracy, the proportion of test races whose winner is correctly predicted, and by ordered top-$k$ accuracy ($k = 2, 3$), the proportion of test races whose first $k$ finishers are predicted in the correct order. We also report the mean negative log-likelihood (NLL) of the observed rankings on the training and test data, computed without the regularization term. The results are summarized in Table~\ref{tab:prediction}.

\begin{table}[t]
\centering
\begin{threeparttable}
\caption{Out-of-sample prediction and goodness of fit.}
\label{tab:prediction}
\footnotesize
\setlength{\tabcolsep}{5.5pt}
\renewcommand{\arraystretch}{0.9}

\begin{tabular*}{0.82\textwidth}{@{\extracolsep{\fill}}l
                S[table-format=2.2]
                S[table-format=1.2]
                S[table-format=1.2]@{}}
\toprule
\multicolumn{4}{@{}l}{\textcolor{rankRed}{\bfseries Panel A}\quad
  \textit{Test accuracy (\%)}} \\
\addlinespace[1pt]
Estimator & {Top-1} & {Ordered Top-2} & {Ordered Top-3} \\
\midrule
$\bu$ only
  & 16.59
  & 4.37
  & 0.87 \\
$\ell_1$-regularized
  & {\bfseries\color{rankRed} 19.21}
  & {\bfseries\color{rankRed} 5.68}
  & {\bfseries\color{rankRed} 1.75} \\
All covariates
  & 16.16
  & 0.87
  & 0.00 \\
\bottomrule
\end{tabular*}

\vspace{0.7ex}

\begin{tabular*}{0.82\textwidth}{@{\extracolsep{\fill}}l
                S[table-format=1.1]
                S[table-format=4.0]
                S[table-format=2.3]
                S[table-format=2.3]@{}}
\multicolumn{5}{@{}l}{\textcolor{rankRed}{\bfseries Panel B}\quad
  \textit{Sparsity and goodness of fit}} \\
\addlinespace[1pt]
Estimator & {$\lambda$} & {$\|\widehat{\bv}\|_0$} & {Train NLL} & {Test NLL} \\
\midrule
$\ell_1$-regularized & 0.2 & 103  & 17.660
                      & {\bfseries\color{rankRed} 24.834} \\
All covariates       & 0   & 2852 & {\bfseries\color{rankRed} 15.197}
                      & 49.195 \\
\bottomrule
\end{tabular*}

\begin{tablenotes}[flushleft]
\item \textit{Note:} Panel~A is based on 229 test races. NLL is the
mean unregularized negative log-likelihood; lower values are better.
Bold red entries mark the best value in each performance column.
\end{tablenotes}
\end{threeparttable}
\end{table}

Panel~A of Table~\ref{tab:prediction} shows that the $\ell_1$-regularized fit achieves the highest accuracy on all three measures. In contrast, the all-covariate fit does no better than the $\bu$-only fit in Top-1 accuracy and is clearly worse on both ordered measures. Covariates therefore help prediction only when they are selected.

Panel~B explains why the all-covariate fit performs poorly. It has the lowest training NLL but a much higher test NLL than the $\ell_1$-regularized fit, which is a clear sign of overfitting. By keeping only a small subset of covariates, the $\ell_1$-regularized fit gives up little in-sample fit and gains substantially out of sample. 

\subsubsection{Unexpected coefficient signs}
Most coefficient signs agree with racing intuition: stronger jockey and trainer--jockey records increase the fitted score, whereas a wider draw and a longer layoff decrease it. Two exceptions are the horse's mean finishing position over all past races (rank~1) and within the current class (rank~5). Their positive coefficients are counterintuitive because larger values indicate poorer performance. We focus on the rank-1 covariate, whose coefficient is substantially larger.

Marginally, the rank-1 covariate behaves as expected: its correlation
with winning is $-0.1178$, and winners' values average $0.434$ lower
than those of their rivals. Its coefficient is also negative without
horse-specific utilities ($-0.3188$) but becomes positive in the full
regularized model ($+0.6488$). The reversal therefore arises from
conditioning on the estimated utilities.

\begin{figure}[!htbp]
\centering
\includegraphics[width=0.9\linewidth]{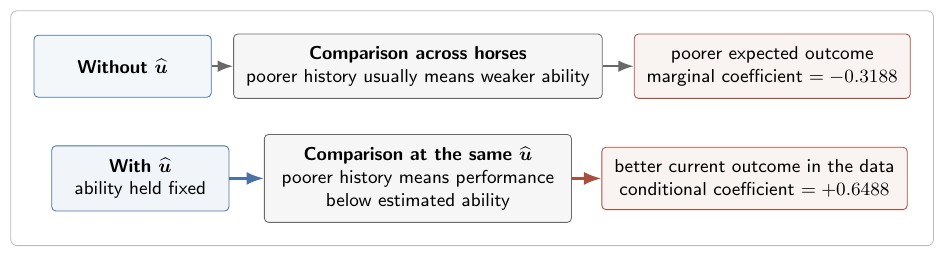}
\caption{Schematic explanation of the coefficient sign reversal.}
\label{fig:sign-reversal}
\end{figure}

This reversal is consistent with the historical measure and
$\widehat{\bm{u}}$ capturing overlapping ability information
(their correlation is $-0.4175$). Without $\widehat{\bm{u}}$, the
historical measure compares horses of different abilities; conditional
on $\widehat{\bm{u}}$, a poorer record instead indicates performance
below the horse's estimated ability. Figure~\ref{fig:sign-reversal}
illustrates this change in interpretation. By contrast, the negative
coefficient on the last-three-start average ($-0.1141$) captures
recent form.

We assess the predictive relevance of the positive coefficients using the same chronological split. Holding the horse utilities and all other coefficients fixed, we set the rank-1 coefficient to zero, assign mildly negative values to both coefficients, or reverse both fitted signs. Table~\ref{tab:sign-sensitivity} reports the resulting Top-1 and exact ordered Top-2 and Top-3 accuracies.

\begin{table}[t]
\centering
\begin{threeparttable}
\caption{Sensitivity of test accuracy to the two unexpected coefficient signs.}
\label{tab:sign-sensitivity}
\footnotesize
\setlength{\tabcolsep}{5.5pt}
\renewcommand{\arraystretch}{0.9}
\begin{tabular*}{0.82\textwidth}{@{\extracolsep{\fill}}l
                S[table-format=-1.4]
                S[table-format=-1.4]
                S[table-format=2.2]
                S[table-format=1.2]
                S[table-format=1.2]@{}}
\toprule
& \multicolumn{2}{c}{\textcolor{rankRed}{\bfseries Coefficient values}}
& \multicolumn{3}{c}{\textcolor{rankRed}{\bfseries Test accuracy (\%)}} \\
\cmidrule(lr){2-3}\cmidrule(l){4-6}
Setting
& \multicolumn{1}{c}{$\widehat{v}_1$}
& \multicolumn{1}{c}{$\widehat{v}_5$}
& \multicolumn{1}{c}{Top-1}
& \multicolumn{1}{c}{\shortstack{Ordered\\Top-2}}
& \multicolumn{1}{c}{\shortstack{Ordered\\Top-3}} \\
\midrule
\textbf{Fitted values}
& 0.6482
& 0.0795
& {\bfseries\color{rankRed} 19.21}
& {\bfseries\color{rankRed} 5.68}
& {\bfseries\color{rankRed} 1.75} \\
\addlinespace[2pt]
Set $\widehat{v}_1=0$
& 0
& 0.0795
& 16.59
& 2.62
& 0.87 \\
Mild negatives
& -0.10
& -0.03
& 16.59
& 2.18
& 0.44 \\
Reverse signs
& -0.6482
& -0.0795
& 15.72
& 3.06
& 0.00 \\
\bottomrule
\end{tabular*}
\end{threeparttable}
\end{table}

The fitted coefficients achieve the highest accuracy under all three criteria, and every perturbation reduces predictive performance. Reversing both signs, for example, lowers Top-1 accuracy from $19.21\%$ to $15.72\%$. The positive coefficients therefore provide useful conditional predictive information.

\section{Conclusion}
\label{conclusion}

In this paper, we investigated the covariate-assisted ranking model under high-dimensional settings. We established necessary and sufficient conditions for model identifiability and rigorously analyzed the theoretical properties of the proposed $\ell_1$-regularized MLE. Specifically, our analysis proved the unique existence of the estimator, derived uniform consistency, and established exact support recovery. To solve the corresponding optimization problem, we developed an alternating minimization algorithm. Finally, numerical experiments and an application to a real-world dataset were conducted to support our theoretical findings and illustrate the practical utility of the proposed method.
 
We conclude with several directions for future research. The first is statistical inference for the proposed estimator. For the item-specific utilities, a possible approach is to isolate the leading linear term and control the nonlinear remainder, as suggested by the recent analysis of \citet{han2026bradley}. For the covariate coefficients, the bias introduced by regularization may dominate and thus necessitates an additional debiasing procedure to establish asymptotic normality. The second direction is to extend our high-dimensional covariate framework to more general ranking models that accommodate intransitive effects; see, e.g., \citet{lee2026pairwise}. We leave these questions for future work.

%\clearpage

\appendix

\numberwithin{equation}{section}
\numberwithin{table}{section}
\numberwithin{figure}{section}

\section{Additional notation}

We begin by introducing some notation. For $i \in [N]$, $j \in [m_i]$, and $j \leq t \leq m_i$, define
$ Z_{(i,j,t)} =(\vec e_{\pi_i(t)\pi_i(j)}^\top, \left(X_{e_i,\pi_i(t)} - X_{e_i,\pi_i(j)}\right)^\top)^\top \in \R^{n+d}$,
where, for $k,l \in [n]$, $\vec e_{kl}$ denotes the difference between the $k$th and $l$th canonical basis vectors of $\R^n$.

In addition, define $\bm{\Lambda}_{ij} (\btheta) = \diag (g_{ij}(\btheta)) - g_{ij}(\btheta) g_{ij}(\btheta)^\top$, where
\begin{equation*}
g_{ij}(\btheta) =
\frac{1}{\sum_{t=j}^{m_i} \exp( Z_{(i,j,t)}^\top \btheta )}
\left[\exp \left( Z_{(i,j,j+1)}^\top \btheta \right),\cdots,\exp \left(Z_{(i,j,m_i)}^\top \btheta\right)\right]^\top \in \R^{m_i-j},
\end{equation*}
and $\diag(g_{ij}(\btheta))$ denotes the diagonal matrix whose diagonal entries are given by $g_{ij}(\btheta)$.

\noindent \emph{Likelihood, gradient, and Hessian.} \\
Since $Z_{(i,j,j)} = 0$, the function $l_n(\btheta)$ and its gradient and Hessian can be written as
\begin{equation}
\label{derivative}
\begin{aligned}
l_n(\btheta)
&= - \frac{1}{N}\sum_{i \in [N]} \sum_{j \in [m_i - 1]} \log \left\{
\sum_{t = j}^{m_i} \exp\left( Z_{(i,j,t)}^\top \btheta \right)
\right\},\\
\nabla l_n(\btheta)
&= - \frac{1}{N}\sum_{i \in [N]} \sum_{j \in [m_i - 1]}\frac{\sum_{t = j+1}^{m_i} \exp\left( Z_{(i,j,t)}^\top \btheta \right) Z_{(i,j,t)}
}{1 + \sum_{t = j+1}^{m_i} \exp\left( Z_{(i,j,t)}^\top \btheta \right)
}, \\
\nabla^2 l_n(\btheta)
&= - \frac{1}{N}\sum_{i \in [N]} \sum_{j \in [m_i - 1]}
\left[ Z_{(i,j,j+1)}, \dots, Z_{(i,j,m_i)} \right]
\bLambda_{ij}(\btheta)
\left[ Z_{(i,j,j+1)}, \dots, Z_{(i,j,m_i)} \right]^\top.
\end{aligned}
\end{equation}

\section{Model identifiability}
\label{app:iden}

\begin{proof}[Proof of Theorem~\ref{thm:iden}]

We first prove the ``if'' part. Suppose, to the contrary, that the model is not identifiable on the restricted parameter space. Then there exist two distinct parameter vectors
\begin{equation*}
\btheta=(\bu^\top,\bv^\top)^\top,\qquad \btheta'=((\bu')^\top,(\bv')^\top)^\top
\end{equation*}
in $\Theta_{d_0}$ such that
\begin{equation*}
\bm{1}_n^\top \bu=\bm{1}_n^\top \bu'=0
\end{equation*}
and they induce the same ranking probabilities on every edge. Let
\begin{equation*}
S=\supp(\bv),\qquad S'=\supp(\bv').
\end{equation*}
Then $|S|\le d_0$ and $|S'|\le d_0$, so $|S\cup S'|\le 2d_0$. Equality of the ranking probabilities on every edge implies that, within each comparison set, the two collections of log-scores differ only by an edge-specific additive constant. 

Fix an observed comparison edge $e_i$. Write
$s_k=u_k+X_{e_i,k}^{\top} \bv$ and
$s'_k=u'_k+X_{e_i,k}^{\top} \bv'$ for $k\in e_i$. If the two parameter vectors induce the same PL ranking probabilities on $e_i$, then their first-stage choice probabilities are equal:
\begin{equation*}
\frac{\exp(s_k)}{\sum_{\ell\in e_i}\exp(s_\ell)} = \frac{\exp(s'_k)}{\sum_{\ell\in e_i}\exp(s'_\ell)}, \qquad k\in e_i.
\end{equation*}
Taking ratios for any $k,\ell\in e_i$ yields
$s_k-s_\ell=s'_k-s'_\ell$. Thus all pairwise log-score differences agree on $e_i$. Because $\W_n$ stacks these differences relative to the fixed reference item across all observed edges, equality of the edgewise PL probabilities is equivalent to
\begin{equation*}
[\W_n]_{\I(S\cup S')} \bigl(\btheta_{\I(S\cup S')}-\btheta'_{\I(S\cup S')}\bigr)=0.
\end{equation*}
Hence $\btheta_{\I(S\cup S')}-\btheta'_{\I(S\cup S')}$ is a nonzero element of $\ker([\W_n]_{\I(S\cup S')})$. The translation direction
\begin{equation*}
\bm{a}\coloneqq(\bm{1}_n^\top,\bm{0}^\top)^\top
\end{equation*}
also belongs to this kernel. Moreover,
\begin{equation*}
\bm{a}^\top \bigl(\btheta_{\I(S\cup S')}-\btheta'_{\I(S\cup S')}\bigr) = \bm{1}_n^\top(\bu-\bu')=0,
\end{equation*}
so this difference vector cannot be a scalar multiple of $\bm{a}$. Thus the kernel has dimension at least two, contradicting \eqref{eq:rank_condition}, which requires rank deficiency exactly one for every support set of size at most $2d_0$.

For the converse direction, suppose that \eqref{eq:rank_condition} fails. Then there exists a set $S\subseteq[d]$ with $|S|\le 2d_0$ such that
\begin{equation*}
    \rank\bigl([\W_n]_{\I(S)}\bigr) < n+|S|-1.
\end{equation*}
Since the translation direction $\bm{a}\coloneqq(\bm{1}_n^\top,\bm{0}^\top)^\top$ always belongs to $\ker([\W_n]_{\I(S)})$, the above rank deficiency implies that there exists a nonzero vector
\begin{equation*}
    \bm{\delta}=(\bm{\delta}_u^\top,\bm{\delta}_{v,S}^\top)^\top \in \ker([\W_n]_{\I(S)})
\end{equation*}
that is not a scalar multiple of $\bm{a}$. Define the normalized direction
\begin{equation*}
    \widetilde{\bm{\delta}}\coloneqq\bm{\delta}-\frac{\bm{1}_n^\top \bm{\delta}_u}{n}\bm{a}.
\end{equation*}
Then $\widetilde{\bm{\delta}}\in \ker([\W_n]_{\I(S)})$ and
$\bm{1}_n^\top\widetilde{\bm{\delta}}_u=0$. Moreover,
$\widetilde{\bm{\delta}}\neq 0$; otherwise, $\bm{\delta}$ would be a scalar multiple of $\bm{a}$.

We can partition $S$ into two disjoint sets $S_1$ and $S_2$ such that
\begin{equation*}
S=S_1\cup S_2,\qquad |S_1|\le d_0,\qquad |S_2|\le d_0.
\end{equation*}
Define
\begin{equation*}
\bu^{(1)}=\frac{1}{2}\widetilde{\bm{\delta}}_u, \qquad
\bu^{(2)}=-\frac{1}{2}\widetilde{\bm{\delta}}_u.
\end{equation*}
For the covariate coefficients, define
\begin{equation*}
v_j^{(1)}= \begin{cases}
\widetilde{\delta}_{v,j}, & j\in S_1,\\
0, & j\notin S_1, \end{cases}
\qquad
v_j^{(2)}= \begin{cases}
-\widetilde{\delta}_{v,j}, & j\in S_2,\\
0, & j\notin S_2. \end{cases}
\end{equation*}
Then both parameter vectors belong to $\Theta_{d_0}$, and
\begin{equation*}
\btheta^{(1)}_{\I(S)}-\btheta^{(2)}_{\I(S)} = \widetilde{\bm{\delta}}.
\end{equation*}

Since $\widetilde{\bm{\delta}}\neq 0$, the two parameter vectors are distinct. On the other hand,
\begin{equation*}
    [\W_n]_{\I(S)} \bigl(\btheta^{(1)}_{\I(S)}-\btheta^{(2)}_{\I(S)}\bigr) = [\W_n]_{\I(S)}\widetilde{\bm{\delta}} = \bm{0}.
\end{equation*}
Thus the two distinct $d_0$-sparse parameter vectors induce the same linear predictors on all observed edges, and hence the same edgewise ranking probabilities. This contradicts $d_0$-sparse identifiability. Therefore, identifiability implies \eqref{eq:rank_condition}.

\end{proof}

\section{~Proof of Theorem~\ref{thm:main_result}}
\label{app:main_proof}
The proof uses a primal-dual witness construction. We first analyze the estimator restricted to the true support: existence is followed by prediction localization, and population profiling then gives separate coefficient and utility rates. We then verify strict dual feasibility, thereby showing that the restricted solution coincides with the full estimator. A unifying feature of the analysis is that the prediction error controls both the Hessian fluctuation and the nonlinear remainder. Throughout, concentration calculations are conditional on the design $\mathscr D_n=\sigma(\mathcal H_n,\mathcal X_n)$.

\subsection{Setup}
Throughout, we work under the assumptions of Theorem~\ref{thm:main_result}, with $\eta$ as in Assumption~\ref{ass:hypergraph_topology}. Recall that
\begin{equation}
\label{eq:pdw_lambda_choice}
\lambda_n=c_{1}\max\left\{
\sqrt{\frac{\log(n+d)}{N}},\;
\diam(\mathcal A_{\mathcal H_n}(\eta))
\frac{n+d_0+\log(n+d)}{N}\right\},
\end{equation}
where $c_{1}$ is a sufficiently large fixed constant. Assumption~\ref{ass:hypergraph_topology} and the sample size condition in Theorem~\ref{thm:main_result} imply, respectively,
\begin{equation}
\label{eq:connect_conditions}
\diam(\mathcal A_{\mathcal H_n}(\eta))
\sqrt{\frac{\log n}{h_{\mathcal H_n}}}=o(1),
\qquad
d_0\lambda_n\diam(\mathcal A_{\mathcal H_n}(\eta))=o(1).
\end{equation}
These results are used throughout the proof. 
%We first establish existence, uniqueness, and error bounds for the oracle estimator without Assumption~\ref{ass:incoherence}. We then use population incoherence to verify strict dual feasibility, which identifies the oracle estimator with the unique solution of the full problem.

To organize the stochastic part of the proof, we will introduce several events $\G_1,\G_2,\ldots,\G_7$ at the points where they are needed. Collectively, these events control the quantities involved in establishing existence, bounding the cut and restricted scores, and handling the active and inactive directional terms. We first derive the deterministic conclusions on their intersection. After completing the deterministic analysis, Lemma~\ref{lem:all_events} collects the corresponding probability bounds and establishes the desired high-probability result via a union bound.

\subsection{KKT conditions and PDW}

We characterize the $\ell_1$-regularized solution through the KKT conditions \citep{boyd2004convex}. A parameter vector $\hbtheta = (\hbu^{\top}, \hbv^{\top})^{\top}$  is an optimal solution to \eqref{eq:mle} if and only if there exist  $\alpha \in \R$ and $\widehat{\bm z} \in \partial \norm{\hbv}_1$ such that
\begin{align}
-\nabla_{\bu} l_n(\hbtheta) + \alpha \bm{1}_n &= \bm{0}, 
\label{eq:kkt_u} \\
-\nabla_{\bv} l_n(\hbtheta) + \lambda_n \widehat{\bm z} &= \bm{0},
\label{eq:kkt_v} \\
\bm{1}^{\top} \hbu &= 0,
\label{eq:kkt_constraint}
\end{align}
where $\widehat{z_j}= \operatorname{sign}(\widehat{v}_j)$ if $\widehat{v}_j \neq 0$, and $|\widehat{z_j}| \leq 1$ otherwise.

>Since the objective is convex and the equality constraint has full row rank, the KKT conditions are necessary and sufficient. We first show that the multiplier $\alpha$ in~\eqref{eq:kkt_u} is zero. Because the likelihood is invariant to a common shift in all item utilities, for any $\btheta$,
\begin{equation*}
    \bm{1}_n^\top \nabla_{\bu} l_n(\btheta)=0.
\end{equation*}
Summing the KKT condition~\eqref{eq:kkt_u} over all utility coordinates gives
\begin{equation*}
    -\bm{1}_n^\top \nabla_{\bu} l_n(\hbtheta)+n\alpha=0.
\end{equation*}
Using the identity above yields $n\alpha=0$, and hence $\alpha=0$. Thus \eqref{eq:kkt_u} reduces to the unregularized score equation
\begin{equation*}
    \nabla_{\bu} l_n(\hbtheta)=\bm{0}.
\end{equation*}

To handle the nonsmooth $\ell_1$ regularization, we adopt the primal-dual witness construction of \citet{wainwright2009sharp}. We first solve the problem restricted to the true support $S^*$ and then verify the full KKT conditions. This yields unique global optimality with high probability. The construction proceeds in three steps:

\begin{enumerate}

    \item \textbf{Restricted primal solution:} Solve the $\ell_1$-regularized objective over covariate vectors supported on the true active set:
    \begin{equation}
        \label{eq:oracle}
        \hbtheta^{\mathrm{oracle}}=
        \bigl((\hbu^{\mathrm{oracle}})^\top,(\hbv^{\mathrm{oracle}})^\top\bigr)^\top\in\argmin_{\substack{\bu\in\R^n,\ \bv\in V(S^*)\\ \bm{1}^{\top}\bu=0}}
        \left\{-l_n(\bu,\bv)+\lambda_n\norm{\bv}_1\right\}.
    \end{equation}

    \item \textbf{Dual variable construction:} Choose $\widehat{\bz}_{S^*} \in\partial\norm{\hbv^{\mathrm{oracle}}_{S^*}}_1$ so that the restricted KKT conditions hold. Since $\hbv^{\mathrm{oracle}}_{S^{*\complement}}=\bm{0}$, the inactive dual variables are determined by the full KKT conditions:
    \begin{equation*}
        \widehat{\bz}_{S^{*\complement}} =
        \frac{1}{\lambda_n} \nabla_{\bv_{S^{*\complement}}}l_n(\hbtheta^{\mathrm{oracle}}).
    \end{equation*}

    \item \textbf{Strict dual feasibility verification:} 
    Verify that the constructed subgradient on the inactive set satisfies $\norm{\widehat{\bz}_{S^{*\complement}}}_\infty < 1$ with high probability. This strict dual feasibility ensures that $\hbtheta^{\mathrm{oracle}}$ is the unique global optimum and that no inactive covariates are selected, that is, $\supp(\hbv^{\mathrm{oracle}}) \subseteq S^*$; the following lemma gives the precise sufficient condition.
\end{enumerate}

\begin{lemma}[PDW sufficient condition]
\label{lem:pdw_condition}
Let $S^*=\operatorname{supp}(\bv^*)$, and suppose that $\hbtheta$ is the unique minimizer of the restricted problem \eqref{eq:oracle}. If there exist a dual vector $\widehat{\bz}\in\partial\norm{\hbv}_1$ and a Lagrange multiplier $\alpha\in\mathbb{R}$ such that $(\hbtheta,\widehat{\bz},\alpha)$ satisfies the KKT conditions in \eqref{eq:kkt_u}--\eqref{eq:kkt_constraint} and
\begin{equation*}
    \norm{\widehat{\bz}_{S^{*\complement}}}_\infty<1,
\end{equation*}
then $\hbtheta$ is the unique minimizer of the full problem
\eqref{eq:mle}. Moreover,
\begin{equation*}
    \operatorname{supp}(\hbv)\subseteq S^*.
\end{equation*}
\end{lemma}
The proof is given in Section~\ref{proof:pdw_condition}.

\subsection{Oracle estimation and dual feasibility}

\begin{lemma}[Existence and consistency of the oracle estimator]
\label{lem:oracle_consistency}
Under Assumptions~\ref{ass:sparsity}--\ref{ass:wellposed}, together with the rate conditions stated above, the oracle estimator exists uniquely with probability at least $1-O(n^{-2})$. On the same event,
\begin{align}
\|\hbv^{\mathrm{oracle}}-\bv^*\|_2&\lesssim\sqrt{d_0}\lambda_n,\label{eq:v-error-final}\\
\|\hbu^{\mathrm{oracle}}-\bu^*\|_\infty
&\lesssim\diam(\mathcal A_{\mathcal H_n}(\eta)) \max
\left\{\sqrt{\frac{\log n}{h_{\mathcal H_n}}}, \,d_0\lambda_n\right\}.\label{eq:refined_u_rate}
\end{align}
\end{lemma}
The proof, given in Section~\ref{proof:oracle_consistency}, proceeds in two stages. It first bounds the squared prediction error, including the \(d_0\lambda_n^2\) contribution from regularization, and then profiles the population Hessian to obtain separate parameter rates.

It remains to control the inactive dual coordinates. The following lemma establishes the strict margin required by the PDW sufficient condition.

\begin{lemma}[Strict dual feasibility]\label{lem:strict_dual_feasibility}
Under the conditions of Lemma~\ref{lem:oracle_consistency} and Assumption~\ref{ass:incoherence}, for sufficiently large $c_{1}$ in~\eqref{eq:pdw_lambda_choice}, the dual vector constructed from the oracle estimator satisfies $\|\widehat{\bz}_{S^{*\complement}}\|_\infty\le1-\alpha/2$ with probability at least $1-O(n^{-2})$ for all sufficiently large $n$.
\end{lemma}

Section~\ref{proof:strict_dual_feasibility} proves this result using the population-profiled dual identity. In the resulting decomposition, the leading term involves $\bm G^*\bm H^{*-1}$, while the fluctuation of the observed Hessian around its population counterpart is controlled separately.

\subsection{Proof of Theorem~\ref{thm:main_result}}
By Lemmas~\ref{lem:oracle_consistency} and~\ref{lem:strict_dual_feasibility}, with probability at least $1-O(n^{-2})$, the oracle estimator exists uniquely and its associated dual vector satisfies $\|\widehat{\bz}_{S^{*\complement}}\|_\infty<1$. On this event, the oracle estimator and the constructed dual vector satisfy the KKT conditions of the full problem. Lemma~\ref{lem:pdw_condition} therefore implies that the full estimator exists uniquely and
\begin{equation*}
\hbtheta=\hbtheta^{\mathrm{oracle}},
\qquad
\supp(\hbv)\subseteq S^*.
\end{equation*}
Consequently, the error bounds in~\eqref{eq:v-error-final}--\eqref{eq:refined_u_rate} also hold for $\hbtheta$. Substituting \eqref{eq:pdw_lambda_choice} gives the rates stated in Theorem~\ref{thm:main_result}, and \eqref{eq:connect_conditions} ensures that both errors converge to zero. This proves the theorem.

To establish Proposition~\ref{prop:exact_support_recovery}, note that, on the same event,
\begin{equation*}
\|\hbv-\bv^*\|_\infty
\le \|\hbv-\bv^*\|_2
\lesssim \sqrt{d_0}\lambda_n.
\end{equation*}
The minimum signal condition, with $C_{\min}$ sufficiently large, thus ensures that
\begin{equation*}   
|\widehat v_j|
\ge |v_j^*|-|\widehat v_j-v_j^*|>0
\qquad\text{for every }j\in S^*.
\end{equation*}
Hence $S^*\subseteq\supp(\hbv)$. Together with the support inclusion established above, this yields $\supp(\hbv)=S^*$ with probability at least $1-O(n^{-2})$.\qed

\section{~Proof of Lemma~\ref{lem:oracle_consistency}}
\label{app:oracle_consistency}
\label{proof:oracle_consistency}

We first establish existence and uniqueness of the restricted estimator, then localize its prediction and parameter errors, and finally profile out the utility coordinates to obtain the stated convergence rates. 

Write $U=[n]$, $S=S^*$, $T=S^{*\complement}$, and $\mathcal I=U\cup S$, using the block-index convention of the main text. Let $\tbtheta=\hbtheta^{\mathrm{oracle}}$, $\bDelta=\tbtheta-\btheta^*$, with inactive coordinates zero. Recall $\W_n=[\Q_n^\top,\K_n]$. Writing $e_i=(j_{i_1},\ldots,j_{i_{m_i}})$ in the fixed ordering and taking $j_{i_1}$ as the reference item, the block of $\W_n$ corresponding to comparison $e_i$ is
\begin{equation*}
\W_i=
\begin{bmatrix}
\vec e_{j_{i_2}j_{i_1}}^\top
& (X_{e_i,j_{i_2}}-X_{e_i,j_{i_1}})^\top\\
\vdots & \vdots\\
\vec e_{j_{i_{m_i}}j_{i_1}}^\top
& (X_{e_i,j_{i_{m_i}}}-X_{e_i,j_{i_1}})^\top
\end{bmatrix}
\in\mathbb R^{(m_i-1)\times(n+d)},
\end{equation*}
where $\vec e_{kl}$ is the difference between the $k$th and $l$th canonical basis vectors in $\mathbb R^n$. Consequently,
\begin{equation*}
\W_i\btheta=
\begin{bmatrix}
s_{j_{i_2}}-s_{j_{i_1}}\\
\vdots\\
s_{j_{i_{m_i}}}-s_{j_{i_1}}
\end{bmatrix},
\qquad
s_k=u_k+X_{e_i,k}^\top\bv.
\end{equation*}
Thus $[\W_n]_{\mathcal I}$ is the restricted design matrix mapping the active parameter vector to all within-comparison score differences.

For comparison $i$, let $\pi_i=(\pi_i(1),\ldots,\pi_i(m_i))$ denote the observed ranking. Its log-likelihood is
\begin{equation*}
\ell_i(\btheta)=
\sum_{r=1}^{m_i-1}\left[ s_{\pi_i(r)}-\log\left\{ \sum_{q=r}^{m_i}\exp(s_{\pi_i(q)}) \right\}\right],
\qquad l_n(\btheta)=\frac1N\sum_{i=1}^N\ell_i(\btheta).
\end{equation*}
Subtracting the reference score $s_{j_{i_1}}$ from all
item scores gives
\begin{equation*}
\ell_i(\btheta) =
\sum_{r=1}^{m_i-1}
\left[ s_{\pi_i(r)}-s_{j_{i_1}} -\log\left\{ \sum_{q=r}^{m_i} \exp\bigl(s_{\pi_i(q)}-s_{j_{i_1}}\bigr) \right\} \right].
\end{equation*}
Since
\begin{equation*}
s_{j_{i_t}}-s_{j_{i_1}} =
\begin{cases}
0, & t=1,\\
[\W_i\btheta]_{t-1}, & t=2,\ldots,m_i,
\end{cases}
\end{equation*}
the log-likelihood depends on $\btheta$ only through
$\W_i\btheta$. View $\ell_i$ as a function of the score differences
$\bm s=\W_i\btheta\in\mathbb R^{m_i-1}$.
We write $\nabla_{\bm s}\ell_i(\btheta)$ and
$\nabla_{\bm s}^2\ell_i(\btheta)$ for its gradient and
Hessian with respect to these score differences,
evaluated at $\bm s=\W_i\btheta$.
In contrast, $\nabla\ell_i(\btheta)$ and
$\nabla^2\ell_i(\btheta)$ denote derivatives with respect
to the parameter vector $\btheta$. The chain rule gives
\begin{equation*}
\nabla\ell_i(\btheta)
=\W_i^\top\nabla_{\bm s}\ell_i(\btheta),
\qquad
\nabla^2\ell_i(\btheta)
=\W_i^\top\nabla_{\bm s}^2\ell_i(\btheta)\W_i.
\end{equation*}

\subsection{Existence and uniqueness}

This subsection establishes the well-posedness of the restricted optimization problem defining the oracle estimator. Uniqueness follows from strict convexity, while existence is established through a coercivity argument based on the observed ordering event.

Retain $\G_1$ for the existence event in Assumption~\ref{ass:wellposed}\ref{ass:exist}. On the event $\G_1$, every nonzero parameter vector $\btheta=(\bu^\top,\bv^\top)^\top\in\mathbb R^{n+d_0}$ satisfying $\bm 1^\top\bu=0$ disagrees with at least one observed ordering. More precisely, there exist a comparison $e\in\mathcal E$ and two items $k_1,k_2\in e$, with $k_1\succ k_2$ on $e$, such that
\begin{equation}
\label{eq:coercive_direction}
u_{k_1}+[X_{e,k_1}]_{S^*}^\top\bv < u_{k_2}+[X_{e,k_2}]_{S^*}^\top\bv.
\end{equation}

We first verify strict convexity. For any $\btheta$ in the restricted parameter space,
\begin{equation*}
\nabla^2 l_{S^*}(\btheta)
=\frac1N\sum_{i\in[N]}[\W_i]_{\mathcal I}^\top
\nabla^2_{\bm s}\ell_i(\btheta)[\W_i]_{\mathcal I}.
\end{equation*}
At each PL stage, the negative Hessian is the covariance matrix of a categorical variable and is therefore positive semidefinite. At the first stage every item in the comparison is present, so after fixing the reference item the negative contrast Hessian is positive definite. Together with \eqref{eq:coercive_direction}, this implies that every nonzero normalized direction is detected by at least one comparison. Consequently,
\begin{equation*}
-\btheta'^\top\nabla^2 l_{S^*}(\btheta)\btheta'>0
\end{equation*}
for every nonzero $\btheta'$ satisfying $\bm1_n^\top\bu'=0$. Thus $-l_{S^*}$ is strictly convex on the restricted parameter space. Adding the convex penalty $\lambda_n\|\bv\|_1$ preserves strict convexity.

We next verify coercivity. Define
\begin{equation*}
\Phi(\btheta)=\max_{i\in[N]}\max_{j\in[m_i-1]}
\max_{j<t\le m_i}[Z_{(i,j,t)}]_{\mathcal I}^\top\btheta.
\end{equation*}
By~\eqref{eq:coercive_direction}, $\Phi(\btheta)>0$ for every nonzero normalized parameter vector. Continuity and compactness of the unit sphere in the normalized space give a constant $c_{2}>0$, for the fixed design and observed rankings, such that
\begin{equation*}
\Phi(\btheta)\ge c_{2}\|\btheta\|_2.
\end{equation*}
Moreover, for every $(i,j,t)$ with $j<t\le m_i$,
\begin{equation*}
-l_{S^*}(\btheta)
\ge\frac1N\log\left(1+\sum_{q=j+1}^{m_i}
\exp\{[Z_{(i,j,q)}]_{\mathcal I}^\top\btheta\}\right)
\ge\frac1N[Z_{(i,j,t)}]_{\mathcal I}^\top\btheta.
\end{equation*}
Taking the maximum yields
\begin{equation*}
-l_{S^*}(\btheta)\ge\frac{c_{2}}{N}\|\btheta\|_2.
\end{equation*}
Hence the restricted objective is coercive and strictly convex, so its minimizer exists uniquely.

\subsection{Prediction bounds and localization}

The utility score equation gives a chaining bound without assuming that the
oracle estimator is already close to the truth. Define the event appearing in
that bound by
\begin{equation}\label{G2}
\G_2=\left\{
\left|\sum_{k\in A}\partial_{u_k} l_n(\btheta^*)\right|
\le\frac{c_{3}}{N}\sqrt{|A|\,|\partial A|\log n}
\text{ for all }0<|A|\le n/2\right\}.
\end{equation}

\begin{lemma}[Chaining upper bound]
\label{chaining}
On $\G_1\cap \G_2$, if $\log n/h_{\mathcal H_n}=o(1)$, there is a fixed $\eta\in(0,1/2]$ such that
\begin{equation}\label{eq:delta-chaining}
\|\bDelta_{\bu}\|_\infty\lesssim
\diam(\mathcal A_{\mathcal H_n}(\eta))
\max \left\{\sqrt{\frac{\log n}{h_{\mathcal H_n}}}, \, \|\bDelta_{\bv}\|_1\right\}.
\end{equation}
\end{lemma}

We combine the chaining inequality with the optimality of
$\tbtheta$ to bound the prediction error and show that
$\|\bDelta_{\bu}\|_\infty+\|\bDelta_{\bv}\|_1=o(1)$,
where $\bDelta=\tbtheta-\btheta^*$ and $\bDelta_T=0$.
This argument requires a bound on the score term
$\nabla l_n(\btheta^*)^\top\bDelta$ and a lower bound
on the curvature of the negative log-likelihood near $\btheta^*$.

For the score term, define
\begin{equation}\label{G3}
\G_3=\left\{
|\nabla l_n(\btheta^*)^\top\bm w|
\le \frac{c_{4}\sqrt{n+d_0+\log(n+d_0)}}{N}
\|\W_n\bm w\|_2
\quad\text{for every $\bm w\in\R^{n+d}$ with $\bm w_T=0$}
\right\}.
\end{equation}
This bound applies to $\bm w=\bDelta$ since $\bDelta_T=0$.
We work on $\G_1\cap\G_2\cap\G_3$ throughout this subsection;
its probability is bounded in Lemma~\ref{lem:all_events}.

We now establish the curvature bound for parameters satisfying
\begin{equation*}
\norm{\bu-\bu^*}_\infty+\norm{\bv_S-\bv_S^*}_1\le1,
\qquad \bv_T=0.
\end{equation*}
We will then use this bound to show that $\tbtheta$
also lies in this neighborhood.
For comparison $i$, write
$s_{i,k}(\btheta)=u_k+X_{e_i,k}^\top\bv$ for the score of item $k\in e_i$.
Since $\bv_T=\bv_T^*=0$,
\begin{equation*}
\abs{s_{i,k}(\btheta)-s_{i,k}(\btheta^*)}
\leq
\norm{\bu-\bu^*}_\infty
+\norm{X_{e_i,k}}_\infty
 \norm{\bv_S-\bv_S^*}_1
\leq
\max\{1,\norm{X_{e_i,k}}_\infty\}.
\end{equation*}
The uniform bounds on the covariates and the true scores therefore imply
that there exists a fixed constant $c_{5}<\infty$ such that
\begin{equation*}
\abs{s_{i,k}(\btheta)}\leq c_{5}
\end{equation*}
for every $i$, every $k\in e_i$, and every $\btheta$ in this neighborhood.

For the observed ranking $\pi_i$, define
\begin{equation*}
p_{i,k}^{(t)}(\btheta) =
\frac{\mathbf1\{k\in\{\pi_i(t),\ldots,\pi_i(m_i)\}\}\exp\{s_{i,k}(\btheta)\}}{\sum_{q=t}^{m_i}\exp\{s_{i,\pi_i(q)}(\btheta)\}},
\qquad k\in e_i,\quad 1\le t\le m_i-1.
\end{equation*}
At the first stage, all items are available. The score bound and $m_i\le M$ therefore imply
\begin{equation*}
p_{i,k}^{(1)}(\btheta)\ge \frac{e^{-2c_5}}{M},
\qquad k\in e_i.
\end{equation*}

Fix $\bm a\in\mathbb R^{m_i-1}$ and set $a_0=0$, so that item $j_{i_q}$ has score perturbation $a_{q-1}$. The second derivative of each stage log-sum-exp is the variance of these perturbations under the stage probabilities:
\begin{align}
&\sum_{q=1}^{m_i}p_{i,j_{i_q}}^{(t)}(\btheta)a_{q-1}^2
-\left\{
\sum_{q=1}^{m_i}p_{i,j_{i_q}}^{(t)}(\btheta)a_{q-1}
\right\}^2 \notag\\
&\qquad=
\sum_{1\le q<r\le m_i}
p_{i,j_{i_q}}^{(t)}(\btheta)
p_{i,j_{i_r}}^{(t)}(\btheta)
(a_{q-1}-a_{r-1})^2.
\label{eq:edgewise_stage_covariance}
\end{align}
Summing over stages gives
\begin{equation*}
\bm a^\top\{-\nabla_{\bm s}^2\ell_i(\btheta)\}\bm a
=
\sum_{t=1}^{m_i-1}
\sum_{1\le q<r\le m_i}
p_{i,j_{i_q}}^{(t)}(\btheta)
p_{i,j_{i_r}}^{(t)}(\btheta)
(a_{q-1}-a_{r-1})^2.
\end{equation*}
For the lower bound, retain only first-stage pairs
containing the reference item:
\begin{equation*}
\bm a^\top\{-\nabla_{\bm s}^2\ell_i(\btheta)\}\bm a
\ge
\sum_{q=2}^{m_i}
p_{i,j_{i_1}}^{(1)}(\btheta)
p_{i,j_{i_q}}^{(1)}(\btheta)a_{q-1}^2
\ge \frac{e^{-4c_5}}{M^2}\|\bm a\|_2^2.
\end{equation*}
For the upper bound, each stage variance is at most its second moment, so
\begin{equation*}
\bm a^\top\{-\nabla_{\bm s}^2\ell_i(\btheta)\}\bm a
\le
\sum_{t=1}^{m_i-1}\sum_{q=1}^{m_i}
p_{i,j_{i_q}}^{(t)}(\btheta)a_{q-1}^2
\le (M-1)\|\bm a\|_2^2.
\end{equation*}
Because these inequalities hold for every
$\bm a\in\mathbb R^{m_i-1}$, we conclude that
\begin{equation}
\label{eq:edge_curvature}
\begin{gathered}
\lambda_{\min}\!\left(-\nabla_{\bm s}^2\ell_i(\btheta)\right)\gtrsim1,\qquad
\left\|\nabla_{\bm s}^2\ell_i(\btheta)\right\|_{\mathrm{op}}\lesssim1
\end{gathered}
\end{equation}
where the implicit constants are independent of $i,n,N,d,d_0$ and of the observed ranking. Moreover, subspace separation and the spectral stability of the active design imply that, for every restricted direction $\bm w$,
\begin{equation}
\label{eq:prediction_identification}
\|\W_n\bm w\|_2^2
\gtrsim
\|\Q_n^\top\bm w_U\|_2^2
+\|[\K_n]_S\bm w_S\|_2^2
\gtrsim
N\|\bm w_S\|_2^2,
\qquad
\bm w_T=0.
\end{equation}
Consequently,
\begin{equation}
\label{eq:active_l1_prediction}
\|\bm w_S\|_1
\le
\sqrt{d_0}\|\bm w_S\|_2
\lesssim
\sqrt{\frac{d_0}{N}}\|\W_n\bm w\|_2.
\end{equation}

To invoke the local curvature without assuming localization in advance, consider the segment joining $\btheta^*$ and $\tbtheta$. Let $0<t\le1$ be such that
\begin{equation}
\label{eq:admissible_scaling}
t\bigl(
\|\bDelta_{\bu}\|_\infty+\|\bDelta_{\bv}\|_1
\bigr)\le1.
\end{equation}
Write
\begin{equation*}
F_n(\btheta)=-l_n(\btheta)+\lambda_n\|\bv\|_1
\end{equation*}
for the restricted regularized objective. Since $F_n$ is convex and is minimized
at $\tbtheta=\btheta^*+\bDelta$,
\begin{equation*}
F_n(\btheta^*+t\bDelta)
\le
(1-t)F_n(\btheta^*)+tF_n(\tbtheta)
\le
F_n(\btheta^*).
\end{equation*}
Rearranging this inequality and using the triangle inequality yields
\begin{align}
l_n(\btheta^*+t\bDelta)-l_n(\btheta^*)
&\ge
\lambda_n
\bigl(
\|\bv^*+t\bDelta_{\bv}\|_1-\|\bv^*\|_1
\bigr) \notag\\
&\ge
-\lambda_n t\|\bDelta_{\bv}\|_1.
\label{eq:basicineq}
\end{align}

Because \eqref{eq:admissible_scaling} ensures that the entire segment $\{\btheta^*+st\bDelta:0\le s\le1\}$ remains in the local-curvature region, Taylor's integral formula and \eqref{eq:edge_curvature} give
\begin{equation*}
l_n(\btheta^*+t\bDelta)-l_n(\btheta^*)
\le
t\nabla l_n(\btheta^*)^\top\bDelta
-\frac{ct^2}{N}\|\W_n\bDelta\|_2^2,
\end{equation*}
where $c=e^{-4c_5}/(2M^2)>0$.
Combining this inequality with \eqref{eq:basicineq}, and then applying the definition of $\G_3$ and
\eqref{eq:active_l1_prediction}, we obtain
\begin{align*}
\frac{ct^2}{N}\|\W_n\bDelta\|_2^2
&\le
t\bigl|\nabla l_n(\btheta^*)^\top\bDelta\bigr|
+\lambda_n t\|\bDelta_{\bv}\|_1\\
&\lesssim
\frac{\sqrt{n+d_0+\log(n+d_0)}}{N}
t\|\W_n\bDelta\|_2
+\lambda_n\sqrt{\frac{d_0}{N}}
t\|\W_n\bDelta\|_2\\
&=
\left\{
\sqrt{\frac{n+d_0+\log(n+d_0)}{N}}
+\sqrt{d_0}\lambda_n
\right\}
\frac{t\|\W_n\bDelta\|_2}{\sqrt N}.
\end{align*}
If $\|\W_n\bDelta\|_2=0$, the desired bound is immediate. Otherwise, dividing both sides by $t\|\W_n\bDelta\|_2/\sqrt N$ gives
\begin{equation}
\label{eq:scaled_prediction_root}
\frac{t\|\W_n\bDelta\|_2}{\sqrt N}
\lesssim
\sqrt{\frac{n+d_0+\log(n+d_0)}{N}}
+\sqrt{d_0}\lambda_n.
\end{equation}
Combining this with \eqref{eq:prediction_identification} further yields
\begin{equation}
\label{eq:scaled_active_l1}
t\|\bDelta_{\bv}\|_1
\lesssim
\sqrt{\frac{d_0\{n+d_0+\log(n+d_0)\}}{N}}
+d_0\lambda_n.
\end{equation}

We now verify that $t=1$ is admissible. The chaining bound
\eqref{eq:delta-chaining}, which applies directly to the actual estimator
$\tbtheta$, implies
\begin{equation*}
t\|\bDelta_{\bu}\|_\infty
\lesssim
\diam(\mathcal A_{\mathcal H_n}(\eta))
\max\left\{
\sqrt{\frac{\log n}{h_{\mathcal H_n}}}
,\ t\|\bDelta_{\bv}\|_1
\right\}.
\end{equation*}
Since $\diam(\mathcal A_{\mathcal H_n}(\eta))\ge1$, adding $t\|\bDelta_{\bv}\|_1$ to both sides and using \eqref{eq:scaled_active_l1} gives
\begin{align}
&t\bigl(
\|\bDelta_{\bu}\|_\infty+\|\bDelta_{\bv}\|_1
\bigr) \notag\\
&\qquad\lesssim
\diam(\mathcal A_{\mathcal H_n}(\eta))
\max\left\{
\sqrt{\frac{\log n}{h_{\mathcal H_n}}}
,\ 
\sqrt{\frac{d_0\{n+d_0+\log(n+d_0)\}}{N}}
+d_0\lambda_n
\right\}.
\label{eq:scaled_localization}
\end{align}
The first and third terms on the right-hand side are $o(1)$ by \eqref{eq:connect_conditions}. For the middle term, \eqref{eq:pdw_lambda_choice} implies
\begin{equation*}
d_0\diam^2(\mathcal A_{\mathcal H_n}(\eta))
\frac{n+d_0+\log(n+d_0)}{N}
\lesssim
d_0\lambda_n
\diam(\mathcal A_{\mathcal H_n}(\eta))
=o(1),
\end{equation*}
and hence
\begin{equation*}
\diam(\mathcal A_{\mathcal H_n}(\eta))
\sqrt{\frac{d_0\{n+d_0+\log(n+d_0)\}}{N}}
=o(1).
\end{equation*}
Thus the right-hand side of \eqref{eq:scaled_localization} is $o(1)$.

Suppose now that
\begin{equation*}
\|\bDelta_{\bu}\|_\infty+\|\bDelta_{\bv}\|_1>1.
\end{equation*}
Then the choice
\begin{equation*}
t=
\bigl(
\|\bDelta_{\bu}\|_\infty+\|\bDelta_{\bv}\|_1
\bigr)^{-1}
\end{equation*}
satisfies \eqref{eq:admissible_scaling}, while the left-hand side of \eqref{eq:scaled_localization} is exactly one. This contradicts \eqref{eq:scaled_localization} for all sufficiently large $n$. Therefore,
\begin{equation*}
\|\bDelta_{\bu}\|_\infty+\|\bDelta_{\bv}\|_1\le1,
\end{equation*}
so $t=1$ is admissible. Substituting $t=1$ into
\eqref{eq:scaled_prediction_root} and \eqref{eq:scaled_localization} gives, on
$\G_1\cap \G_2\cap \G_3$,
\begin{align}
\frac{\|\W_n\bDelta\|_2}{\sqrt N}
&\lesssim
\sqrt{\frac{n+d_0+\log(n+d_0)}{N}}
+\sqrt{d_0}\lambda_n,
\label{eq:prediction_root}\\
\|\bDelta_{\bu}\|_\infty+\|\bDelta_{\bv}\|_1
&=o(1).
\label{eq:prediction_localization}
\end{align}
Finally, squaring \eqref{eq:prediction_root} and using $(a+b)^2\lesssim a^2+b^2$ yields
\begin{equation}
\label{eq:prediction_energy}
\frac{\|\W_n\bDelta\|_2^2}{N}
\lesssim
\frac{n+d_0+\log(n+d_0)}{N}
+d_0\lambda_n^2.
\end{equation}

\subsection{Population profiling and convergence rates}

Building on the prediction-error and local-curvature bounds, we now derive separate rates for the covariate and utility coordinates. Since the utility Hessian is singular along the constant direction, we profile out the utilities from the population KKT equation to obtain an exact equation for $\bDelta_{\bv}$ with four error terms. Uniform invertibility of the profiled curvature matrix, together with the prediction-error bound, yields the covariate rate. The utility rate then follows from the chaining inequality.

Retain $\bm M=\nabla^2l_n(\btheta^*)$, $\bm M^*=\expect\bm M$ and
$\bm Q=\nabla l_n(\btheta^*)$, and define the Taylor remainder, with its sign
fixed by the identity below, as
\begin{equation}
\label{eq:Rl_integral}
\bm R=\int_0^1\{\nabla^2l_n(\btheta^*+t\bDelta)-\bm M\}\bDelta\,dt,
\end{equation}
so that Taylor's formula reads
$\nabla l_n(\btheta^*+\bDelta)=\bm Q+\bm M\bDelta+\bm R$. Combining this
identity with the restricted KKT equation and then adding and subtracting
$\bm M^*\bDelta$ yields the exact decomposition
\begin{equation}
\label{eq:population_kkt}
\bm M^*_{\mathcal I\mathcal I}\bDelta_{\mathcal I}
=\lambda_n(\bm0_U^\top,\widehat{\bz}_S^\top)^\top-\bm Q_{\mathcal I}
-(\bm M-\bm M^*)_{\mathcal I\mathcal I}\bDelta_{\mathcal I}
-\bm R_{\mathcal I}.
\end{equation}

By shift invariance, every vector in the $U$-block of~\eqref{eq:population_kkt} is orthogonal to $\bm1_n$, and $\bDelta_{\bu}$ is centered; the $U$-block may therefore be solved on $\bm1_n^\perp$, giving
\begin{equation*}
\bDelta_{\bu}=-(\bm M^*_{UU})^\dagger\bigl\{
\bm M^*_{US}\bDelta_{\bv}+\bm Q_U
+[(\bm M-\bm M^*)_{\mathcal I\mathcal I}\bDelta_{\mathcal I}]_U
+\bm R_U\bigr\}.
\end{equation*}
Substituting this expression into the $S$-block of~\eqref{eq:population_kkt} and
recalling that
$\bm H^*=\bm M^*_{SS}-\bm M^*_{SU}(\bm M^*_{UU})^\dagger\bm M^*_{US}$
produces the exact profiled equation
\begin{equation}\label{eq:profiled_kkt}
\begin{aligned}
\bm H^*\bDelta_{\bv}
={}&\lambda_n\widehat{\bz}_S
-\left\{\bm Q_S-\bm M^*_{SU}(\bm M^*_{UU})^\dagger\bm Q_U\right\}\\
&-\left\{[(\bm M-\bm M^*)_{\mathcal I\mathcal I}\bDelta_{\mathcal I}]_S
-\bm M^*_{SU}(\bm M^*_{UU})^\dagger
[(\bm M-\bm M^*)_{\mathcal I\mathcal I}\bDelta_{\mathcal I}]_U\right\}\\
&-\left\{\bm R_S-\bm M^*_{SU}(\bm M^*_{UU})^\dagger\bm R_U\right\}.
\end{aligned}
\end{equation}
We refer to the four terms on the right-hand side, in order, as (i) the penalty bias, (ii) the profiled score noise, (iii) the profiled Hessian fluctuation, and (iv) the profiled nonlinear remainder.

We next verify that the profiled population curvature is uniformly invertible. There are three ingredients: the variational representation of the Schur complement, separation of the utility and active-covariate design spaces, and the spectral stability of the active design.

First, shift invariance of the likelihood implies
\begin{equation*}
\Q_n^\top\bm1_n=\bm0,
\qquad
\bm M^*_{UU}\bm1_n=\bm0,
\qquad
\bm M^*_{SU}\bm1_n=\bm0.
\end{equation*}
Because the comparison hypergraph is connected and the contrast curvature at $\btheta^*$ is positive, the constant direction is the only null direction of the utility block:
\begin{equation*}
\ker(\bm M^*_{UU})=\operatorname{span}(\bm1_n).
\end{equation*}
Thus $-\bm M^*_{UU}$ is positive definite on $\bm1_n^\perp$. With the Moore--Penrose convention in the main text,
\begin{equation*}
(\bm M^*_{UU})^\dagger
=(\bm M^*_{UU}+\bm1_n\bm1_n^\top/n)^{-1}
-\bm1_n\bm1_n^\top/n.
\end{equation*}
It agrees with the inverse of $\bm M^*_{UU}$ on $\bm1_n^\perp$ and annihilates $\bm1_n$. All utility vectors to which it is applied in the profiled equations are centered.

For any $\bm b\in\mathbb R^{d_0}$, consider the population curvature in the joint direction $(\bm a^\top,\bm b^\top)^\top$, where $\bm1_n^\top\bm a=0$. Completing the square in $\bm a$ shows that the minimum is attained at
\begin{equation*}
\bm a
=
-(\bm M^*_{UU})^\dagger\bm M^*_{US}\bm b.
\end{equation*}
Substitution of this minimizer gives the Schur-complement identity
\begin{equation}
\label{eq:profile_curvature_variational}
\bm b^\top(-\bm H^*)\bm b
=
\inf_{\bm1_n^\top\bm a=0}
(\bm a^\top,\bm b^\top)
(-\bm M^*_{\mathcal I\mathcal I})
(\bm a^\top,\bm b^\top)^\top.
\end{equation}
This identity has a useful interpretation: after fixing an active-covariate direction $\bm b$, profiling chooses the utility direction $\bm a$ that explains as much of its effect as possible; $-\bm H^*$ measures the curvature remaining after this optimal adjustment.

At $\btheta^*$, the contrast-curvature bound~\eqref{eq:edge_curvature} and the chain rule imply, uniformly over $\bm a$ and $\bm b$,
\begin{equation}
\label{eq:population_design_curvature}
(\bm a^\top,\bm b^\top)
(-\bm M^*_{\mathcal I\mathcal I})
(\bm a^\top,\bm b^\top)^\top
\asymp
\frac1N
\norm{\Q_n^\top\bm a+[\K_n]_S\bm b}_2^2.
\end{equation}
Combining the lower inequality in~\eqref{eq:population_design_curvature} with \eqref{eq:profile_curvature_variational} gives
\begin{align}
\bm b^\top(-\bm H^*)\bm b
&\gtrsim
\frac1N
\inf_{\bm1_n^\top\bm a=0}
\norm{\Q_n^\top\bm a+[\K_n]_S\bm b}_2^2\notag\\
&=
\frac1N
\operatorname{dist}^2\!\left(
[\K_n]_S\bm b,
\operatorname{range}(\Q_n^\top)
\right).
\label{eq:profile_curvature_distance}
\end{align}
The equality holds because centering $\bm a$ does not change
$\Q_n^\top\bm a$, as $\Q_n^\top\bm1_n=\bm0$.

Assumption~\ref{ass:wellposed}\ref{subspace} states that the maximal correlation between $\operatorname{range}(\Q_n^\top)$ and $\operatorname{range}([\K_n]_S)$ is at most $1-\epsilon$. Since $[\K_n]_S\bm b\in\operatorname{range}([\K_n]_S)$, its projection onto $\operatorname{range}(\Q_n^\top)$ therefore has norm at most $(1-\epsilon)\norm{[\K_n]_S\bm b}_2$. By the Pythagorean identity,
\begin{align*}
\operatorname{dist}^2\!\left(
[\K_n]_S\bm b,
\operatorname{range}(\Q_n^\top)
\right)
&\geq
\{1-(1-\epsilon)^2\}
\norm{[\K_n]_S\bm b}_2^2.
\end{align*}
The lower active-design spectral bound in
Assumption~\ref{ass:wellposed}\ref{nondegeneracy} then gives
\begin{equation*}
\frac1N\norm{[\K_n]_S\bm b}_2^2
\gtrsim\norm{\bm b}_2^2.
\end{equation*}
Substituting these two bounds into~\eqref{eq:profile_curvature_distance} yields
\begin{equation}
\label{eq:profile_curvature_lower}
\bm b^\top(-\bm H^*)\bm b
\gtrsim
\norm{\bm b}_2^2.
\end{equation}

For the upper bound, take $\bm a=\bm0$ in the infimum in~\eqref{eq:profile_curvature_variational}. The upper inequality in~\eqref{eq:population_design_curvature} and the upper active-design spectral
bound give
\begin{equation}
\label{eq:profile_curvature_upper}
\bm b^\top(-\bm H^*)\bm b
\leq
\bm b^\top(-\bm M^*_{SS})\bm b
\lesssim
\frac1N\norm{[\K_n]_S\bm b}_2^2
\lesssim
\norm{\bm b}_2^2.
\end{equation}
Since~\eqref{eq:profile_curvature_lower} and
\eqref{eq:profile_curvature_upper} hold for every
$\bm b\in\mathbb R^{d_0}$, we conclude that
\begin{equation}
\label{eq:inc_population_profile_curvature}
\lambda_{\min}(-\bm H^*)\gtrsim1,\qquad\|\bm H^*\|_{\mathrm{op}}\lesssim1.
\end{equation}

We turn to the four terms. Terms (ii)--(iv) share the same residualized form, and it is more economical to bound them through a single family of directions than one by one. The next result shows that these directions are controlled at the level of each individual comparison, which is what allows the concentration arguments below to be applied blockwise.
\begin{lemma}
\label{lem:graph_residualization}
For every covariate coordinate $r\in S\cup T$, let $\bm b=(\bm M^*_{UU})^\dagger\bm M^*_{Ur}$. Then, uniformly in $r$,
\begin{equation}\label{eq:residual_oscillation}
\max_k b_k-\min_k b_k\lesssim\diam(\mathcal A_{\mathcal H_n}(\eta)).
\end{equation}
Moreover, each comparison block of the residualized column
$[\K_n]_r-\Q_n^\top(\bm M^*_{UU})^\dagger\bm M^*_{Ur}$
has Euclidean norm $\lesssim\diam(\mathcal A_{\mathcal H_n}(\eta))$.
\end{lemma}

Accordingly, for each $r\in S$ we define the profiled direction
$\bm w_{\mathrm{prof}}^{(r)}\in\R^{n+d}$ by
\begin{equation*}
(\bm w_{\mathrm{prof}}^{(r)})_U=-(\bm M^*_{UU})^\dagger\bm M^*_{US}\bm e_r,
\qquad
(\bm w_{\mathrm{prof}}^{(r)})_S=\bm e_r,
\qquad
(\bm w_{\mathrm{prof}}^{(r)})_T=0,
\end{equation*}
where $\bm e_r\in\mathbb R^{d_0}$ is the canonical basis vector corresponding to coordinate $r\in S$.  $\W_i\bm w_{\mathrm{prof}}^{(r)}$ is the $i$th comparison block of the residualized covariate column in Lemma~\ref{lem:graph_residualization}, and the lemma supplies the required uniform blockwise bound. The expected contrast curvature controls the average squared size of these directions as well, because
$(\bm w_{\mathrm{prof}}^{(r)})^\top(-\bm M^*)\bm w_{\mathrm{prof}}^{(r)}=(-\bm H^*)_{rr}\lesssim1$. Together,
\begin{equation}
\label{eq:direction_design}
\max_i\|\W_i\bm w_{\mathrm{prof}}^{(r)}\|_2\lesssim\diam(\mathcal A_{\mathcal H_n}(\eta)),
\qquad
\frac1N\sum_i\|\W_i\bm w_{\mathrm{prof}}^{(r)}\|_2^2\lesssim1.
\end{equation}
Indeed, by the definition of $\bm M^*$ and the local curvature bound at
$\btheta^*$,
\begin{equation*}
(\bm w_{\mathrm{prof}}^{(r)})^\top(-\bm M^*)\bm w_{\mathrm{prof}}^{(r)}
=
\frac1N\sum_{i=1}^N
(\W_i\bm w_{\mathrm{prof}}^{(r)})^\top
\expect\!\left[
-\nabla^2_{\bm s}\ell_i(\btheta^*)
\mid\mathscr D_n
\right]
(\W_i\bm w_{\mathrm{prof}}^{(r)})
\gtrsim
\frac1N\sum_{i=1}^N\norm{\W_i\bm w_{\mathrm{prof}}^{(r)}}_2^2.
\end{equation*}

The value of this construction is that the $r$th coordinate of each of (ii), (iii) and (iv) becomes a scalar directional quantity; for instance
\begin{equation*}
(\bm w_{\mathrm{prof}}^{(r)})^\top\bm Q
=\bm e_r^\top\left\{\bm Q_S
-\bm M^*_{SU}(\bm M^*_{UU})^\dagger\bm Q_U\right\}.
\end{equation*}
We control these quantities on two events, defined using the active directions
$\bm w_{\mathrm{prof}}^{(r)}$ constructed above:
\begin{equation}\label{G4}
\G_4=\left\{
|(\bm w_{\mathrm{prof}}^{(r)})^\top\bm Q|\le c_{6}\sqrt{\frac{\log(n+d_0)}N}
\text{ for every }r\in S
\right\}
\end{equation}
and
\begin{equation}\label{G5}
\G_5=\left\{
\begin{aligned}
|(\bm w_{\mathrm{prof}}^{(r)})^\top&(\bm M-\bm M^*)\bDelta|
\le c_{7}\diam(\mathcal A_{\mathcal H_n}(\eta))
\sqrt{\frac{n+d_0+\log(n+d_0)}N}
\frac{\|\W_n\bDelta\|_2}{\sqrt N} \\
&\text{for every $r\in S$ and every $\bDelta$ with $\bDelta_T=0$}
\end{aligned}
\right\}.
\end{equation}
In the definition of $\G_5$, the bound is uniform over all restricted error vectors, so it applies in particular to the data-dependent oracle error. For the rest of the oracle argument, we work on $\G_1\cap\cdots\cap \G_5$. Lemma~\ref{lem:all_events} establishes its probability at the end of appendix. The remainder estimate below is deterministic and requires no additional event.

The penalty bias (i) is immediate from the subgradient constraint: since $\|\widehat{\bz}_S\|_\infty\le1$, we have $\|\lambda_n\widehat{\bz}_S\|_2\le\sqrt{d_0}\lambda_n$.

For the profiled score noise (ii), the $r$th coordinate is $(\bm w_{\mathrm{prof}}^{(r)})^\top\bm Q$, so using the definition of $\G_4$ over the $d_0$ active coordinates and passing from the maximum norm to the Euclidean norm gives
\begin{equation*}
\left\|\bm Q_S-\bm M^*_{SU}(\bm M^*_{UU})^\dagger\bm Q_U\right\|_2
\lesssim\sqrt{\frac{d_0\log(n+d_0)}N}.
\end{equation*}

The profiled Hessian fluctuation (iii) involves the data-dependent error $\bDelta$, which is precisely why the uniformity in the definition of $\G_5$ is needed; using $\G_5$ coordinatewise gives
\begin{equation*}
\begin{aligned}
&\left\|[(\bm M-\bm M^*)_{\mathcal I\mathcal I}\bDelta_{\mathcal I}]_S
-\bm M^*_{SU}(\bm M^*_{UU})^\dagger
[(\bm M-\bm M^*)_{\mathcal I\mathcal I}\bDelta_{\mathcal I}]_U\right\|_2\\
&\qquad\lesssim
\sqrt{d_0}\diam(\mathcal A_{\mathcal H_n}(\eta))
\sqrt{\frac{n+d_0+\log(n+d_0)}N}
\frac{\|\W_n\bDelta\|_2}{\sqrt N}.
\end{aligned}
\end{equation*}

The profiled nonlinear remainder (iv) is quadratic in the prediction error. Indeed, each PL stage is a linear term minus a log-sum-exp function of a uniformly bounded number of coordinates. Its third derivatives are therefore uniformly bounded, which implies the global Lipschitz bound
\begin{equation}
\label{eq:contrast_hessian_lipschitz}
\norm{
\nabla^2_{\bm s}\ell_i(\btheta)
-\nabla^2_{\bm s}\ell_i(\btheta')
}_{\mathrm{op}}
\leq
L\norm{\W_i(\btheta-\btheta')}_2.
\end{equation}
Thus the change of the contrast Hessian along the segment from $\btheta^*$ to $\btheta^*+\bDelta$ is proportional to the corresponding contrast error, and the chain rule together with the first bound in~\eqref{eq:direction_design} gives
\begin{align}
|(\bm w_{\mathrm{prof}}^{(r)})^\top\bm R|
&\le\frac1N\sum_i\int_0^1\|\W_i\bm w_{\mathrm{prof}}^{(r)}\|_2
\|\nabla^2_{\bm s}\ell_i(\btheta^*+t\bDelta)
-\nabla^2_{\bm s}\ell_i(\btheta^*)\|_{\mathrm{op}}
\|\W_i\bDelta\|_2\,dt\notag\\
&\lesssim\frac1N\sum_i\|\W_i\bm w_{\mathrm{prof}}^{(r)}\|_2\|\W_i\bDelta\|_2^2
\;\lesssim\;\diam(\mathcal A_{\mathcal H_n}(\eta))
\frac{\|\W_n\bDelta\|_2^2}{N}.\label{eq:directional_remainder}
\end{align}
The remainder is therefore controlled by the squared prediction error alone, without any factor involving the maximal utility error, which is what keeps the final sample size requirement from deteriorating. Applying~\eqref{eq:directional_remainder} to all $d_0$ active coordinates gives
\begin{equation*}
\left\|\bm R_S-\bm M^*_{SU}(\bm M^*_{UU})^\dagger\bm R_U\right\|_2
\lesssim\sqrt{d_0}\diam(\mathcal A_{\mathcal H_n}(\eta))
\frac{\|\W_n\bDelta\|_2^2}{N}.
\end{equation*}

Since $\|\bm H^{*-1}\|_{\mathrm{op}}\lesssim1$ by~\eqref{eq:inc_population_profile_curvature}, we may solve the exact profiled equation~\eqref{eq:profiled_kkt} for $\bDelta_{\bv}$ without changing the orders of the four bounds just obtained. Combining them yields
\begin{align}
\label{eq:oracle_profile_bound}
\|\bDelta_{\bv}\|_2\lesssim\sqrt{d_0}\bigg[&
\lambda_n+\sqrt{\frac{\log(n+d_0)}N}
+\diam(\mathcal A_{\mathcal H_n}(\eta))
\sqrt{\frac{n+d_0+\log(n+d_0)}N}
\frac{\|\W_n\bDelta\|_2}{\sqrt N}\notag\\
&+\diam(\mathcal A_{\mathcal H_n}(\eta))
\frac{\|\W_n\bDelta\|_2^2}{N}\bigg].
\end{align}

Substituting~\eqref{eq:prediction_root} and~\eqref{eq:prediction_energy} into~\eqref{eq:oracle_profile_bound}, and using $\sqrt{\log(n+d_0)/N}\lesssim\lambda_n$, gives
\begin{align*}
\|\bDelta_{\bv}\|_2\lesssim\sqrt{d_0}\bigg[
&\lambda_n+ \diam(\mathcal A_{\mathcal H_n}(\eta)) \frac{n+d_0+\log(n+d_0)}{N}\\
&+\lambda_n\diam(\mathcal A_{\mathcal H_n}(\eta)) \sqrt{\frac{d_0\{n+d_0+\log(n+d_0)\}}{N}} +d_0\lambda_n^2\diam(\mathcal A_{\mathcal H_n}(\eta)) \bigg].
\end{align*}
The second term is $O(\lambda_n)$ by~\eqref{eq:pdw_lambda_choice}. Moreover,
\begin{equation*}
d_0\diam^2(\mathcal A_{\mathcal H_n}(\eta))
\frac{n+d_0+\log(n+d_0)}{N}
\lesssim
d_0\lambda_n\diam(\mathcal A_{\mathcal H_n}(\eta))
=o(1)
\end{equation*}
by~\eqref{eq:connect_conditions}, so the last two terms are $o(\lambda_n)$. Hence
$\|\bDelta_{\bv}\|_2\lesssim\sqrt{d_0}\lambda_n$.

The utility rate now follows by chaining. The likelihood controls comparison-level utility differences rather than the individual utility coordinates, and the chaining bound~\eqref{eq:delta-chaining} is what converts this local comparison information into a uniform coordinatewise bound, the geometry of the comparison hypergraph entering through
$\diam(\mathcal A_{\mathcal H_n}(\eta))$. Since $\|\bDelta_{\bv}\|_1\le\sqrt{d_0}\|\bDelta_{\bv}\|_2\lesssim d_0\lambda_n$, substitution into~\eqref{eq:delta-chaining} gives
\begin{equation*}
\|\bDelta_{\bu}\|_\infty
\lesssim\diam(\mathcal A_{\mathcal H_n}(\eta))
\max \left\{\sqrt{\frac{\log n}{h_{\mathcal H_n}}}
, \, d_0\lambda_n\right\},
\end{equation*}
Thus both asserted rates hold on $\G_1\cap\cdots\cap \G_5$.
Lemma~\ref{lem:all_events} gives the probability of this intersection and
completes the proof of Lemma~\ref{lem:oracle_consistency}.

\section{~Proof of Lemma~\ref{lem:strict_dual_feasibility}}
\label{proof:strict_dual_feasibility}

We first sketch the proof. To begin with, we construct, for each inactive coordinate, a population-profiled direction that annihilates the active and utility columns of $\bm M^*$. Multiplying the gradient expansion by this direction removes the active estimation error entirely and leaves an exact identity in which the leading term is the population incoherence quantity. We then bound the three stochastic terms surviving on the right-hand side, and finally verify that their total contribution is at most $\frac{\alpha}{2}$.

For each $k\in T$, define $\bm w_{\mathrm{prof}}^{(k)}\in\R^{n+d}$ by
\begin{equation}\label{eq:pdw_direction}
(\bm w_{\mathrm{prof}}^{(k)})_U=-(\bm M^*_{UU})^\dagger
(\bm M^*_{US}(\bm w_{\mathrm{prof}}^{(k)})_S+\bm M^*_{UT}\bm e_k),
\ \ 
(\bm w_{\mathrm{prof}}^{(k)})_S=-(\bm G^*\bm H^{*-1})_{k:}^\top,
\ \ 
(\bm w_{\mathrm{prof}}^{(k)})_T=\bm e_k.
\end{equation}
Here $\bm e_k\in\mathbb R^{d-d_0}$ is the canonical basis vector corresponding to coordinate $k\in T$. The direction is deterministic conditional on the design, and the bounds below hold uniformly over $k\in T$.
Three properties of $\bm w_{\mathrm{prof}}^{(k)}$ are used below: it is orthogonal to the columns of
$\bm M^*$ indexed by $\mathcal I$,
\begin{equation}\label{eq:pdw_orthogonality}
(\bm w_{\mathrm{prof}}^{(k)})^\top[\bm M^*]_{\mathcal I}=0,
\end{equation}
its active block reproduces the incoherence coefficients, and it satisfies the
design bounds~\eqref{eq:direction_design} without the factor $\sqrt{d_0}$ that
appeared in the active case. The second property holds by construction, and we
verify the other two.

Both follow from the observation that~\eqref{eq:pdw_direction} solves
\begin{equation*}
\min\bigl\{\bm w^\top(-\bm M^*)\bm w:\ \bm w_T=\bm e_k\bigr\},
\end{equation*}
the minimum being taken over the utility and active blocks. Indeed, eliminating
the utilities and solving the resulting active quadratic problem gives the
first-order equation $\bm H^*(\bm w_{\mathrm{prof}}^{(k)})_S+(\bm G^*)_{k:}^\top=0$, whose solution is
the active block in~\eqref{eq:pdw_direction}; the full set of first-order
equations for this problem is exactly~\eqref{eq:pdw_orthogonality}. For the
design bounds, note first that under Assumption~\ref{ass:incoherence} the
absolute sum of the coefficients in $(\bm w_{\mathrm{prof}}^{(k)})_S$ is at most one, so
Lemma~\ref{lem:graph_residualization} yields the first bound
in~\eqref{eq:direction_design}, that is,
$\max_i\|\W_i\bm w_{\mathrm{prof}}^{(k)}\|_2\lesssim\diam(\mathcal A_{\mathcal H_n}(\eta))$. Next, the
minimum value above is nonnegative and no larger than the value at
$\bm w_U=\bm w_S=0$, namely $-M^*_{kk}\lesssim1$; since the expected contrast
curvature controls the average squared size of $\W_i\bm w_{\mathrm{prof}}^{(k)}$, the second bound
in~\eqref{eq:direction_design} follows as well.

For these inactive directions, define
\begin{equation}\label{G6}
\G_6=\left\{
|(\bm w_{\mathrm{prof}}^{(k)})^\top\bm Q|\le c_{8}\sqrt{\frac{\log(n+d)}N}
\text{ for every }k\in T
\right\}
\end{equation}
and
\begin{equation}\label{G7}
\G_7=\left\{
\begin{aligned}
|(\bm w_{\mathrm{prof}}^{(k)})^\top(\bm M-\bm M^*)&\bDelta|
\le c_{9}\diam(\mathcal A_{\mathcal H_n}(\eta))
\sqrt{\frac{n+d_0+\log(n+d)}N}
\frac{\|\W_n\bDelta\|_2}{\sqrt N}\\
&\text{for every $k\in T$ and every $\bDelta$ with $\bDelta_T=0$}
\end{aligned}
\right\}.
\end{equation}
For the remainder of this proof, we work on $\G_1\cap\cdots\cap \G_7$.The first five events give the oracle bounds; $\G_6$ and $\G_7$ control the inactive score and Hessian terms.

We now derive the dual identity. Because $\bDelta_T=0$, the full gradient expansion reads
\begin{equation*}
\lambda_n(\bm0_U^\top,\widehat{\bz}_S^\top,
\widehat{\bz}_T^\top)^\top
=\bm Q+[\bm M^*]_{\mathcal I}\bDelta_{\mathcal I}
+[\bm M-\bm M^*]_{\mathcal I}\bDelta_{\mathcal I}+\bm R.
\end{equation*}
Multiplying on the left by $(\bm w_{\mathrm{prof}}^{(k)})^\top$ annihilates the second term on the right
by~\eqref{eq:pdw_orthogonality}, while on the left it produces the incoherence
combination of subgradients. This gives the exact identity
\begin{equation}
\label{eq:master_dual}
\lambda_n\{\widehat z_k-(\bm G^*\bm H^{*-1})_{k:}
\widehat{\bz}_S\}
=(\bm w_{\mathrm{prof}}^{(k)})^\top\bm Q
+(\bm w_{\mathrm{prof}}^{(k)})^\top(\bm M-\bm M^*)\bDelta
+(\bm w_{\mathrm{prof}}^{(k)})^\top\bm R.
\end{equation}
Two features of~\eqref{eq:master_dual} deserve emphasis. The active estimation
error $\bDelta_{\mathcal I}$ has been eliminated at the population level, so the
leading term on the left is controlled by Assumption~\ref{ass:incoherence}
alone. Moreover, the difference between the observed and expected Hessians
remains explicitly on the right-hand side, where it is handled by concentration;
no incoherence condition on the sample Hessian is needed.

It remains to bound the three terms on the right of~\eqref{eq:master_dual}. The
first two follow directly from $\G_6$ and $\G_7$, respectively:
\begin{align}
\max_{k\in T}|(\bm w_{\mathrm{prof}}^{(k)})^\top\bm Q|
&\lesssim\sqrt{\frac{\log(n+d)}N},
\label{eq:Q_inf_bound}\\
\max_{k\in T}|(\bm w_{\mathrm{prof}}^{(k)})^\top(\bm M-\bm M^*)\bDelta|
&\lesssim\diam(\mathcal A_{\mathcal H_n}(\eta))
\sqrt{\frac{n+d_0+\log(n+d)}N}
\frac{\|\W_n\bDelta\|_2}{\sqrt N}.
\label{eq:pdw_fluctuation}
\end{align}
The event on which~\eqref{eq:pdw_fluctuation} holds is uniform over all
restricted prediction vectors, so the bound applies to the data-dependent oracle
error $\bDelta$ even though $\bDelta$ is not independent of the observed
Hessian. The third term is deterministic given $\bDelta$: combining the
remainder bound~\eqref{eq:directional_remainder} with the first bound
in~\eqref{eq:direction_design} gives
\begin{equation}
\label{eq:pdw_projected_remainder}
\max_{k\in T}|(\bm w_{\mathrm{prof}}^{(k)})^\top\bm R|
\lesssim\diam(\mathcal A_{\mathcal H_n}(\eta))
\frac{\|\W_n\bDelta\|_2^2}{N}.
\end{equation}
Here the remainder is measured by the average squared contrast error rather than
by a maximal utility error multiplied by an inverse utility-curvature factor.

Divide~\eqref{eq:master_dual} by $\lambda_n$,
insert the three bounds~\eqref{eq:Q_inf_bound}--\eqref{eq:pdw_projected_remainder},
and use Assumption~\ref{ass:incoherence} together with
$\|\widehat{\bz}_S\|_\infty\le1$ to bound the incoherence combination on the left
by $1-\alpha$. Substituting the prediction bounds
\eqref{eq:prediction_root} and~\eqref{eq:prediction_energy} for
$\|\W_n\bDelta\|_2$ in the resulting expressions, exactly as in the proof of
Lemma~\ref{lem:oracle_consistency}, and using $\log(n+d_0)\le\log(n+d)$, yields
\begin{align*}
\|\widehat{\bz}_T\|_\infty-(1-\alpha)
\lesssim\bigg[&
\frac1{\lambda_n}\sqrt{\frac{\log(n+d)}N}
+\frac{\diam(\mathcal A_{\mathcal H_n}(\eta))}{\lambda_n}
\frac{n+d_0+\log(n+d)}N\\
&+\diam(\mathcal A_{\mathcal H_n}(\eta))
\sqrt{\frac{d_0\{n+d_0+\log(n+d)\}}N}
+d_0\lambda_n\diam(\mathcal A_{\mathcal H_n}(\eta))\bigg].
\end{align*}
The first two bracketed terms are at most a constant multiple of $1/c_{1}$ by~\eqref{eq:pdw_lambda_choice}; choosing $c_{1}$ large enough makes their joint contribution at most $\alpha/4$. The last two tend to zero by~\eqref{eq:connect_conditions} and~\eqref{eq:pdw_lambda_choice}, hence contribute at most $\alpha/4$ for all $n$ sufficiently large. Therefore
\begin{equation*}
\|\widehat{\bz}_T\|_\infty\le1-\alpha/2<1.
\end{equation*}
We have proved strict dual feasibility on $\G_1\cap\cdots\cap \G_7$.
Its probability is bounded in Lemma~\ref{lem:all_events},completing the proof of Lemma~\ref{lem:strict_dual_feasibility}.

\begin{lemma}[Probability of the good events]\label{lem:all_events}
Under the conditions of Lemma~\ref{lem:oracle_consistency}, with sufficiently
large fixed event thresholds $c_{3},c_{4},c_{6},c_{7},c_{8},c_{9}$,
\begin{equation*}
\Prob\left(\bigcap_{j=1}^5 \G_j\,\middle|\,\mathscr D_n\right)
\ge 1-O(n^{-3}).
\end{equation*}
If Assumption~\ref{ass:incoherence} also holds, then
\begin{equation}\label{eq:all_events_probability}
\Prob\left(\bigcap_{j=1}^7 \G_j\,\middle|\,\mathscr D_n\right)
\ge 1-O(n^{-3}).
\end{equation}
\end{lemma}

\section{~Proofs of auxiliary lemmas}
\label{app:auxlemma}
This section proves the deterministic PDW and graph residualization lemmas.
It concludes with a single probability lemma for $\G_1,\ldots,\G_7$, whose proof
bounds each failure probability before applying a union bound.

\subsection{Proof of Lemma~\ref{lem:pdw_condition}}\label{proof:pdw_condition}
\begin{proof}
Suppose there exists another optimal solution $\hbtheta'=(\hbu^{'\top},\hbv^{'\top})^\top$. By the optimality condition $-l_n(\hbtheta)+\lambda_n\norm{\hbv}_1 = -l_n(\hbtheta')+\lambda_n\norm{\hbv'}_1$, and letting $\widehat{\bz}\in\partial\norm{\hbv}_1$ be the subgradient from the KKT conditions (where $\langle\widehat{\bz},\hbv\rangle=\norm{\hbv}_1$), we can rearrange the terms to obtain:
\begin{equation*}
    -l_n(\hbtheta) + \lambda_n\langle\widehat{\bz},\hbv-\hbv'\rangle + l_n(\hbtheta')
    = \lambda_n\left(\norm{\hbv'}_1-\langle\widehat{\bz},\hbv'\rangle\right).
\end{equation*}

Next, we evaluate the left-hand side. Using the KKT conditions $\nabla_{\bu}(-l_n(\hbtheta))= \bm{0}$, $\nabla_{\bv}(-l_n(\hbtheta))=-\lambda_n\widehat{\bz}$ and the convexity of $-l_n$, we have:
\begin{equation*}
    -l_n(\hbtheta') \ge -l_n(\hbtheta) + \left\langle\nabla(-l_n(\hbtheta)), \hbtheta'-\hbtheta\right\rangle = -l_n(\hbtheta) - \lambda_n\langle\widehat{\bz}, \hbv'-\hbv\rangle.
\end{equation*}
This implies that the left-hand side of the first equation is non-positive. Consequently, the right-hand side must satisfy:
\begin{equation*}
    \norm{\hbv'}_1 - \langle\widehat{\bz},\hbv'\rangle \le 0.
\end{equation*}

On the other hand, Hölder's inequality guarantees $\langle\widehat{\bz},\hbv'\rangle \le \norm{\widehat{\bz}}_\infty\norm{\hbv'}_1 \le \norm{\hbv'}_1$. Combining these inequalities yields the strict equality $\norm{\hbv'}_1=\langle\widehat{\bz},\hbv'\rangle$. Given the strict dual feasibility $\norm{\widehat{\bz}_{S^{*\complement}}}_\infty < 1$, this equality can hold only if $\hbv'_j=0$ for all $j\in S^{*\complement}$. Thus, $\supp(\hbv')\subseteq S^*$, so $\hbtheta'$ is feasible for the restricted problem~\eqref{eq:oracle}; because $\hbtheta'$ attains the same objective value as $\hbtheta$ and $\hbtheta$ is the unique minimizer of the restricted problem, we have $\hbtheta'=\hbtheta$, proving that $\hbtheta$ is the unique minimizer of the full problem.
\end{proof}

\subsection{Proof of Lemma~\ref{chaining}}
\label{chainingproof}

\begin{lemma}[{\citet[Lemma 3]{dong2026statistical}}]
\label{sumlemma}
Let $a_1, a_2, b_1, b_2 > 0$. Suppose there exist $0 < \varepsilon_1, \varepsilon_2 \le 1$ such that
\begin{equation*}
a_2 \ge \varepsilon_1 a_1 \quad\text{and}\quad \frac{a_2}{b_2} \ge (1+\varepsilon_2)\frac{a_1}{b_1}.
\end{equation*}
Then,
\begin{equation*}
\frac{a_1+a_2}{b_1+b_2} \ge \left(1+\frac{\varepsilon_1\varepsilon_2}{3}\right)\frac{a_1}{b_1}.
\end{equation*}
\end{lemma}

Throughout this proof, $\tbv\in\R^d$ denotes the full oracle coefficient vector. We abbreviate $\bv^*_0 = [\bv^*]_{S^*}$, $\btheta^*_0 = (\bu^{* \top}, \bv^{* \top}_0)^\top$, and $\tbtheta_0=(\tbu^\top,(\tbv_{S^*})^\top)^\top\in\R^{n+d_0}$. Since $\tbv_{S^{*\complement}}=\bv^*_{S^{*\complement}}=\bm0$, we have $\norm{\tbv_{S^*}-\bv_0^*}_1=\norm{\tbv-\bv^*}_1$. We write $l_{S^*}$ for the log-likelihood as a function of the restricted parameter, with inactive covariate coordinates set to zero. Noting that $\bm{1}^\top (\tbu - \bu^*) = 0$, we have
\begin{equation*}
\norm{\tbu - \bu^*}_\infty \leq \max_{k \in [n]} (\tu_k - u^*_k) - \min_{k \in [n]} (\tu_k - u^*_k).
\end{equation*}
Define the indices attaining the maximum and minimum deviations as
\begin{equation*}
\alpha \in \arg\max_{k \in [n]} (\tu_k - u^*_k), \quad \beta \in \arg\min_{k \in [n]} (\tu_k - u^*_k).
\end{equation*}
For $t \geq 1$, define the level sets:
\begin{equation*}
\begin{aligned}
B_t &= \{k: (\tu_\alpha - u^*_\alpha) - (\tu_k - u^*_k) \leq (t-1)\eta_n \},\\
D_t &= \{k: (\tu_k - u^*_k) - (\tu_\beta - u^*_\beta) \leq (t-1)\eta_n \},
\end{aligned}
\end{equation*}
where the threshold $\eta_n$ is defined as
\begin{equation*} 
\eta_n = \max \left\{4C \norm{\tbv_{S^*}-\bv_0^*}_1 \, , \, c\sqrt{\frac{\log n }{h_{\mathcal{H}_n}}}  \right\}.
\end{equation*}

Let $\I_{n,1}$ and $\I_{n,2}$ be the stopping times defined as
\begin{equation*}
\I_{n,1} = \min \left\{ t \, :\, \abs{B_t} > \frac{n}{2}\right\}, \quad \quad \I_{n,2} = \min \left\{ t \, :\, \abs{D_t} > \frac{n}{2}\right\}.
\end{equation*}
If we can show that both $\{ B_t\}_{t \in [\I_{n,1}]}$ and $\{ D_t\}_{t \in [\I_{n,2}]}$ are $\eta$-weakly admissible sequences for some fixed $\eta\in(0,1/2)$, then standard chaining arguments imply
\begin{equation*}
 \max_{k \in [n]} (\tu_k - u^*_k) - \min_{k \in [n]} (\tu_k - u^*_k) \leq 2\diam (\A_{\mathcal{H}_n}(\eta)) \cdot \eta_n.
\end{equation*}

\paragraph{Upper bound via concentration.}
By the first-order optimality condition for the restricted MLE, the score with respect to each item parameter vanishes at $\tbtheta_0$:
\begin{equation*}
    \partial_k l_{S^*}(\tbtheta_0) = \frac{1}{N}\sum_{i:k\in e_i}\psi(k;e_i,\pi_i,\tbtheta_0)=0,\qquad k\in[n],
\end{equation*}
where, for $k\in e_i$,
\begin{equation*}
    \psi(k;e_i,\pi_i,\btheta_0) = 1-\sum_{j\in[r_i(k)]}\frac{\exp\{s^\flat_{i r_i(k)}\}}{\sum_{t=j}^{m_i}\exp\{s^\flat_{it}\}} .
\end{equation*}
Here $\btheta_0=(\bu^\top,\bv_{S^*}^\top)^\top \in \R^{n+d_0}$, $\pi_i\in \mathfrak{S}(e_i)$, and $r_i(k)$ denotes the rank of item $k$ in $\pi_i$.

On the event $\mathcal{G}_{2}$, the empirical score over any subset is controlled by its boundary size. In particular, with probability at least $1-n^{-2}$,
\begin{equation}
\label{upperbound}
    \sum_{k\in B_t}\partial_k l_{S^*}(\btheta_0^*) - \sum_{k\in B_t}\partial_k l_{S^*}(\tbtheta_0) \le \frac{c_2}{N} \sqrt{|B_t|\,|\partial B_t|\log n}, \qquad t<\I_{n,1}.
\end{equation}

\paragraph{Lower bound via curvature.}
We now derive a lower bound for the left-hand side of (\ref{upperbound}), from which we deduce the desired weakly admissible property.
We start by rewriting $\sum_{k \in B_t} \partial_k l_{S^*}(\btheta_0^*) - \sum_{k \in B_t} \partial_k l_{S^*}(\tbtheta_0)$ as
\begin{align*}
&\sum_{k \in B_t} \partial_k l_{S^*}(\btheta_0^*) - \sum_{k \in B_t} \partial_k l_{S^*}(\tbtheta_0) = \frac{1}{N}\sum_{i: e_i \in \partial B_t} \sum_{k \in B_t \cap e_i} \sum_{j \in [r_i(k)]} \left(\frac{\exp\{\widetilde{s}^\flat_{i r_i(k)}\}}{\sum_{q=j}^{m_i}\exp\{\widetilde{s}^\flat_{iq}\}} - \frac{\exp\{s^*_{i r_i(k)}\}}{\sum_{q=j}^{m_i}\exp\{s^*_{iq}\}} \right) \notag \\
&= \frac{1}{N}\sum_{i:e_i\in\partial B_t} \sum_{j=1}^{\max_{k\in B_t\cap e_i}r_i(k)}
\left( \frac{\sum_{q\ge j:\,\pi_i(q)\in B_t}\exp\{\widetilde{s}^\flat_{iq}\}}{\sum_{q=j}^{m_i}\exp\{\widetilde{s}^\flat_{iq}\}} - \frac{\sum_{q\ge j:\,\pi_i(q)\in B_t}\exp\{s^*_{iq}\}}{\sum_{q=j}^{m_i}\exp\{s^*_{iq}\}} \right).
\end{align*}

We consider two cases depending on whether $e_i \cap (B_{t+1}\setminus B_t)=\emptyset$ or not.

\noindent\textsc{Case 1: $e_i \cap (B_{t+1}\setminus B_t)=\emptyset$.}
By definition, for all objects $k \in e_i \cap B_t^c \subseteq B_{t+1}^c$,
\begin{equation*}
\tu_\alpha-u_\alpha^*-(\tu_k-u_k^*)\ge t \eta_n.
\end{equation*}
Meanwhile, for the remaining vertices $j \in e_i \cap B_t$,
\begin{equation*}
\tu_\alpha-u_\alpha^*-(\tu_j-u_j^*)\le (t-1)\eta_n.
\end{equation*}
Combining both yields $\tu_j-u_j^*-(\tu_k-u_k^*)\ge \eta_n$.
Therefore,
\begin{align*}
\widetilde{s}^\flat_{i r_i(j)} - s^*_{i r_i(j)} - (\widetilde{s}^\flat_{i r_i(k)} - s^*_{i r_i(k)}) =& \tu_j-u_j^*-(\tu_k-u_k^*)+ [(X_{e_i,j}-X_{e_i,k})]_{S^*}^\top(\tbv_{S^*}-\bv_0^*) \notag \\
\ge& \eta_n - 2C\norm{\tbv_{S^*}-\bv_0^*}_1 \geq \frac{1}{2}\eta_n,
\end{align*}
which implies
\begin{equation*}
\frac{\exp\{s^*_{i r_i(k)}\}}{\exp\{\widetilde{s}^\flat_{i r_i(k)}\}} \ge \exp\left\{\frac{\eta_n}{2}\right\} \times \frac{\exp\{s^*_{i r_i(j)}\}}{\exp\{\widetilde{s}^\flat_{i r_i(j)}\}} \ge \left(1+\frac{\eta_n}{2}\right) \frac{\exp\{s^*_{i r_i(j)}\}}{\exp\{\widetilde{s}^\flat_{i r_i(j)}\}}.
\end{equation*}

We first consider a stage $j$ at which the remaining set $\{\pi_i(q):q\ge j\}$ intersects both $B_t$ and $B_t^c$. In Case 1, the remaining items outside $B_t$ all belong to $B_{t+1}^c$, so both groups of sums below are nonempty. Taking the ratio yields
\begin{align*}
&\frac{\sum_{q\ge j:\,\pi_i(q)\in B_t}\exp\{\widetilde{s}^\flat_{iq}\}}
{\sum_{q\ge j:\,\pi_i(q)\in B_t}\exp\{s^*_{iq}\}} \times \frac{\sum_{q=j}^{m_i}\exp\{s^*_{iq}\}} {\sum_{q=j}^{m_i}\exp\{\widetilde{s}^\flat_{iq}\}} \\
=& \frac{\sum_{q\ge j:\,\pi_i(q)\in B_t}\exp\{\widetilde{s}^\flat_{iq}\}}
{\sum_{q\ge j:\,\pi_i(q)\in B_t}\exp\{s^*_{iq}\}} \times \frac{\sum_{q\ge j:\,\pi_i(q)\in B_t}\exp\{s^*_{iq}\} + \sum_{q\ge j:\,\pi_i(q)\in B_{t+1}^c}\exp\{s^*_{iq}\}}
{\sum_{q\ge j:\,\pi_i(q)\in B_t}\exp\{\widetilde{s}^\flat_{iq}\} +
\sum_{q\ge j:\,\pi_i(q)\in B_{t+1}^c}\exp\{\widetilde{s}^\flat_{iq}\}}.
\end{align*}

To analyze the last term, we appeal to Lemma~\ref{sumlemma} to obtain
\begin{equation*}
\frac{\sum_{q\ge j:\,\pi_i(q)\in B_t}\exp\{s^*_{iq}\} + \sum_{q\ge j:\,\pi_i(q)\in B_{t+1}^c}\exp\{s^*_{iq}\}} {\sum_{q\ge j:\,\pi_i(q)\in B_t}\exp\{\widetilde{s}^\flat_{iq}\} + \sum_{q\ge j:\,\pi_i(q)\in B_{t+1}^c}\exp\{\widetilde{s}^\flat_{iq}\}} \ge \left(1+c_3\min\{\eta_n,1\}\right) \frac{\sum_{q\ge j:\,\pi_i(q)\in B_t}\exp\{s^*_{iq}\}}{\sum_{q\ge j:\,\pi_i(q)\in B_t}\exp\{\widetilde{s}^\flat_{iq}\}}.
\end{equation*}
Bounded true scores and comparison size ensure that both groups have true probability mass uniformly bounded away from zero at such stages. Hence,
\begin{align*}
&\frac{\sum_{q\ge j:\,\pi_i(q)\in B_t}\exp\{\widetilde{s}^\flat_{iq}\}}
{\sum_{q=j}^{m_i}\exp\{\widetilde{s}^\flat_{iq}\}}-\frac{\sum_{q\ge j:\,\pi_i(q)\in B_t}\exp\{s^*_{iq}\}}{\sum_{q=j}^{m_i}\exp\{s^*_{iq}\}}\\
&\ge c_3\min\{\eta_n,1\} \times \frac{\sum_{q\ge j:\,\pi_i(q)\in B_t}\exp\{s^*_{iq}\}}
{\sum_{q=j}^{m_i}\exp\{s^*_{iq}\}} \ge c_4\min\{\eta_n,1\}.
\end{align*}

If the remaining set lies entirely inside or outside $B_t$, the difference is zero. Thus all stages contribute nonnegatively, and since $e_i\in\partial B_t$, the first stage satisfies the positive lower bound above. Consequently,
\begin{align*}
&\sum_{j=1}^{\max_{k\in B_t\cap e_i}r_i(k)}
\left( \frac{\sum_{q\ge j:\,\pi_i(q)\in B_t}
      \exp\{\widetilde{s}^\flat_{iq}\}}
     {\sum_{q=j}^{m_i}\exp\{\widetilde{s}^\flat_{iq}\}}
-
\frac{\sum_{q\ge j:\,\pi_i(q)\in B_t}\exp\{s^*_{iq}\}}
     {\sum_{q=j}^{m_i}\exp\{s^*_{iq}\}} \right)
\ge c_4\min\{\eta_n,1\}.
\end{align*}

\noindent\textsc{Case 2: $e_i \cap (B_{t+1}\setminus B_t)\neq \emptyset$.}
In this case, we do not know whether the summand is positive or negative; however, we can also bound it from below using $\min\{\eta_n,1\}$.
To see this, note for any $k \in e_i \cap (B_{t+1}\setminus B_t)$ and $j \in e_i \cap B_t$,
\begin{equation*}
\widetilde{s}^\flat_{i r_i(j)} - s^*_{i r_i(j)} - (\widetilde{s}^\flat_{i r_i(k)} - s^*_{i r_i(k)}) \ge -2C\norm{\tbv_{S^*}-\bv_0^*}_1 \ge -\frac{\eta_n}{2},
\end{equation*}
which implies
\begin{equation*}
\frac{\exp\{s^*_{i r_i(k)}\}}{\exp\{\widetilde{s}^\flat_{i r_i(k)}\}} \ge \exp\left\{-\frac{\eta_n}{2}\right\} \times \frac{\exp\{s^*_{i r_i(j)}\}}{\exp\{\widetilde{s}^\flat_{i r_i(j)}\}} \ge \left(1-\frac{\eta_n}{2}\right) \frac{\exp\{s^*_{i r_i(j)}\}} {\exp\{\widetilde{s}^\flat_{i r_i(j)}\}}.
\end{equation*}
Thus, if $\eta_n\le 1$, taking the ratio of the corresponding summand yields
\begin{equation*}
\frac{\sum_{q\ge j:\,\pi_i(q)\in B_t}\exp\{\widetilde{s}^\flat_{iq}\}}
{\sum_{q\ge j:\,\pi_i(q)\in B_t}\exp\{s^*_{iq}\}} \times
\frac{\sum_{q=j}^{m_i}\exp\{s^*_{iq}\}}{\sum_{q=j}^{m_i}\exp\{\widetilde{s}^\flat_{iq}\}}
\ge 1-\frac{\eta_n}{2}.
\end{equation*}
As a result,
\begin{align*}
&\frac{\sum_{q\ge j:\,\pi_i(q)\in B_t} \exp\{\widetilde{s}^\flat_{iq}\}} {\sum_{q=j}^{m_i}\exp\{\widetilde{s}^\flat_{iq}\}} - \frac{\sum_{q\ge j:\,\pi_i(q)\in B_t} \exp\{s^*_{iq}\}} {\sum_{q=j}^{m_i}\exp\{s^*_{iq}\}}\\
&\ge -\frac{\eta_n}{2} \frac{\sum_{q\ge j:\,\pi_i(q)\in B_t}\exp\{s^*_{iq}\}} {\sum_{q=j}^{m_i}\exp\{s^*_{iq}\}} \ge -\eta_n
= -\min\{\eta_n,1\},
\end{align*}
If $\eta_n>1$, the last step still holds.

By defining the proportion of edges in $\partial B_t$ that intersect $B_{t+1}\setminus B_t$ as
\begin{equation*}
\kappa \coloneqq \frac{\left|\left\{e \in \partial B_t : e \cap (B_{t+1}\setminus B_t)\neq \emptyset\right\}\right|}{\abs{\partial B_t}},
\end{equation*}
we have
\begin{equation*}
\left[c_4(1-\kappa)-M\kappa\right] \min\{1,\eta_n\}|\partial B_t| \le c_2\sqrt{|B_t|\,|\partial B_t|\log n}.
\end{equation*}
Set $\eta\coloneqq c_4/(2c_4+2M)\in(0,1/2)$. If $\kappa\le\eta$, then
\begin{equation*}
\frac{c_4}{2}\min\{1,\eta_n\} \le c_2\sqrt{\frac{\log n}{h_{\mathcal{H}_n}}} \to 0, \quad n\to\infty,
\end{equation*}
which implies $\eta_n<1$. Hence,
\begin{equation*}
\eta_n = \max\left\{ c\sqrt{\frac{\log n}{h_{\mathcal H_n}}},
\, 4C\norm{\tbv_{S^*}-\bv_0^*}_1 \right\} \le \frac{2c_2}{c_4} \sqrt{\frac{\log n}{h_{\mathcal H_n}}}.
\end{equation*}
Since
\begin{equation*} 
\eta_n\ge c\sqrt{\frac{\log n}{h_{\mathcal H_n}}},
\end{equation*}
this is a contradiction whenever $c>2c_2/c_4$.
Therefore, $\kappa>\eta$.
This implies that $\{B_t\}_{t=1}^{\I_{n,1}}$ is a $\eta$-weakly admissible sequence. 

By an entirely symmetric argument applied to $\{D_t\}_{t=1}^{\I_{n,2}}$, with the maximizing item $\alpha$ replaced by the minimizing item $\beta$ and all score inequalities reversed, we obtain that $\{D_t\}_{t=1}^{\I_{n,2}}$ is also $\eta$-weakly admissible. Therefore, both $\{B_t\}_{t=1}^{\I_{n,1}}$ and $\{D_t\}_{t=1}^{\I_{n,2}}$ are $\eta$-weakly admissible sequences, and the claimed chaining bound follows.

\subsection{Proof of Lemma~\ref{lem:graph_residualization}}
\label{proof:graph_residualization}
\begin{proof}
Fix $r\in S\cup T$ and write
\begin{equation*}
\bm b=(\bm M^*_{UU})^\dagger\bm M^*_{Ur}.
\end{equation*}
All expectations below are conditional on the comparison design and
covariates. The constants depend only on the uniform bounds in
Assumptions~\ref{ass:size} and~\ref{ass:unibound}; in particular, they are
independent of $r$ and $d_0$.

\emph{Step 1: represent the population Hessian as a weighted Laplacian.}
For comparison $i$, recall that
$\pi_i=(\pi_i(1),\ldots,\pi_i(m_i))$ is the observed ranking and define the
true score of item $k\in e_i$ by
\begin{equation*}
s_{i,k}^*=u_k^*+X_{e_i,k}^\top\bv^*.
\end{equation*}
At stage $t\in\{1,\ldots,m_i-1\}$, the available items are
$\pi_i(t),\ldots,\pi_i(m_i)$. Embed the corresponding choice-probability
vector into $\mathbb R^{m_i}$ by setting
\begin{equation}
\label{eq:stage_choice_probability}
p_{i,k}^{(t)}(\pi_i)
=
\mathbf 1\!\left\{k\in\{\pi_i(t),\ldots,\pi_i(m_i)\}\right\}
\frac{\exp(s_{i,k}^*)}
{\displaystyle\sum_{q=t}^{m_i}\exp(s_{i,\pi_i(q)}^*)},
\qquad k\in e_i.
\end{equation}
Thus $p_{i,k}^{(t)}(\pi_i)=0$ if item $k$ has already been selected, and the
coordinates in~\eqref{eq:stage_choice_probability} sum to one.

Direct differentiation of the $t$th stage log-likelihood with respect to the
item scores gives the negative Hessian
\begin{equation*}
\diag(\bm p_i^{(t)})
-\bm p_i^{(t)}\bm p_i^{(t)\top}.
\end{equation*}
For every vector $\bm y=(y_k)_{k\in e_i}$, its quadratic form is
\begin{align}
\bm y^\top
\{\diag(\bm p_i^{(t)})
-\bm p_i^{(t)}\bm p_i^{(t)\top}\}\bm y
&=
\frac12\sum_{k,l\in e_i}
p_{i,k}^{(t)}(\pi_i)p_{i,l}^{(t)}(\pi_i)(y_k-y_l)^2\notag\\
&=
\sum_{k<l:\,k,l\in e_i}
p_{i,k}^{(t)}(\pi_i)p_{i,l}^{(t)}(\pi_i)(y_k-y_l)^2.
\label{eq:stage_laplacian_identity}
\end{align}

For each pair $k<l$ contained in $e_i$, define
\begin{equation}
\label{eq:omega_definition}
\omega_{i,kl}
=
\expect\!\left[
\left.
\sum_{t=1}^{m_i-1}
p_{i,k}^{(t)}(\pi_i)p_{i,l}^{(t)}(\pi_i)
\,\right|\,
\mathscr D_n
\right].
\end{equation}
Because the probabilities in~\eqref{eq:stage_choice_probability} are set to
zero after an item is selected, the sum in~\eqref{eq:omega_definition}
automatically includes exactly the stages at which both $k$ and $l$ remain
available. Thus $\omega_{i,kl}$ is the total conditional population-curvature
weight assigned to the pair $(k,l)$ by comparison $i$.

Now let $\bm w=(\bm w_U^\top,\bm w_{S\cup T}^\top)^\top$. The score perturbation
of item $k\in e_i$ in direction $\bm w$ is
\begin{equation*}
w_{U,k}+X_{e_i,k}^\top\bm w_{S\cup T}.
\end{equation*}
Since
\begin{equation*}
\bm M^*
=
\expect\!\left[
\left.\nabla^2l_n(\btheta^*)\,\right|\,\mathscr D_n
\right],
\end{equation*}
summing~\eqref{eq:stage_laplacian_identity} over the stages and comparisons and
then taking the conditional expectation gives
\begin{align}
\bm w^\top(-\bm M^*)\bm w
&=
\frac1N\sum_{i=1}^N
\expect\!\left[
\left.
\sum_{t=1}^{m_i-1}
\sum_{k<l:\,k,l\in e_i}
p_{i,k}^{(t)}(\pi_i)p_{i,l}^{(t)}(\pi_i)
\right.\right.\notag\\
&\hspace{42mm}\left.\left.
\times
\left\{
w_{U,k}-w_{U,l}
+(X_{e_i,k}-X_{e_i,l})^\top\bm w_{S\cup T}
\right\}^2
\,\right|\,\mathscr D_n
\right]\notag\\
&=
\frac1N\sum_{i=1}^N\sum_{k<l:\,k,l\in e_i}
\omega_{i,kl}
\left\{
w_{U,k}-w_{U,l}
+(X_{e_i,k}-X_{e_i,l})^\top\bm w_{S\cup T}
\right\}^2.
\label{eq:population_laplacian}
\end{align}

The true item scores are uniformly bounded because
\begin{equation*}
\abs{u_k^*+X_{e_i,k}^\top\bv^*}
\leq
\norm{\bu^*}_\infty
+\norm{X_{e_i,k}}_\infty\norm{\bv^*}_1
\lesssim1.
\end{equation*}
Since the comparison size is bounded, every first-stage choice probability is
bounded below by a positive constant. The first stage therefore contributes a
fixed positive amount to every pair in $e_i$, while the total number of stages
is uniformly bounded. Hence
\begin{equation}
\label{eq:weight_bounds}
1\lesssim\omega_{i,kl}\lesssim1
\qquad
\text{for all $i$ and all $k<l$ in $e_i$}.
\end{equation}

\emph{Step 2: derive the equation satisfied by $\bm b$.}
The utility block of~\eqref{eq:population_laplacian} is a weighted graph
Laplacian. Because the comparison hypergraph is connected, its kernel is
$\operatorname{span}(\bm1_n)$. Moreover, $\bm M^*_{Ur}$ is centered. Hence
$\bm b$ is the unique centered solution of
\begin{equation*}
(-\bm M^*_{UU})\bm b=-\bm M^*_{Ur}.
\end{equation*}
Writing this equation coordinatewise and cancelling the common factor $N^{-1}$
gives
\begin{equation}
\label{eq:residual_laplacian_equation}
\sum_{i:k\in e_i}\sum_{l\in e_i\setminus\{k\}}
\omega_{i,kl}(b_k-b_l)
=
\sum_{i:k\in e_i}\sum_{l\in e_i\setminus\{k\}}
\omega_{i,kl}(X_{e_i,k,r}-X_{e_i,l,r}),
\qquad k\in[n].
\end{equation}
Thus $\bm b$ solves a discrete Poisson equation whose forcing term consists of
bounded covariate differences. The remainder of the proof converts this local
balance equation into a global oscillation bound.

\emph{Step 3: grow a level set from the maximizers of $\bm b$.}
Choose a sufficiently large fixed constant $c_{10}>0$ and define
\begin{equation*}
A_j=\left\{k\in[n]:\max_l b_l-b_k\leq jc_{10}\right\},
\qquad j=0,1,2,\ldots.
\end{equation*}
For a set $A\subset[n]$, let
\begin{equation*}
\partial_{\rm pair}A
=
\{(i,k,l):k,l\in e_i,\ k\in A,\ l\notin A\}
\end{equation*}
denote its boundary-pair occurrences. Thus $(i,k,l)$ records a pair in
comparison $i$, oriented so that $k\in A$ and $l\notin A$.
Summing~\eqref{eq:residual_laplacian_equation} over $k\in A_j$ cancels all
pairs internal to $A_j$ and yields
\begin{equation}
\label{eq:boundary_balance}
\sum_{(i,k,l)\in\partial_{\rm pair}A_j}
\omega_{i,kl}(b_k-b_l)
=
\sum_{(i,k,l)\in\partial_{\rm pair}A_j}
\omega_{i,kl}(X_{e_i,k,r}-X_{e_i,l,r}).
\end{equation}
Every term on the left is nonnegative, because $k\in A_j$ and
$l\notin A_j$ imply $b_k>b_l$. The absolute value of the right-hand side is at
most a fixed multiple of $\abs{\partial_{\rm pair}A_j}$. If a boundary pair has
$l\notin A_{j+1}$, then $b_k-b_l>c_{10}$. It follows from
\eqref{eq:boundary_balance}--\eqref{eq:weight_bounds} that
\begin{equation*}
c_{10}\abs{\{(i,k,l)\in\partial_{\rm pair}A_j:l\notin A_{j+1}\}}
\lesssim
\abs{\partial_{\rm pair}A_j}.
\end{equation*}
Taking $c_{10}$ larger than twice the implicit constant gives
\begin{equation}
\label{eq:half_boundary}
\abs{\{(i,k,l)\in\partial_{\rm pair}A_j:
l\in A_{j+1}\setminus A_j\}}
\geq
\frac12\abs{\partial_{\rm pair}A_j}.
\end{equation}

We now translate this pair count into the hyperedge count used in the
definition of weak admissibility. Every boundary hyperedge contains at least
one and at most $\binom{M}{2}$ boundary pairs. Therefore,
\begin{equation*}
\frac{
\abs{\{e\in\partial A_j:
e\cap(A_{j+1}\setminus A_j)\neq\emptyset\}}
}{\abs{\partial A_j}}
\geq
\frac{1}{2\binom{M}{2}}.
\end{equation*}
Choose the fixed $\eta$ in the main argument so that
$\eta\leq\{2\binom{M}{2}\}^{-1}$. Then $\{A_j\}$ is
$\eta$-weakly admissible. Moreover, connectivity ensures that
$A_{j+1}\supsetneq A_j$ whenever $A_j\neq[n]$: otherwise the left-hand side of
\eqref{eq:half_boundary} would be zero, forcing $\partial A_j=\emptyset$.

\emph{Step 4: bound the oscillation and the residualized blocks.}
Let $J$ be the first index such that $\abs{A_J}>n/2$. By weak admissibility,
\begin{equation*}
J\leq\diam(\mathcal A_{\mathcal H_n}(\eta)).
\end{equation*}
Applying the same argument to $-\bm b$ gives lower level sets
\begin{equation*}
B_j=\left\{k\in[n]:b_k-\min_l b_l\leq jc_{10}\right\},
\end{equation*}
and an index $J'\leq\diam(\mathcal A_{\mathcal H_n}(\eta))$ for which
$\abs{B_{J'}}>n/2$. Hence $A_J\cap B_{J'}\neq\emptyset$. For any
$k_0\in A_J\cap B_{J'}$,
\begin{equation*}
\max_l b_l-\min_l b_l
\leq
(J+J')c_{10}
\leq
2c_{10}\diam(\mathcal A_{\mathcal H_n}(\eta)),
\end{equation*}
which proves
\begin{equation*}
\max_k b_k-\min_k b_k
\lesssim
\diam(\mathcal A_{\mathcal H_n}(\eta)).
\end{equation*}

Finally, every entry in the $i$th comparison block of
$[\K_n]_r-\Q_n^\top\bm b$ has the form
\begin{equation*}
(X_{e_i,k,r}-X_{e_i,l,r})-(b_k-b_l)
\end{equation*}
for two items $k,l\in e_i$. Its absolute value is at most
\begin{equation*}
2\max_{i,k}\norm{X_{e_i,k}}_\infty
+\max_k b_k-\min_k b_k
\lesssim
\diam(\mathcal A_{\mathcal H_n}(\eta)).
\end{equation*}
Each block has at most $M-1=O(1)$ rows, so its Euclidean norm satisfies the
same bound. This proves both conclusions uniformly in $r$.
\end{proof}

\subsection{Proof of Lemma~\ref{lem:all_events}}
\label{proof:all_events}
The following elementary projection inequality will be used for the Hessian bounds.

\begin{lemma}
    \label{lem:projection_blocks}
Let $\bz$ have independent centered blocks, each with Euclidean norm at most $b$. For a deterministic linear subspace $V$ of dimension $m$ and every $t\ge1$, there is a universal constant $c_{11}$ such that
\begin{equation}
\label{eq:projection_blocks}
\Prob\left\{\|\operatorname{Proj}_V\bz\|_2>c_{11}b\sqrt{m+t}\right\} \le 2e^{-t}.
\end{equation}
\end{lemma}

\begin{proof}[Proof of Lemma~\ref{lem:projection_blocks}]
For every deterministic unit vector $\bm a$, the independent centered terms
$\bm a_i^\top\bz_i$ are bounded in absolute value by $b\|\bm a_i\|_2$.
Since $\sum_i\|\bm a_i\|_2^2=1$, Hoeffding's inequality gives
\begin{equation*}
\Prob\{|\bm a^\top\bz|>s\}\le2\exp\{-s^2/(2b^2)\},\qquad s>0.
\end{equation*}
A $1/2$-net of the unit sphere of $V$ has at most $5^m$ points, and
$\|\operatorname{Proj}_V\bz\|_2$ is at most twice the maximum of
$|\bm a^\top\bz|$ over this net. A union bound at
$s=(c_{11}/2)b\sqrt{m+t}$ proves~\eqref{eq:projection_blocks} for a sufficiently
large universal $c_{11}$. If $m=0$ or $b=0$, the assertion is immediate.
\end{proof}

\begin{proof}
All probability calculations below are conditional on $\mathscr D_n$.
We bound the failure probability of each event and then take a union bound.

\underline{The event $\G_1$.} 
Assumption~\ref{ass:wellposed}\ref{ass:exist} directly gives $\Prob(\G_1^c\mid\mathscr D_n)\le n^{-3}$.

\underline{The event $\G_2$.} 
We use the fact that only comparisons crossing a cut contribute to
its total utility score.
Fix a nonempty set $A\subset[n]$ with $|A|\le n/2$ and write
\begin{equation*}
N\sum_{k\in A}\partial_{u_k} l_n(\btheta^*)
=\sum_{i=1}^N Y_{i,A},
\qquad
Y_{i,A}=\sum_{k\in A}\partial_{u_k}\ell_i(\btheta^*).
\end{equation*}
Given $\mathscr D_n$, the variables $Y_{i,A}$ are independent and centered.
Shift invariance of the edgewise likelihood gives $Y_{i,A}=0$ whenever $e_i$
is contained in either $A$ or $A^c$. On each edge in $\partial A$, bounded
comparison size gives $|Y_{i,A}|\lesssim1$. Hence Hoeffding's inequality, with
the constant $c_{3}$ in~\eqref{G2} chosen sufficiently large, yields
\begin{equation*}
\Prob\left\{
\left|\sum_{i=1}^N Y_{i,A}\right|
>c_{3}\sqrt{|A|\,|\partial A|\log n}
\,\middle|\,\mathscr D_n
\right\}
\le 2n^{-6|A|}.
\end{equation*}
If $\partial A$ is empty, the sum is identically zero and the same conclusion
is immediate. A union bound over all eligible subsets gives
\begin{align}
\Prob(\G_2^c\mid\mathscr D_n)
&\le 2\sum_{a=1}^{\lfloor n/2\rfloor}\binom na n^{-6a}
\le 2\sum_{a=1}^{\lfloor n/2\rfloor}n^{-5a}
=O(n^{-5}).
\label{eq:G2_probability}
\end{align}

\underline{The event $\G_3$.} 
Stack the contrast scores
$\bm\xi_i=\nabla_{\bm s}\ell_i(\btheta^*)$ into $\bm\xi$.
Conditional on the design, these blocks are independent, centered, and
uniformly bounded. Let $\bm P$ be the orthogonal projection onto
$\operatorname{range}([\W_n]_{\mathcal I})$, whose dimension is at most $n+d_0$.
Lemma~\ref{lem:projection_blocks}, with $t=4\log(n+d_0)$, gives
\begin{equation*}
\Prob\left\{\|\bm P\bm\xi\|_2>c_{4}\sqrt{n+d_0+\log(n+d_0)}
\,\middle|\,\mathscr D_n\right\}\le2(n+d_0)^{-4}.
\end{equation*}
For every restricted direction $\bm w$, the chain rule and Cauchy--Schwarz give
\begin{equation*}
N|\nabla l_n(\btheta^*)^\top\bm w|
=|\langle\bm P\bm\xi,\W_n\bm w\rangle|
\le\|\bm P\bm\xi\|_2\|\W_n\bm w\|_2.
\end{equation*}
Consequently,
\begin{equation}\label{eq:G3_probability}
\Prob(\G_3^c\mid\mathscr D_n)\le2(n+d_0)^{-4}=O(n^{-4}).
\end{equation}

For the remaining events, we first record the common comparison-level
representation. In the proofs of $\G_4$ and $\G_5$, $r$ ranges over $S$, and $\bm w_{\mathrm{prof}}^{(r)}$ denotes the corresponding
active directions constructed in Section~\ref{proof:oracle_consistency}.
Since $\ell_i$ depends on $\btheta$ only through the contrast vector
$\W_i\btheta$, the chain rule gives
$\nabla\ell_i(\btheta)=\W_i^\top\nabla_{\bm s}\ell_i(\btheta)$ and
$\nabla^2\ell_i(\btheta)=\W_i^\top\nabla^2_{\bm s}\ell_i(\btheta)\W_i$.
Because $l_n=N^{-1}\sum_i\ell_i$, this yields the two identities needed here:
for any $\bm w\in\R^{n+d}$ and any $\bDelta$,
\begin{align}
\bm w^\top\bm Q
&=\frac1N\sum_{i=1}^N(\W_i\bm w)^\top
\nabla_{\bm s}\ell_i(\btheta^*),
\label{eq:score_blockwise}\\
\bm w^\top(\bm M-\bm M^*)\bDelta
&=\frac1N\sum_{i=1}^N(\W_i\bm w)^\top
\{\nabla^2_{\bm s}\ell_i(\btheta^*)
-\expect\nabla^2_{\bm s}\ell_i(\btheta^*)\}
(\W_i\bDelta).
\label{eq:hessian_blockwise}
\end{align}
Both quantities are averages of blockwise inner products in which $\bm w$
enters only through
$\W_i\bm w$, so that, for $\bm w=\bm w_{\mathrm{prof}}^{(r)}$, the design bounds~\eqref{eq:direction_design} are exactly
what is needed. Throughout we condition on the design. Conditional independence
of the rankings gives independence across $i$. The score terms are centered by
differentiating the normalization identity for the conditional likelihood, and
the Hessian-difference terms are centered by the definition of $\bm M^*$.
Finally, bounded comparison size makes both the contrast scores and contrast
Hessians uniformly bounded. Hence the summands in
\eqref{eq:score_blockwise} and~\eqref{eq:hessian_blockwise} satisfy
\begin{equation}\label{eq:blockwise_boundedness}
\|\nabla_{\bm s}\ell_i(\btheta^*)\|_2\lesssim1,
\qquad
\|\nabla^2_{\bm s}\ell_i(\btheta)
-\expect\nabla^2_{\bm s}\ell_i(\btheta^*)\|_{\mathrm{op}}\lesssim1
\quad\text{for all }\btheta .
\end{equation}

\underline{The event $\G_4$.}
By~\eqref{eq:score_blockwise} and Cauchy--Schwarz, the absolute value of the $i$th summand is $\lesssim\|\W_i\bm w_{\mathrm{prof}}^{(r)}\|_2$. The second bound in~\eqref{eq:direction_design} therefore bounds the sum of squared summand bounds by a fixed multiple of $N$. Hoeffding's inequality, with $c_{6}$ chosen sufficiently large, gives
\begin{equation*}
\Prob\left( |(\bm w_{\mathrm{prof}}^{(r)})^\top\bm Q|>c_{6}\sqrt{\frac{\log(n+d_0)}N}\,\middle|\,\mathscr D_n\right)
\le2(n+d_0)^{-5}.
\end{equation*}
A union bound over the at most $d_0$ active directions gives
\begin{equation*}
\Prob\{\G_4^c\mid\mathscr D_n\}
\le 2d_0(n+d_0)^{-5}=O(n^{-3}).
\end{equation*}
Only the second bound
in~\eqref{eq:direction_design} was used, so the score bound carries no diameter
factor.

\underline{The event $\G_5$.}
Let $\bm\xi\in\R^{\sum_i(m_i-1)}$ denote the vector obtained by stacking the $N$
blocks
\begin{equation*}
\{\nabla^2_{\bm s}\ell_i(\btheta^*)
-\expect\nabla^2_{\bm s}\ell_i(\btheta^*)\}\W_i\bm w_{\mathrm{prof}}^{(r)},
\qquad i=1,\dots,N,
\end{equation*}
so that~\eqref{eq:hessian_blockwise} reads
$(\bm w_{\mathrm{prof}}^{(r)})^\top(\bm M-\bm M^*)\bDelta=N^{-1}\langle\bm\xi,\W_n\bDelta\rangle$. The
blocks of $\bm\xi$ are independent, centered, and by
\eqref{eq:blockwise_boundedness} and the first bound
in~\eqref{eq:direction_design} satisfy
$\|\bm\xi_i\|_2\lesssim\|\W_i\bm w_{\mathrm{prof}}^{(r)}\|_2\lesssim\diam(\mathcal A_{\mathcal H_n}(\eta))$
for every $i$. Since $\bDelta_T=0$ we have
$\W_n\bDelta=[\W_n]_{\mathcal I}\bDelta_{\mathcal I}
\in\operatorname{range}([\W_n]_{\mathcal I})$, and therefore
\begin{equation}\label{eq:projection_identity}
\langle\bm\xi,\W_n\bDelta\rangle
=\langle\bm P\bm\xi,\W_n\bDelta\rangle,
\end{equation}
where $\bm P$ is the same restricted-design projection used for $\G_3$. Choosing $c_{7}$ sufficiently large and applying Lemma~\ref{lem:projection_blocks} to $\bm\xi$ with
$t=5\log(n+d_0)$ and taking a union bound over the $d_0$ active directions gives,
outside an event of conditional probability $O(n^{-3})$ and simultaneously
over the active directions,
\begin{equation*}
\|\bm P\bm\xi\|_2\le
c_{7}\diam(\mathcal A_{\mathcal H_n}(\eta))\sqrt{n+d_0+\log(n+d_0)} .
\end{equation*}
Combining this with~\eqref{eq:projection_identity} and Cauchy--Schwarz,
\begin{equation*}
|(\bm w_{\mathrm{prof}}^{(r)})^\top(\bm M-\bm M^*)\bDelta|
\le\frac1N\|\bm P\bm\xi\|_2\|\W_n\bDelta\|_2
\le c_{7}\diam(\mathcal A_{\mathcal H_n}(\eta))
\sqrt{\frac{n+d_0+\log(n+d_0)}N}\,
\frac{\|\W_n\bDelta\|_2}{\sqrt N},
\end{equation*}
This proves
$\Prob\{\G_5^c\mid\mathscr D_n\}=O(n^{-3})$.
The event is
determined by $\bm\xi$ and by the design alone; $\bDelta$ enters only through
the deterministic inequality~\eqref{eq:projection_identity} and
Cauchy--Schwarz. The bound is therefore valid simultaneously for every
restricted error vector, in particular for the data-dependent oracle error,
and no independence between $\bDelta$ and $\bm M$ is required.

\underline{The event $\G_6$.} 
Use the inactive directions $\bm w_{\mathrm{prof}}^{(k)}$, $k\in T$, from~\eqref{eq:pdw_direction}.
They satisfy the same average squared design bound, so the score calculation
above, with threshold $c_8\sqrt{\log(n+d)/N}$, and a union bound over the $d-d_0$ inactive coordinates give
\begin{equation*}
\Prob(\G_6^c\mid\mathscr D_n)\le 2(d-d_0)(n+d)^{-5}=O(n^{-3}).
\end{equation*}

\underline{The event $\G_7$.} 
The inactive directions also satisfy the same maximal block bound.
The projection calculation for $\G_5$ therefore applies with $\bm w_{\mathrm{prof}}^{(r)}$ replaced by $\bm w_{\mathrm{prof}}^{(k)}$ for each $k\in T$.
Taking $t=5\log(n+d)$ in the projection bound and a union bound over the inactive coordinates gives
\begin{equation*}
\Prob(\G_7^c\mid\mathscr D_n)\le 2(d-d_0)(n+d)^{-5}=O(n^{-3}).
\end{equation*}
If there are no inactive coordinates, both events hold automatically.

Finally, combining the seven failure bounds yields
\begin{equation*}
\Prob\left(\left(\bigcap_{j=1}^7 \G_j\right)^c
\,\middle|\,\mathscr D_n\right)
\le\sum_{j=1}^7\Prob(\G_j^c\mid\mathscr D_n)=O(n^{-3}).
\end{equation*}
The same argument restricted to $j=1,\ldots,5$ proves the first assertion
without incoherence. No independence among the events is required.
\end{proof}

\section{~Proof of Theorem~\ref{thm:randomcon}}
\label{app:random}

We discuss the assumptions of Theorem~\ref{thm:main_result} in order, followed by its sample-size condition. For Assumption~\ref{ass:incoherence}, we establish its required consequence, strict dual feasibility, directly from the independent noise features.

\begin{itemize}

\item \textbf{Assumption~\ref{ass:hypergraph_topology}: hypergraph connectivity.}
By the random-hypergraph results of \citet{dong2026statistical}, for the fixed $\eta$ used in the oracle analysis, the following bounds hold eventually almost surely:
\begin{align*}
\text{NURHM:}\quad
\diam(\mathcal A_{\mathcal H_n}(\eta))
&\lesssim\frac{\xi_{n,+}}{\xi_{n,-}}\log n,
&h_{\mathcal H_n}&\gtrsim\xi_{n,-},\\
\text{HSBM:}\quad
\diam(\mathcal A_{\mathcal H_n}(\eta))
&\lesssim\log n,
&h_{\mathcal H_n}&\gtrsim\zeta_{n,-}.
\end{align*}
No separate degree-concentration lemma is needed. Indeed, applying the definition of the Cheeger constant to singleton sets and summing gives
\begin{equation*}
n h_{\mathcal H_n}
\le\sum_{k=1}^n|\partial\{k\}|
=\sum_{e\in\mathcal E_n}|e|
\le MN.
\end{equation*}
Thus $N\gtrsim n\xi_{n,-}$ under NURHM and $N\gtrsim n\zeta_{n,-}$ under HSBM on the same graph events.

Substitution into Assumption~\ref{ass:hypergraph_topology} gives
\begin{equation*}
\frac{\diam^2(\mathcal A_{\mathcal H_n}(\eta))\log n}{h_{\mathcal H_n}}
\lesssim
\begin{cases}
\xi_{n,+}^2\log^3 n/\xi_{n,-}^3,&\text{NURHM},\\
\log^3 n/\zeta_{n,-},&\text{HSBM}.
\end{cases}
\end{equation*}
The right-hand side tends to zero under the corresponding scaling condition stated below, which verifies Assumption~\ref{ass:hypergraph_topology} on the graph events.

\item \textbf{Assumption~\ref{ass:unibound}: boundedness.}
The parameter bounds are retained explicitly in Theorem~\ref{thm:randomcon}. The active- and noise-feature conditions in Assumption~\ref{ass:random_design} provide the uniform coordinatewise covariate bound.

\item \textbf{Assumption~\ref{ass:wellposed}: oracle well-posedness.}
The following lemma records the dimension dependence needed to extend the fixed-dimensional argument to the active design.
\begin{lemma}
\label{lem:random_wellposed}
Conditional on a connected comparison hypergraph, suppose the active features satisfy Assumption~\ref{ass:random_design}. If
\begin{equation}
\label{eq:random_wellposed_scaling}
h_{\mathcal H_n}\gg d_0\log(d_0+1)+\log n,
\qquad N\gg nd_0,
\end{equation}
then Assumption~\ref{ass:wellposed} holds with probability
$1-O(n^{-3})$ for all sufficiently large $n$.
\end{lemma}

The scaling conditions in Assumption~\ref{ass:random_design}, together with $h_{\mathcal H_n}\gtrsim\xi_{n,-}$ and $N\gtrsim n\xi_{n,-}$ under NURHM, or $h_{\mathcal H_n}\gtrsim\zeta_{n,-}$ and $N\gtrsim n\zeta_{n,-}$ under HSBM, imply \eqref{eq:random_wellposed_scaling} because $1\le d_0<n$.  Thus Lemma~\ref{lem:random_wellposed} verifies oracle well-posedness under the stated random-design assumptions.

\item \textbf{Assumption~\ref{ass:incoherence}: strict dual feasibility.}
Under the random-design noise-feature condition, we verify strict dual feasibility directly instead of imposing the population incoherence inequality. Thus we use the oracle lemma and the PDW sufficient condition.
\begin{lemma}[Inactive oracle-score concentration]
\label{lem:random_dual}
Under the noise-feature condition in Assumption~\ref{ass:random_design} and bounded comparison size, conditional on the hypergraph, active covariates, and rankings for which the oracle estimator exists,
\begin{equation*}
\left\|\nabla_{\bv_{S^{*\complement}}}
 l_n(\hbtheta^{\rm oracle})\right\|_\infty
\le C\sqrt{\frac{\log(n+d)}N}
\end{equation*}
with probability at least $1-n^{-3}$, for a fixed constant $C>0$.
With $\lambda_n$ as in~\eqref{eq:pdw_lambda_choice} and $c_1$ sufficiently large, the resulting inactive dual vector satisfies
\begin{equation*}
\|\widehat{\bz}_{S^{*\complement}}\|_\infty\le\frac12<1
\end{equation*}
on the same event.
\end{lemma}

\item \textbf{Regularization and sample size.}
Choose $\lambda_n$ as in~\eqref{eq:pdw_lambda_choice}, with $c_1$ sufficiently large for Lemma~\ref{lem:random_dual}. For $1\le d_0<n$, the sample-size condition in Theorem~\ref{thm:main_result} is bounded by
\begin{align*}
\frac{d_0\diam^2(\mathcal A_{\mathcal H_n}(\eta))
\{n+d_0\log(n+d)\}}{N} \lesssim
\begin{cases}
\displaystyle\frac{d_0\xi_{n,+}^2\log^2 n}{\xi_{n,-}^3}
\left\{1+\frac{d_0\log(n+d)}n\right\},&\text{NURHM},\\
\displaystyle\frac{d_0\log^2 n}{\zeta_{n,-}}
\left\{1+\frac{d_0\log(n+d)}n\right\},&\text{HSBM}.
\end{cases}
\end{align*}
Consequently, given the graph estimates above, sufficient scaling conditions for the current theorem are
\begin{align}
\frac{\xi_{n,+}^2\log^2 n}{\xi_{n,-}^3}
\left\{\log n+d_0+\frac{d_0^2\log(n+d)}n\right\}&=o(1)
&&\text{under NURHM},\label{eq:random_nurhm_scaling}\\
\frac{\log^2 n}{\zeta_{n,-}}
\left\{\log n+d_0+\frac{d_0^2\log(n+d)}n\right\}&=o(1)
&&\text{under HSBM}.\label{eq:random_hsbm_scaling}
\end{align}
These are precisely the scaling conditions in Assumption~\ref{ass:random_design}. Together with the graph bounds, they verify the connectivity and sample-size requirements.

\end{itemize}

Together with well-posedness, the graph bounds, Lemma~\ref{lem:random_dual}, and Lemma~\ref{lem:pdw_condition} identify the oracle estimator with the unique full estimator. Substituting the graph bounds and $N\gtrsim nh_{\mathcal H_n}$ into the oracle rates gives, eventually almost surely,
\begin{align*}
\text{NURHM:}\quad
\|\hbv-\bv^*\|_2
&\lesssim\sqrt{d_0}\max\left\{
\sqrt{\frac{\log(n+d)}{n\xi_{n,-}}},
\frac{\xi_{n,+}\log n\{n+d_0+\log(n+d)\}}{n\xi_{n,-}^2}
\right\},\\
\|\hbu-\bu^*\|_\infty
&\lesssim\frac{\xi_{n,+}\log n}{\xi_{n,-}}
\max\biggl\{
\sqrt{\frac{\log n}{\xi_{n,-}}},
 d_0\sqrt{\frac{\log(n+d)}{n\xi_{n,-}}},\\
&\qquad\qquad
\frac{d_0\xi_{n,+}\log n\{n+d_0+\log(n+d)\}}{n\xi_{n,-}^2}
\biggr\},\\
\text{HSBM:}\quad
\|\hbv-\bv^*\|_2
&\lesssim\sqrt{d_0}\max\left\{
\sqrt{\frac{\log(n+d)}{n\zeta_{n,-}}},
\frac{\log n\{n+d_0+\log(n+d)\}}{n\zeta_{n,-}}
\right\},\\
\|\hbu-\bu^*\|_\infty
&\lesssim\log n\max\biggl\{
\sqrt{\frac{\log n}{\zeta_{n,-}}},
 d_0\sqrt{\frac{\log(n+d)}{n\zeta_{n,-}}},\\
&\qquad\qquad
\frac{d_0\log n\{n+d_0+\log(n+d)\}}{n\zeta_{n,-}}
\biggr\}.
\end{align*}
Both errors vanish under~\eqref{eq:random_nurhm_scaling} or~\eqref{eq:random_hsbm_scaling}, respectively. The eventual almost-sure statement follows from summability and the Borel--Cantelli lemma. The minimum signal condition in Proposition~\ref{prop:exact_support_recovery} then yields exact support recovery.

\begin{proof}[Proof of Lemma~\ref{lem:random_wellposed}]
We give the dimension-dependent parts of the argument.  Write
$[\K_n]_{S^*}$ as the stack of its independent comparison blocks.  Uniform
coordinatewise boundedness and bounded comparison size imply that the
operator norm of each block Gram matrix is $O(d_0)$.  Its expectation has
eigenvalues of order one after normalization by $N$, by the covariance bounds
in Assumption~\ref{ass:random_design}.  Matrix Chernoff bounds therefore give
\begin{equation*}
c\le \lambda_{\min}\!\left(
N^{-1}[\K_n]_{S^*}^{\top}[\K_n]_{S^*}\right)
\le \lambda_{\max}\!\left(
N^{-1}[\K_n]_{S^*}^{\top}[\K_n]_{S^*}\right)\le C
\end{equation*}
with failure probability at most $O(d_0e^{-cN/d_0})$.

For subspace separation, apply Lemma~\ref{lem:projection_blocks} to each of
the $d_0$ centered columns of $[\K_n]_{S^*}$ and take a union bound.  The
operator norm of its projection onto $\operatorname{range}(\Q_n^\top)$ is at
most its Frobenius norm, and hence is
$O\{\sqrt{d_0(n+\log n)}\}$.  Combining this with the preceding lower
singular-value bound yields
\begin{equation*}
\sup_{\substack{
\bm x\in\operatorname{range}(\Q_n^\top),\;
\bm y\in\operatorname{range}([\K_n]_{S^*})\\
\|\bm x\|_2=\|\bm y\|_2=1}}
|\bm x^\top\bm y|
\lesssim \sqrt{\frac{d_0(n+\log n)}{N}}=o(1).
\end{equation*}

It remains to prove the existence condition. Cover the unit sphere in $\mathbb R^{d_0}$ by a sufficiently fine net. Uniform boundedness of the active features implies that a net with
\[
\log|\mathcal N|\lesssim d_0\log(d_0+1)
\]
is sufficient to preserve the fixed margin in Assumption~\ref{ass:random_design}. For each net point and each vertex cut,the balancing condition and bounded true scores imply that every boundary comparison has a fixed positive probability of producing the required ordering. Independence across comparisons therefore bounds the failure probability by $\exp(-c|\partial U|)$. A union bound over all net points and vertex cuts is $O(n^{-6})$ because $h_{\mathcal H_n}\gg d_0\log(d_0+1)+\log n$.

The net argument extends the result to every direction in $\mathbb R^{d_0}$. Markov's inequality then converts the joint failure bound into the conditional bound required by Assumption~\ref{ass:wellposed}\ref{ass:exist}, outside a design event of probability $O(n^{-3})$. Together with the preceding spectral and subspace bounds, this proves Assumption~\ref{ass:wellposed}.
\end{proof}

\begin{proof}[Proof of Lemma~\ref{lem:random_dual}]
Let
\begin{equation*}
\mathscr F_n=\sigma\!\left(\mathcal H_n,\mathcal X_{n,S^*},\pi_1,\ldots,\pi_N\right).
\end{equation*}
Because $\bv^*_{S^{*\complement}}=\bm0$, the conditional distribution of the rankings depends only on the hypergraph and active covariates. Hence the inactive covariates remain independent, coordinatewise mean-zero, and uniformly bounded conditional on $\mathscr F_n$. The oracle estimator is $\mathscr F_n$-measurable because its inactive coefficients are fixed at zero.

For $k\in S^{*\complement}$, differentiation of the log-likelihood at the oracle estimator gives
\begin{equation}
\label{eq:random_inactive_score}
\partial_{v_k}l_n(\hbtheta^{\rm oracle})
=\frac1N\sum_{i=1}^N\sum_{a\in e_i}c_{ia}
[X_{e_i,a}]_k,
\end{equation}
where, conditional on $\mathscr F_n$, the coefficients $c_{ia}$ are fixed. Since the comparison size is bounded, the PL choice probabilities lie in $[0,1]$, and each incidence enters at most a fixed number of ranking stages, $|c_{ia}|\le C_M$ for a constant depending only on the maximum comparison size. The variables in~\eqref{eq:random_inactive_score} are independent across incidences under Assumption~\ref{ass:random_design}. Conditional Hoeffding's inequality therefore yields
\begin{equation*}
\Prob\!\left(
\left|\partial_{v_k}l_n(\hbtheta^{\rm oracle})\right|>t
\,\middle|\,\mathscr F_n
\right)
\le2\exp(-cNt^2).
\end{equation*}
A union bound over $k\in S^{*\complement}$, with
$t=C\sqrt{\log(n+d)/N}$ and $C$ sufficiently large, proves
\begin{equation*}
\norm{\nabla_{\bv_{S^{*\complement}}}l_n(\hbtheta^{\rm oracle})}_\infty
\lesssim\sqrt{\frac{\log(n+d)}{N}}
\end{equation*}
with conditional probability at least $1-n^{-3}$ on the oracle existence event. Finally,
the KKT condition for the covariate block of the full regularized problem is
\begin{equation*}
-\nabla_{\bv}l_n(\hbtheta^{\rm oracle})
+\lambda_n\widehat{\bz}=\bm0,
\qquad
\widehat{\bz}\in\partial\norm{\hbv^{\rm oracle}}_1.
\end{equation*}
The inactive dual coordinates are defined by extending the preceding KKT equation to
$S^{*\complement}$:
\begin{equation*}
\widehat{\bz}_{S^{*\complement}}
=\frac{1}{\lambda_n}
\nabla_{\bv_{S^{*\complement}}}l_n(\hbtheta^{\rm oracle}).
\end{equation*}
Consequently,
\begin{equation*}
\norm{\widehat{\bz}_{S^{*\complement}}}_\infty
=\frac{\norm{\nabla_{\bv_{S^{*\complement}}}l_n(\hbtheta^{\rm oracle})}_\infty}{\lambda_n}
\le\frac{C}{c_1}\le\frac12<1,
\end{equation*}
where we used $\lambda_n\ge c_1\sqrt{\log(n+d)/N}$ and chose $c_1\ge2C$. This proves strict dual feasibility.
\end{proof}

%\clearpage
\bibliographystyle{JASA-style}
\bibliography{paper-ref}
\end{document}